\documentclass[10pt,aps,prx,twocolumn,superscriptaddress]{revtex4-2}
\usepackage{braket}
\usepackage{amsfonts, bm, mathtools}
\usepackage{booktabs}
\usepackage{siunitx}
\usepackage{xcolor}
\usepackage{xurl}
\usepackage[breaklinks=true,colorlinks,citecolor=blue,linkcolor=blue,urlcolor=blue]{hyperref}
\usepackage{txfonts}
\usepackage{orcidlink}

\newcommand{\trace}[1]{\text{tr}#1}
\newcommand{\eqlabel}[1]{Eq.~\eqref{#1}}
\newcommand{\figlabel}[1]{Fig.~\ref{#1}}
\newcommand{\supnote}[1]{Supplementary Note #1}
\newcommand{\seclabel}[2]{\hyperref[#1]{#2}%
}
\newcommand{\tablabel}[1]{Table~\ref{#1}}
\DeclareMathOperator{\sign}{sgn}



\begin{document}

\title{Bias-field digitized counterdiabatic quantum algorithm for higher-order binary optimization}
\author{Sebastián V. Romero$^{\orcidlink{0000-0002-4675-4452}}$}
\affiliation{Kipu Quantum GmbH, Greifswalderstrasse 212, 10405 Berlin, Germany}
\affiliation{Department of Physical Chemistry, University of the Basque Country UPV/EHU, Apartado 644, 48080 Bilbao, Spain}

\author{Anne-Maria Visuri$^{\orcidlink{0000-0002-4167-7769}}$}
\affiliation{Kipu Quantum GmbH, Greifswalderstrasse 212, 10405 Berlin, Germany}

\author{Alejandro Gomez Cadavid$^{\orcidlink{0000-0003-3271-4684}}$}
\affiliation{Kipu Quantum GmbH, Greifswalderstrasse 212, 10405 Berlin, Germany}
\affiliation{Department of Physical Chemistry, University of the Basque Country UPV/EHU, Apartado 644, 48080 Bilbao, Spain}

\author{Anton Simen$^{\orcidlink{0000-0001-8863-4806}}$}
\affiliation{Kipu Quantum GmbH, Greifswalderstrasse 212, 10405 Berlin, Germany}
\affiliation{Department of Physical Chemistry, University of the Basque Country UPV/EHU, Apartado 644, 48080 Bilbao, Spain}

\author{Enrique Solano$^{\orcidlink{0000-0002-8602-1181}}$}
\affiliation{Kipu Quantum GmbH, Greifswalderstrasse 212, 10405 Berlin, Germany}

\author{Narendra N. Hegade$^{\orcidlink{0000-0002-9673-2833}}$}
\email[]{narendrahegade5@gmail.com}
\affiliation{Kipu Quantum GmbH, Greifswalderstrasse 212, 10405 Berlin, Germany}
\date{\today}

\begin{abstract}
 Combinatorial optimization plays a crucial role in many industrial applications. While classical computing often struggles with complex instances, quantum optimization emerges as a promising alternative. Here, we present an enhanced bias-field digitized counterdiabatic quantum optimization (BF-DCQO) algorithm to address higher-order unconstrained binary optimization (HUBO). We apply BF-DCQO to a HUBO problem featuring three-local terms in the Ising spin-glass model, validated experimentally using 156 qubits on an IBM quantum processor. In the studied instances, our results outperform standard methods such as the quantum approximate optimization algorithm, quantum annealing, simulated annealing, and Tabu search. Furthermore, we provide numerical evidence of the feasibility of a similar HUBO problem on a 433-qubit Osprey-like quantum processor. Finally, we solve denser instances of the MAX 3-SAT problem in an IonQ emulator. Our results show that BF-DCQO offers an effective path for solving large-scale HUBO problems on current and near-term quantum processors.
\end{abstract}

\maketitle

\section*{Introduction}

Combinatorial optimization problems arise in a multitude of situations in science and industry, from logistics and scheduling to computational chemistry and biology. In combinatorial optimization, the best or near-optimal solution is searched within a finite but large discrete configuration space.
Theoretical tools from statistical physics have given insights into these problems, as their connection to disordered systems was identified early on~\cite{fu1986application}.
These problems generally belong to the NP-hard complexity class, requiring a computation time that grows exponentially with the problem size on classical computers. 
While complex problem instances test the limits of classical computing, quantum computing has emerged as a suitable counterpart to drive the state of the art in this field. This is possible due to the rapid development of quantum technologies combined with the natural mapping of combinatorial optimization problems to the widely studied Ising spin glass model, whose ground state encodes the optimal solution~\cite{lucas2014ising}. Several methods have been proposed so far, including adiabatic quantum optimization~\cite{albash2018adiabatic} and the quantum approximate optimization algorithm (QAOA)~\cite{farhi2014quantumapproximateoptimizationalgorithm}, which can be thought as a digitized version of adiabatic quantum computing~\cite{barends2016digitized}.

In addition to the capacity of quantum hardware to embed large-scale systems, the potential of quantum computing to surpass classical optimization methods has been demostrated through dedicated quantum algorithms~\cite{durr1999quantumalgorithmfindingminimum, montanaro2018quantum-walk, montanaro2020quantumspeedup, chakrabarti2022universalquantumspeedupbranchandbound, somma2008quantum, wocjan2008speedup, hastings2018shortpathquantum, dalzell2023mind}. Among these, QAOA has emerged as one of the front-runner candidates for showing quantum speedup in optimization. Recent studies on 8-SAT problems with up to 20 variables~\cite{boulebnane2022solvingbooleansatisfiabilityproblems} show that the time to solution for QAOA outperforms best-known classical methods. Similar evidence of quantum speedup for solving the low autocorrelation binary sequence (LABS) problem has also been found~\cite{boehmer1967binary,schroeder1970synthesis,shaydulin2024evidence}.

In parallel with quantum algorithms, quantum-inspired algorithms run on classical hardware, especially those based on tensor networks (TNs)~\cite{orus2014practical, silvi2019tensor, ran2020tensor, cirac2021matrix, banuls2023tensor}, have evolved rapidly in the recent years.
They were developed over the past three decades mostly for the study of condensed-matter physics~\cite{orus2019tensor, cirac2021matrix} and have been adopted as efficient simulators of quantum hardware, as classical benchmarks for quantum algorithms, and in hybrid approaches combining TNs and quantum algorithms~\cite{lin2021real, rudolph2023synergistic, martin2024combining, berry2024rapid}. 
The recent kicked-Ising-model experiment of IBM on a 127-qubit device~\cite{kim2023evidence, tindall2024efficient, begusic2024fast, patra2024efficient, liao2023simulation} and D-Wave's simulation of quantum quench dynamics across a spin-glass phase transition~\cite{king2023quantum, king2024computationalsupremacyquantumsimulation} highlighted the interplay of quantum simulations and TNs.
For combinatorial optimization, new approaches have combined generative machine learning models with TNs~\cite{han2018unsupervised, lopez2023symmetric, alcazar2024enhancing}, and ground-state search methods have been applied to problems with up to quadratic terms on higher-dimensional graphs~\cite{mugel2022dynamic, hao2022quantum, patra2024projected}.

Compared to quadratic problems, higher-order ones are significantly more difficult to solve. While a number of experiments have recently demonstrated higher-order unconstrained binary optimization (HUBO) implementations on quantum hardware~\cite{pelofske2023quantum, pelofske2024short-depth,barron2023provableboundsnoisefreeexpectation,sachdeva2024quantum}, there are hardware limitations that complicate the effective implementation of these protocols on large- and even intermediate-scale problems. These include short decoherence times, qubit connectivities, and the presence of noise. Several alternative methods, such as counterdiabatic driving (CD) protocols, have been proposed to overcome these issues~\cite{demirplak2003adiabatic,berry2009transitionless,chen2010fast,campo2013shortcuts,chandarana2022digitized,hegade2022digitized}. By suppressing diabatic transitions, CD protocols can find better solutions in different scenarios.

In this work, we develop and exploit an improved version of the recently introduced bias-field digitized counterdiabatic quantum optimization (BF-DCQO) formulation~\cite{cadavid2024biasfielddigitizedcounterdiabaticquantum} to address HUBO problems on both ideal simulators and hardware. We make use of the entire new 156-qubit IBM quantum computer \textsc{ibm\_fez} and a matrix product state (MPS) simulation of the upcoming IBM Osprey processor with 433 qubits. To test BF-DCQO on denser instances, we consider a MAX 3-SAT problem at the critical density and employ the 25-qubit \textsc{IonQ Aria 1} emulator with a noise model. All results are compared with those obtained through well-established classical optimization methods as well as the D-Wave quantum annealing platform. Unlike these methods, BF-DCQO does not require additional qubits for mapping the initial HUBO problem into a quadratic one. Even though our work addresses problems with up to three-body terms, our method is equally applicable to higher-order optimization problems, including the LABS problem~\cite{shaydulin2024evidence}, MAX $k$-SAT~\cite{Battiti2009}, factorization~\cite{hegade2021factorization} and protein folding~\cite{robert2021resource, chandarana2024digitized}, among others.

\section*{Methods}\label{sec:methods}

\subsection*{Digitized counterdiabatic quantum optimization (DCQO)}

\begin{figure*}[!tb]
    \centering
    \includegraphics[width=\linewidth]{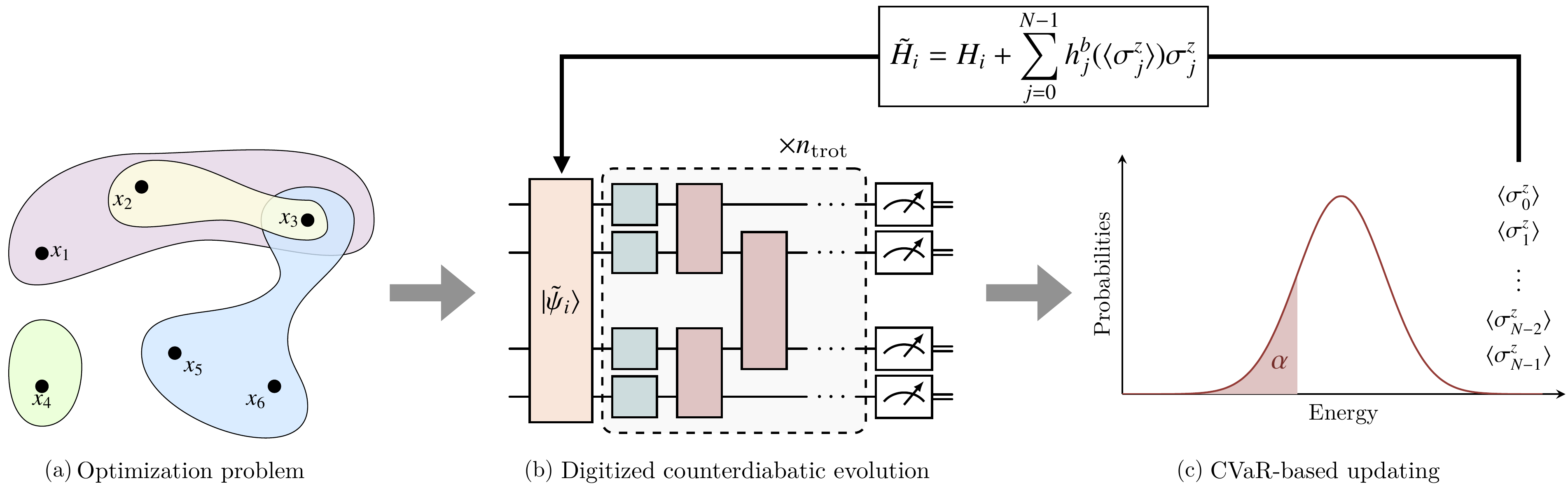}%
    \caption{\textbf{Schematic of the BF-DCQO protocol.} Given a binary optimization problem of arbitrary order (depicted in \textbf{a}) encoded as a $p$-spin glass [\eqlabel{eq:hubo}], the BF-DCQO algorithm performs DCQO iteratively, updating the input bias $h_j^{b}$ from the solutions obtained from the previous step. Each iteration consists of running the corresponding DCQO circuit (depicted in \textbf{b}), sampling the final state, and updating the bias fields. It amounts to the use of (i) the CVaR-based approach, which considers only a subset of the low-energy samples, and (ii) the updating strategy as a function of the expectation value, i.e., $h^b_j(\braket{\sigma^z_j})$ (depicted in \textbf{c}).}\label{fig:schematic} 
\end{figure*}%

Adiabatic quantum optimization aims to prepare the ground state of a given problem Hamiltonian. An initial state is evolved within the time window $t\in[0,T]$ under the adiabatic protocol $H_\text{ad}(\lambda)=(1-\lambda)H_i + \lambda H_f$. Here, $\lambda(t)$ is a time-dependent scheduling function that describes the adiabatic evolution of the system, starting from an initial Hamiltonian $H_i$, whose corresponding ground state is easy to prepare, towards the problem Hamiltonian $H_f$. Relying on the adiabatic theorem, this protocol reaches the desired ground state of $H_f$ in the adiabatic limit $\dot{\lambda}(t)\to 0$. We consider the disordered Ising Hamiltonian with up to three-body terms:
\begin{equation}\label{eq:hubo}
    H_f=\sum_i h_i^z \sigma^z_{i} + \sum_{i<j} J_{ij} \sigma^z_{i} \sigma^z_{j}+ \sum_{i<j<k} K_{ijk} \sigma^z_{i} \sigma^z_{j} \sigma^z_{k},
\end{equation}
in contrast to Ref.~\cite{cadavid2024biasfielddigitizedcounterdiabaticquantum}, which only considers up to quadratic terms. We use here natural units with $\hbar=1$. The ground state of this Hamiltonian is called a $p$-spin glass with $p=3$~\cite{gardner1985spin}, and the model is commonly known in quantum optimization as the higher-order Ising chain~\cite{pelofske2023quantum,pelofske2024short-depth}. In the following, we use $\lambda(t)=\sin^2(\pi\sin^2(\pi t/2T)/2)$ as the scheduling function. The initial Hamiltonian is
\begin{equation}
    H_i = \sum_j h_j^x \sigma^x_{j} + \sum_j h^b_j \sigma^z_{j}
\end{equation}
with $h_j^x$ ($h_j^b$) the transverse- (longitudinal bias-) field contributions acting on the $j$th spin. We set $h_j^x=-1$ and $h^b_j=0$, thus $H_i=-\sum_{j=0}^{N-1} \sigma^x_j$, so that the $N$-qubit ground state of $H_i$ is
\begin{equation}
    \ket{+}^{\otimes N} = \left( \frac{\ket{0} + \ket{1}}{\sqrt{2}} \right)^{\otimes N}.
\end{equation}

In order to overcome the intrinsic slow adiabatic evolution, it is possible to introduce an auxiliary counterdiabatic driving contribution which accelerates the original protocol and suppresses diabatic transitions~\cite{demirplak2003adiabatic,berry2009transitionless}. The transitionless protocol takes the form 
\begin{equation}\label{eq:counterdiabatic_hamiltonian}
 H_\text{cd}(\lambda)=H_\text{ad}(\lambda) + \dot{\lambda}A_{\lambda},
\end{equation}
where $A_{\lambda}$ is the adiabatic gauge potential~\cite{kolodrubetz2017geometry}. Nevertheless, its exact implementation is severely impractical due to its many-body structure and the necessity for the knowledge of the full energy spectrum. To overcome these issues, several approximate implementations have been proposed~\cite{kolodrubetz2017geometry,sels2017minimizing,claeys2019floquet,hatomura2021controlling,takahashi2024shortcuts}. One such proposal is the approximation of the adiabatic gauge potential by a nested-commutator series expansion 
\begin{equation}\label{eq:nc}
 A_\lambda^{(l} = i\sum_{k=1}^l \alpha_k(t)\mathcal{O}_{2k-1}(t),
\end{equation}
where $\mathcal{O}_0(t) = \partial_\lambda H_\text{ad}$ and $\mathcal{O}_k(t) = [H_\text{ad}, \mathcal{O}_{k-1}(t)]$. In the limit $l\to\infty$, this expansion converges to the exact gauge potential. The coefficients $\alpha_k$ are obtained by minimizing the action $S_l=\trace{[G_l^2]}$ with $G_l=\partial_\lambda H_\text{ad} - i\big[H_\text{ad},A^{(l}_\lambda\big]$. For simplicity, we consider the first-order ($l=1$) nested-commutator term (see~\supnote{1} for a step-by-step derivation), which takes the form
\begin{equation}\label{eq:1nc_general}
    \begin{split}
        \mathcal{O}_1 &=-2i\Big[\sum_{i} h_{i}^{z} \sigma^y_{i} + \sum_{i<j} J_{i j}(\sigma^y_{i} \sigma^z_{j}+ \sigma^z_{i} \sigma^y_{j}) \\
        &+ \sum_{i<j<k}K_{ijk}(\sigma^y_{i}\sigma^z_{j}\sigma^z_{k} + \sigma^z_{i}\sigma^y_{j}\sigma^z_{k} + \sigma^z_{i}\sigma^z_{j}\sigma^y_{k})\Big].
    \end{split}
\end{equation}
Higher-order terms demand significantly more computational resources, and their complexity increases with the system size.

In the fast evolution regime, the adiabatic term \( H_\text{ad}(\lambda) \) in~\eqlabel{eq:counterdiabatic_hamiltonian} can be omitted, reducing the number of required quantum gates and thereby decreasing hardware noise, while still maintaining the quality of the solution~\cite{romero2024optimizing, chandarana2024digitized, cadavid2024biasfielddigitizedcounterdiabaticquantum, dalal2024digitizedcounterdiabaticquantumalgorithms}. This approach is adopted in the present study, where only the CD contribution is considered without compromising performance. The time evolution of such Hamiltonians remains a challenge on current analog quantum platforms, and may even be highly difficult due to the nonstoquasticity~\cite{hormozi2017nonstoquastic}. To address this issue, digitized counterdiabatic quantum protocols have been proposed for implementation on digital quantum computers~\cite{hegade2021shortcuts}. The resulting time-evolution operator can be decomposed using a first-order Trotter-Suzuki decomposition~\cite{SUZUKI1990319} and digitized with a gate-based approach,
\begin{equation}\label{eq:trotter}
    U(T,0) = \prod_{k=1}^{n_\text{trot}} \prod_{j=1}^{n_\text{terms}} \exp\left[-i\gamma_j(k\Delta t)\Delta tH_j\right],
\end{equation}
where $H_\text{cd}(t)=\sum_{j=1}^{n_\text{terms}}\gamma_j(t)H_j$ is encoded in $n_\text{terms}$ different Pauli operators $H_j$ and $n_\text{trot}$ are the Trotter steps considered with $\Delta t = T/n_\text{trot}$. We use $T=1$ in units of the inverse energy scale relevant for each problem as defined below, and $n_\text{trot}=3$, thus one effective Trotter step is implemented. The number of gates required for implementation is reduced by optimally decomposing the desired circuit following Refs.~\cite{Sriluckshmy_2023, Algaba2024lowdepthsimulations}. Furthermore, we set a gate-cutoff threshold $\theta_\text{cutoff}$ in our experiments, such that any term of~\eqlabel{eq:trotter} with an angle with absolute value below this threshold is discarded, thus if $|\gamma_j(k\Delta t)\Delta t|\text{ mod }2\pi < \theta_\text{cutoff}$ is met. The larger the $\theta_\text{cutoff}$, the lower is the amount of resources that are required, but the obtained circuit expressivity is also lower. Therefore, it is crucial to select an appropriate value of this parameter to significantly reduce the amount of gates to implement without compromising the quality of the outcomes.

\subsection*{Bias-field updating protocol}\label{sec:update}

State initialization is crucial for the performance of quantum optimization routines, where an initial state with a nonzero overlap with the desired final state is advantageous~\cite{grass2019quantum,grass2022quantum}. Several warm-starting techniques have been proposed, where a relaxed version of the problem is solved classically and its solution fed into the quantum algorithm, leveraging the complexity and cost of the quantum routine while ensuring a better performance~\cite{Egger2021warmstartingquantum,truger2024warm}.

In BF-DCQO, the standard DCQO is performed iteratively, so that the solution from each step serves as the input bias for the next iteration~\cite{hegade2022digitized, cadavid2024biasfielddigitizedcounterdiabaticquantum, grass2019quantum}. See~\figlabel{fig:schematic} for a schematic diagram of the protocol.
The initial Hamiltonian is changed after each bias-field update, 
\begin{equation}
    \tilde{H}_i = H_i + \sum_{j=0}^{N-1} h^b_j(\braket{\sigma^z_j})\sigma^z_j,
\end{equation}
so that its ground state also changes and has to be prepared for the circuit execution. The smallest eigenvalue of the single-body operator \( \left[h_i^x \sigma^x_i - h_i^b \sigma^z_i \right]  \) is given by \( \lambda^{\min}_i = -\sqrt{(h^b_i)^2 + (h^x_i)^2} \), and its associated eigenvector is \( \ket{\tilde{\phi}}_i = R_y(\theta_i) \ket{0}_i \) where \( \theta_i = 2\tan^{-1}\left(\frac{h^b_i + \lambda_{min}}{h^x_i}\right) \). Therefore, the ground state of \( \tilde{H_i} \) can be prepared using $N$ y-axis rotations as 
\begin{equation}
\ket{\tilde{\psi_i}} = \bigotimes_{i=1}^{N} \ket{\tilde{\phi}}_i = \bigotimes_{i=1}^{N} R_y(\theta_i)\ket{0}_i.
\end{equation}
 
In each iteration, the bias field \( h^b_j \) for the \( j \)th qubit is updated based on the measured longitudinal component of the qubit \( \braket{\sigma^z_j} \). This update follows the rule \( h^b_j = h^b_j(\braket{\sigma^z_j}) \), where the functional form of \( h^b_j \) represents the specific bias-field strategy employed. In this work, we explore the following strategies:
\begin{itemize}
    \item Unsigned antibias($+$)/bias($-$): $h^b_j = \pm\braket{\sigma^z_{j}}$. This strategy offers bias fields whose strength is exactly the longitudinal component. This is a sensible choice when a qubit is effectively close to an equal superposition, such that there is no clear direction of bias.
    \item Signed antibias($+$)/bias($-$): $h^b_j = \pm\sign\braket{\sigma^z_{j}}$. This strategy offers bias fields that are equivalent to the effective orientation of the qubit, regardless the magnitude. This is a sensible choice when a qubit has a clear preferred direction of bias.
\end{itemize}

Once a suitable updating method has been selected, BF-DCQO iteratively measures the final state after time evolution in the computational basis and the longitudinal bias fields $h^b_j$ are updated accordingly. In the following, we set $h^b_j=0$ for the first bias-field iteration. According to the findings of Ref.~\cite{cadavid2024biasfielddigitizedcounterdiabaticquantum}, modifying the selected updating strategy throughout the entire routine can circumvent local minima and lead to faster convergence. For our results, we use the unsigned bias followed by a final iteration with a weighted signed bias. Specifically, the strength of $h^b_j$ is increased by rescaling $h^b_j \mapsto \kappa h^b_j$ with $\kappa>0$. In our experiments, we rescale by a factor of five to ensure that the magnitude of the bias is significantly larger than the transverse term, restricting the search space near the previous solution. Since the solution provided by DCQO, from the first iteration, samples low-energy states of the spin-glass Hamiltonian [\eqlabel{eq:hubo}], the expectation values of local Pauli-$z$ operators can be used as an effective bias in subsequent iterations, facilitating convergence. 

Although this updating strategy is employed in our experiments, alternative approaches might be more convenient depending on the nature of the problem at hand. Using $\kappa\in(0,1)$ grants more importance to the transverse contribution in $\tilde{H}_i$, whose ground state is an equal superposition of the computational basis. Thereby, a broader energy landscape is explored by flattening the expected sampling distribution. The algorithm is less prone to getting stuck in local minima with this choice, though it may significantly slow down convergence. On the other hand, using $\kappa>1$ provides greater strength to the longitudinal bias-field update, i.e. the optimal solution obtained from the last BF-DCQO iteration, constraining the search space around this solution. This accelerates the convergence but may lead to an undesired local minimum.

Once the circuit has been executed, all qubits are measured $n_\text{shots}$ times, returning an $N$-bitstring $x_0\dots x_{N-1}$. Each measurement trivially translates into a sample of $H_f$ due to its construction as a weighted sum of Pauli-$z$ operators. Denoting each of these samples as $E_k$, the sample mean energy is defined as $E= (1/n_\text{shots})\sum_{k=1}^{n_\text{shots}} E_k$. It can be used as an estimator of the expectation value of $H_f$ after ground-state preparation. 

It has been demonstrated that incorporating Conditional Value-at-Risk (CVaR) techniques into quantum optimization routines~\cite{Barkoutsos2020improving} can lead to significant improvements. Recent studies have explored the impact of noise when sampling bitstrings from quantum computers. They determined upper and lower bounds for noiseless expectation values using CVaR noisy samples and quantified the sampling overhead needed to achieve good solutions~\cite{barron2023provableboundsnoisefreeexpectation}. Building upon this knowledge, we propose a CVaR-inspired method for iteratively updating the bias fields. Accordingly, let $\alpha$ be the ratio between the number of measurements considered and the total number of shots $n_\text{shots}$. Rather than taking the entire distribution of samples into account ($\alpha=1$), we first sort them in terms of energy so that $E_k\le E_{k+1}$. Next, the bias is updated using a fraction $0<\alpha<1$ of the lowest-energy outcomes taken from
\begin{equation}
 E(\alpha)= \frac{1}{\lceil \alpha n_\text{shots}\rceil}\sum_{k=1}^{\lceil \alpha n_\text{shots}\rceil} E_k,
\end{equation}
thus taking only a part of the total distribution into account. From the results drawn from both noiseless simulators and hardware, we find the fastest convergence to the best solutions for values of $\alpha\sim 1\%$. This value is found empirically and the optimal value will depend on the problem and the number of shots. For example, we find that using smaller (larger) values of $\alpha$ shows faster convergence if the number of shots is large (small) with respect to the numbers used here.

Apart from seeking a reduction in the resources required and depth of the circuits considered, it is important to quantify the quality of the results obtained by the routines. In order to evaluate the performance of our routines, we use two metrics: the approximation ratio
\begin{equation}\label{eq:AR}
    \text{AR} = \frac{E(\alpha=1)}{E_0},
\end{equation}
and the distance to solution
\begin{equation}\label{eq:DS}
    \text{DS} = 1 - \frac{\min\{E_k\}_{k=1}^{n_\text{shots}}}{E_0}.
\end{equation}
In all our studies of 156 and 433 qubits, the exact ground-state energies $E_0$ are obtained numerically by an iterative method similar to Ref.~\cite{gardner1985zero} (see~\supnote{2} for details). For 25-qubit instances, we obtain the exact solution using a brute force method.

\subsection*{HUBO instances considered}

In this section, we present the HUBO Hamiltonians that are employed for testing BF-DCQO. Based on the gate error rates and on the coupling map connectivities of the selected platforms, we work with a nearest-neighbour three-body spin glass. In order to tackle industry-related use cases, we furthermore consider the weighted MAX 3-SAT problem~\cite{Battiti2009}. 
To the best of our knowledge, this is the first work to tackle a binary satisfiability problem with the aid of CD protocols. 
For both cases, the corresponding circuit decomposition is shown in~\supnote{3}. 

\emph{Nearest-neighbour spin glass.}---For the first case, the nearest-neighbour (NN) three-body Hamiltonian takes the form of
\begin{equation}\label{eq:nearest_hubo}
    H^\text{NN}_f =\sum_i h_i^z \sigma^z_{i} + \sum_{\braket{ij}} J_{ij} \sigma^z_{i} \sigma^z_{j}+ \sum_{\braket{ijk}} K_{ijk} \sigma^z_{i} \sigma^z_{j} \sigma^z_{k},
\end{equation}
where $\braket{\cdot}$ indicates that the enclosed qubit indices are nearest neighbours in the one-dimensional chain (thus $j=i+1$ and $k=i+2$). The coupling constants $h_i$, $J_{ij}$ and $K_{ijk}$ are taken as random Gaussian-distributed numbers with zero mean and unit variance. The first-order nested-commutator expansion for this Hamiltonian is given by~\eqlabel{eq:1nc_general} but it only contains the nearest-neighbour qubit indices as in~\eqlabel{eq:nearest_hubo}. Due to its linearity, it perfectly fits the heavy-hex IBM qubit coupling maps~\cite{chamberland2020topological}. Furthermore, the ease with which this and related models can be simulated using tensor networks makes their simulation an ideal benchmark for evaluating the performance of quantum platforms.
In addition to the 156-qubit experiment, we conduct a similar test consider a 433-qubit NN HUBO instance with the objective of simulating ideally (without noise) the expected performance of the upcoming IBM Osprey quantum platform via MPS.

\emph{Weighted MAX 3-SAT problem.}---The maximum $k$-satisfiability (MAX $k$-SAT) problem belongs to the NP-complete complexity class~\cite{impagliazzo2001complexity} and its weighted variant is defined as follows: given a Boolean formula in conjunctive normal form (thus conjunction of disjuncted clauses), the weighted maximum $k$-satisfiability problem (MAX W-$k$-SAT) aims to find an assignment of truth-valued variables that maximizes the sum of weights of all the $k$-variable clauses met~\cite{Battiti2009}. The MAX $k$-SAT problem can be recovered simply by setting the weights equal to one. If we let $k=3$, this problem can be written as
\begin{equation}
    C^\text{MW3S}(l) = \bigwedge_{ijk}\omega_{ijk}l_i\vee l_j \vee l_k,
\end{equation}
with $l_i$ a literal representing a propositional variable, either $u_i$ or its negation $\bar{u}_i$, and $\omega_{ijk}$ are the weights. The problem can be mapped into binary variables by performing the substitution $u_i\mapsto 1-x_i$ and $\bar{u}_i\mapsto x_i$ with $x_i\in\{0,1\}$. For each clause, the OR product of literals can be expanded into products of binary variables, e.g.
\begin{equation}
 u_i\vee \bar{u}_j \vee u_k\mapsto (1-x_i)x_j(1-x_k).   
\end{equation}
In order to address this problem with the aid of quantum computers, we perform an additional map turning the binary variables into Ising variables using the transformation $x_i\mapsto (I-\sigma^z_{i})/2$.

Setting $c_i=1$ if $l_i$ is negated and $c_i=0$ otherwise, we finally define the problem Hamiltonian as
\begin{equation}\label{eq:mw3s}
    H^\text{MW3S}_f = \sum_{ijk} \frac{\omega_{ijk}}{8} [I+(-1)^{c_i}\sigma^z_{i}][I+(-1)^{c_j}\sigma^z_{j}][I+(-1)^{c_k}\sigma^z_{k}].
\end{equation}
Here, $\omega_{ijk}$ is chosen as a random uniformly distributed variable between zero and one. As this is a maximization problem, it can be transformed into a minimization one by performing the map $H^\text{MW3S}_f \mapsto -H^\text{MW3S}_f$. The first-order nested-commutator of~\eqlabel{eq:mw3s} yields
\begin{equation}
\begin{split}
    \mathcal{O}^\text{MW3S}_1 &= -i\sum_{ijk} \frac{\omega_{ijk}}{4} \Big[(-1)^{c_i} \sigma^y_{i}[I+(-1)^{c_j}\sigma^z_{j}][I+(-1)^{c_k}\sigma^z_{k}] \\
    &+ (-1)^{c_j} [I+(-1)^{c_i}\sigma^z_{i}]\sigma^y_{j}[I+(-1)^{c_k}\sigma^z_{k}] \\
    &+ (-1)^{c_k} [I+(-1)^{c_i}\sigma^z_{i}][I+(-1)^{c_j}\sigma^z_{j}]\sigma^y_{k}\Big].
\end{split}
\end{equation}

It can be seen that MAX W-3-SAT is in general a highly nonlocal Hamiltonian and thus is challenging to implement. We consider here the NN version of the MAX W-3-SAT Hamiltonian of~\eqlabel{eq:mw3s}, where consecutive indices are summed over as in~\eqlabel{eq:nearest_hubo}. This model is similar to the nearest-neighbour spin-glass Hamiltonian $H^\text{NN}_f$ but with additional next-to-nearest neighbour terms, namely $\sigma^z_i\sigma^z_{i+2}$, which makes it more complex to implement. In essence, this is due to the fact that such terms are seen as three-body interactions for a heavy-hex lattice, where an additional number of entangling gates are needed to couple distant qubits.

In contrast to physical phase transitions where variations of the system parameters lead to qualitative changes in its properties, in computer science, computational phase transitions occur at critical points where algorithms require an increasing amount of computational resources, becoming less and less tractable~\cite{Philathong_2021}. The computational phase transitions of $k$-SAT have been the subject of several works~\cite{Philathong_2021,zhang2022quantum,philathong2023computational}. They depend critically on the density of the problem, i.e., the ratio between its clauses and variables. For random 3-SAT instances, it has been proven that the critical point is located around a density of $\alpha_C=4.3$.
\begin{figure*}[!tb]
    \centering
    \includegraphics[width=\linewidth]{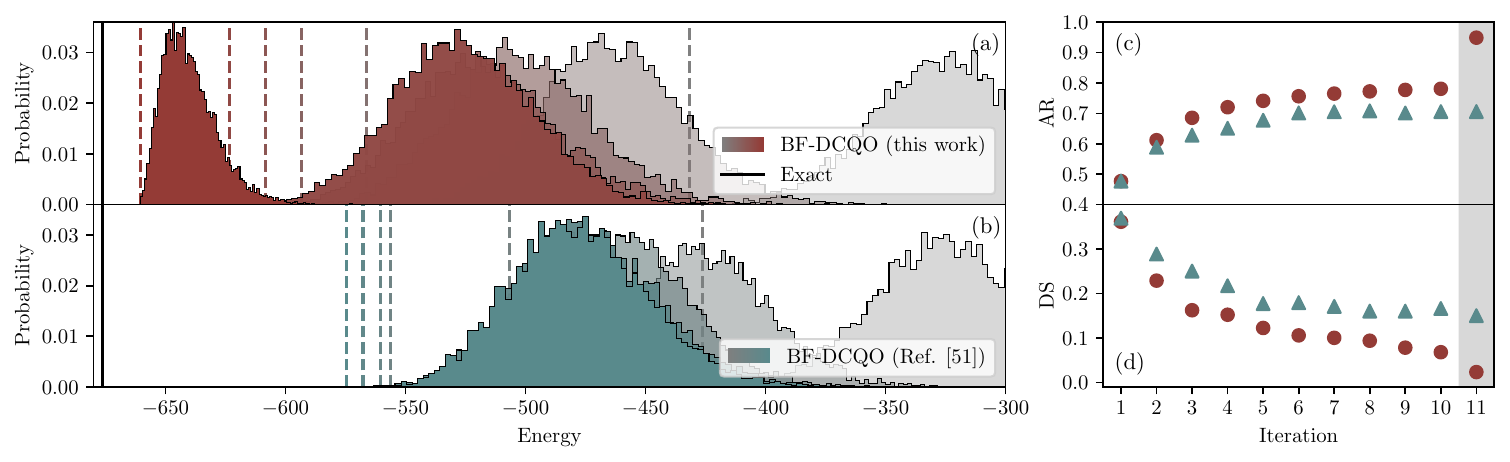}%
    \caption{\textbf{Performance comparison of the BF-DCQO variants for the 433-qubit NN HUBO results simulated via MPS.} \textbf{a,} Energy distributions for iterations 1, 3, 5, 7, 9 and 11; from grey to maroon using our novel BF-DCQO variant. The continuous line corresponds to the exact ground-state energy $E_0$ given by an iterative method [\supnote{2}], and the dashed lines correspond to the minima obtained from the distributions. Here, we report energies in units of the standard deviation of the coupling constants in~\eqlabel{eq:nearest_hubo}. \textbf{b,} Same plot as \textbf{a}, going from grey to teal, but considering the original BF-DCQO approach~\cite{cadavid2024biasfielddigitizedcounterdiabaticquantum}. \textbf{c-d,} Approximation ratios and distances to solution obtained for each iteration for both approaches. For both panels and only for our variant (maroon markers), the shaded area indicates that the bias update is changed from unsigned to signed bias in the last iteration. See~\tablabel{tab:nn_osprey_resources}.}\label{fig:ibm_nn_osprey}
\end{figure*}%
\begin{table*}[!tb]
    \caption{\textbf{433-qubit NN HUBO resources needed and results simulated via MPS.} We compare both our BF-DCQO variant (left columns) and the original approach (right columns) with $n_\text{shots}=10000$ and $\theta_\text{cutoff}=0.06$. For our variant, we used $\alpha=0.01$ and we perform a signed bias in the last iteration.}\label{tab:nn_osprey_resources}
    \begin{ruledtabular}\begin{tabular}{rrrrrrrrrrrrrrr}
       \multicolumn{1}{c}{Iteration}  & \multicolumn{2}{c}{$X$} & \multicolumn{2}{c}{$\sqrt{X}$} & \multicolumn{2}{c}{$\text{RZ}(\theta)$} & \multicolumn{2}{c}{CZ} & \multicolumn{2}{c}{Depth} & \multicolumn{2}{c}{AR} & \multicolumn{2}{c}{DS} \\ \midrule
        1 & 497 & 497 & 5910 & 5910 & 5934 & 5934 & 2275 & 2275 & 99 & 99 & 0.477 & 0.477 & 0.361 & 0.370 \\
        2 & 470 & 453 & 5563 & 5421 & 5525 & 5379 & 2156 & 2108 & 98 & 98 & 0.612 & 0.589 & 0.229 & 0.289 \\
        3 & 466 & 446 & 5412 & 5310 & 5365 & 5247 & 2106 & 2066 & 98 & 98 & 0.686 & 0.628 & 0.163 & 0.251 \\
        4 & 436 & 427 & 5314 & 5314 & 5254 & 5261 & 2063 & 2065 & 98 & 98 & 0.720 & 0.651 & 0.152 & 0.218 \\
        5 & 455 & 441 & 5283 & 5310 & 5240 & 5242 & 2052 & 2063 & 98 & 98 & 0.741 & 0.678 & 0.123 & 0.178 \\
        6 & 479 & 450 & 5287 & 5310 & 5238 & 5251 & 2052 & 2063 & 98 & 98 & 0.756 & 0.702 & 0.106 & 0.180 \\
        7 & 474 & 448 & 5287 & 5301 & 5247 & 5207 & 2052 & 2061 & 99 & 98 & 0.765 & 0.705 & 0.101 & 0.171 \\
        8 & 455 & 435 & 5283 & 5301 & 5198 & 5226 & 2052 & 2061 & 98 & 99 & 0.772 & 0.708 & 0.094 & 0.161 \\
        9 & 464 & 456 & 5283 & 5310 & 5210 & 5251 & 2052 & 2063 & 98 & 98 & 0.777 & 0.701 & 0.078 & 0.161 \\
        10 & 447 & 447 & 5284 & 5277 & 5210 & 5214 & 2052 & 2052 & 99 & 98 & 0.781 & 0.706 & 0.068 & 0.166 \\
        11 & 100 & 456 & 1879 & 5283 & 1653 & 5225 & 496 & 2052 & 80 & 97 & 0.949 & 0.706 & 0.023 & 0.150
    \end{tabular}\end{ruledtabular}
\end{table*}%

\subsection*{von Neumann entanglement entropy}

The role of entanglement in quantum optimization is an interesting question that has been discussed recently in the context of adiabatic~\cite{hauke2015probing, bauer2015entanglement} and variational~\cite{dupont2022entanglement, dupont2022calibrating, sreedhar2022quantumapproximate, nakhl2024calibrating, santra2024genuine} algorithms, with no clear consensus. Here, we study how entanglement changes with each bias-field iteration, by means of MPS simulations, to support our understanding of how BF-DCQO converges to the final solution. 

The entanglement between two subsystems, in the case of a composite pure state, can be quantified by the von Neumann entropy of the reduced density matrix of either subsystem. In an MPS representation of the wavefunction~\cite{Schollwock2011}, the entanglement entropy for any bipartition is given by the singular values corresponding to the bond that connects the two parts. For bond index $i$,
we denote the $j$th-largest singular value by $\lambda_{ij}$. Here, $j$ has values $j \in \{ 1, 2, \dots, \chi_i \}$ up to the bond dimension $\chi_i$ and the bonds between qubits (tensors in the MPS) are indexed by $i \in \{ 1, \dots, N-1 \}$. We consider the entanglement entropy averaged over bonds
\begin{equation}\label{eq:vN_entropy}
    S = -\frac{1}{N-1}\sum_{i=1}^{N-1}\sum_{j=1}^{\chi_i} \lambda^2_{ij}\log_2 \lambda^2_{ij}.
\end{equation}
This quantity is minimally zero when the state is a product state and has an upper bound $\sim \log_2 2^N = N$ when the reduced density matrix of either subsystem is maximally mixed~\cite{nielsen2001quantum}. While entanglement generally scales with the subsystem volume, in special cases, stricter bounds such as area laws can be derived~\cite{eisert2010colloquium, Srivastava2024}. Nonequilibrium states resulting from global parameter quenches are expected to follow a volume law, though recent quantum simulations of paramagnetic-to-spin-glass quench dynamics exhibited area-law scaling~\cite{king2024computationalsupremacyquantumsimulation}. 
Here, we do not analyze the scaling behavior but rather focus on the entanglement generated in the short-time dynamics of the counterdiabatically driven systems, as discussed in~\seclabel{sec:entanglement}{Entanglement in the iterative optimization}.

\section*{Results and discussion}\label{sec:results}

\subsection*{Enhanced BF-DCQO}

In the following lines, we perform a comparison of the original BF-DCQO approach~\cite{cadavid2024biasfielddigitizedcounterdiabaticquantum} and our proposed variant. We study a 433-qubit NN HUBO instance [\eqlabel{eq:nearest_hubo}], using MPS to simulate the expected performance of the upcoming IBM Osprey quantum platform without noise. In this case, the values $n_\text{shots}=10000$ and $\theta_\text{cutoff}=0.06$ are set for both approaches and $\alpha=0.01$ for our variant. The results can be seen in~\figlabel{fig:ibm_nn_osprey} and the amount of resources needed per iteration in~\tablabel{tab:nn_osprey_resources}.

Figure~\ref{fig:ibm_nn_osprey} shows that before applying the signed bias, we obtain a $10.6\%$ enhancement of the AR and $59.1\%$ for the DS, demonstrating that our new CVaR-based approach for updating the bias fields provides a better performance than standard BF-DCQO. After the signed-bias method is applied in the last iteration for our variant, these values increase to $34.4\%$ and $84.4\%$, respectively, showing that this additional feature remarkably contributes to improving the results obtained from standard BF-DCQO.

In addition, we find as an intrinsic signature of our BF-DCQO variant that the circuits implemented for each iteration require less and less resources even while the outcomes are more optimal. Conceptually, the goal of BF-DCQO is to encode the ground state of $H_f$, which can be either a product state in the computational basis or a superposition of them if the optimal solution is degenerate. For the first case---a separable state with no entanglement---BF-DCQO pushes the energy distribution towards the preparation of this state, reducing the amount of entanglement required per iteration. This feature indicates that BF-DCQO returns a lower-resource-demanding circuit per iteration, yielding a more classically tractable instance. Therefore, the combination of both quantum and classical computation paradigms appears like a promising route to address optimization problems by means of our BF-DCQO protocol. In the case of a superposition ground state, entanglement might still be reduced depending on the instance. We discuss the reduction of entanglement in more detail in~\seclabel{sec:entanglement}{Entanglement in the iterative optimization}.
\begin{figure*}[!tb]
    \centering
    \includegraphics[width=\linewidth]{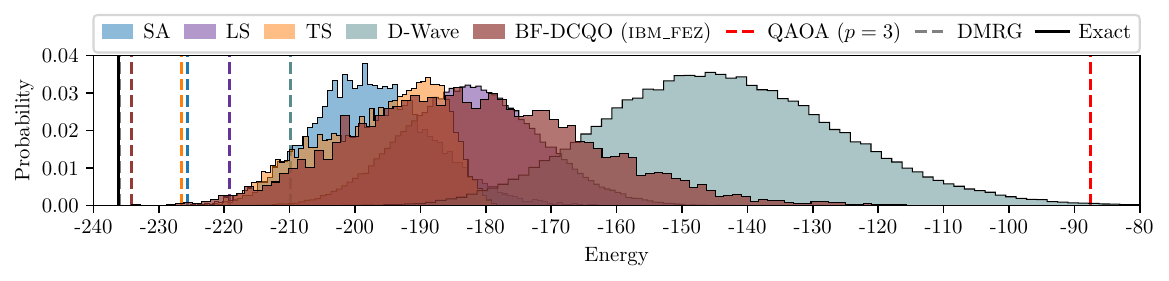}%
    \caption{\textbf{156-qubit NN HUBO instance under different approaches.} We solve~\eqlabel{eq:nearest_hubo} considering simulated annealing (SA, blue), local search algorithm (LS, purple), Tabu search (TS, orange), D-Wave (green), BF-DCQO on \textsc{ibm\_fez} (maroon), the best outcome of QAOA up to $p=3$ layers (dashed red), DMRG solution (dashed grey) and the exact solution (black line). The last two mostly overlap as DMRG reaches a nearly optimal solution [\supnote{4}].} The dashed vertical lines indicate the minimum energy values obtained for each approach. The D-Wave and BF-DCQO distributions correspond to experiments run on quantum hardware. Energies are in units the standard deviation of the coupling constants in~\eqlabel{eq:nearest_hubo}.\label{fig:comparison}
\end{figure*}

\subsection*{Comparison of methods}

In order to assess the potential of our proposed BF-DCQO variant, we conduct a comprehensive benchmark study comparing it against several well-established classical optimization techniques as well as the D-Wave platform. In particular, we optimize the nearest-neighbour three-body spin glass of~\eqlabel{eq:nearest_hubo} using a 156-qubit instance.

The results are presented in~\figlabel{fig:comparison}, where the methods and parameters are given by:
\begin{itemize}
    \item Simulated annealing (SA)~\cite{kirkpatrick1983optimization} with $100000$ number of reads, initial (final) temperature $T_i=2$ ($T_f=0.05$) using a geometric cooling scheduling with $1000$ sweeps.%
    \item Local search (LS) algorithm, which can be seen as SA with zero-temperature Metropolis-Hastings criterion~\cite{metropolis1953equation,hastings1970monte}, with $100000$ reads and $10000$ sweeps.
    \item Tabu search (TS)~\cite{tabu} with $100000$ reads starting from randomly generated initial states.
    \item D-Wave quantum annealers using \textsc{Advantage\_system4.1} with an annealing time of $\SI{2000}{\micro\second}$ and $n_\text{shots}=100000$. Based on Ref.~\cite{pelofske2024short-depth}, we set a coupling strength that is two times the largest coupling of~\eqlabel{eq:hubo} in absolute value. We thereby introduce a penalty term after the HUBO-into-QUBO mapping when a product constraint is not met in the corresponding reduced spin glass model.
    \item Tenth iteration of BF-DCQO~\cite{cadavid2024biasfielddigitizedcounterdiabaticquantum}, with $\alpha=0.02$ run on \textsc{ibm\_fez} using $\theta_\text{cutoff}=0.06$, $n_\text{trot}=3$ and $n_\text{shots}=10000$, ten times less than SA, TS and D-Wave to compensate the number of iterations. Previously, the ten BF-DCQO iterations were simulated using MPS without noise.
    \item CVaR-QAOA~\cite{farhi2014quantumapproximateoptimizationalgorithm, Barkoutsos2020improving} with $p=3$ layers, $n_\text{shots}=10000$ and $\alpha=0.1$ via MPS using the SPSA optimizer.
    \item The DMRG implementation of the ITensor library~\cite{itensor}. We find convergence with 10 sweeps, maximum bond dimension $10$, truncation error cutoff of $10^{-10}$, and a noise parameter that is gradually reduced from $10^{-3}$ to zero through the sweeps (see~\supnote{4}).%
    \item Exact solution using a numerical iterative method [\supnote{2}].
\end{itemize}%
\begin{figure*}[!tb]
    \centering
    \includegraphics[width=\linewidth]{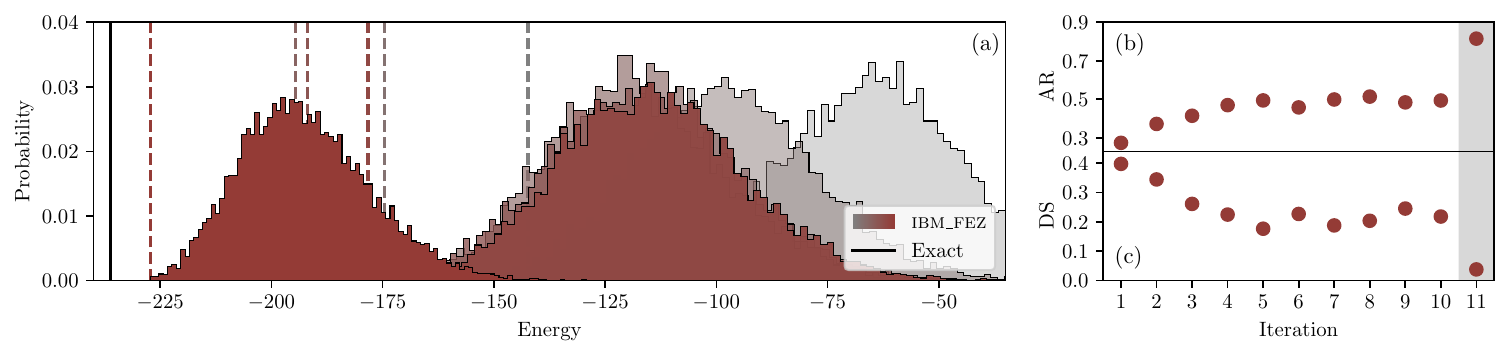}%
    \caption{\textbf{156-qubit NN HUBO results on \textsc{ibm\_fez}.} \textbf{a,} Energy distributions for iterations 1, 3, 5, 7, 9 and 11; from grey to maroon respectively. The continuous line corresponds to the exact ground state energy $E_0$ [\supnote{2}]. The dashed lines correspond to the minima obtained from the distributions. Energies are in units the standard deviation of the coupling constants in~\eqlabel{eq:nearest_hubo}. \textbf{b-c,} Approximation ratios and distances to solution for each iteration. For both panels, the shaded area indicates that the bias update is changed from unsigned to signed bias in the last iteration. See~\tablabel{tab:nn_resources}.}\label{fig:ibm_nn}
\end{figure*}%
\begin{figure*}[!tb]
    \centering
    \includegraphics[width=\linewidth]{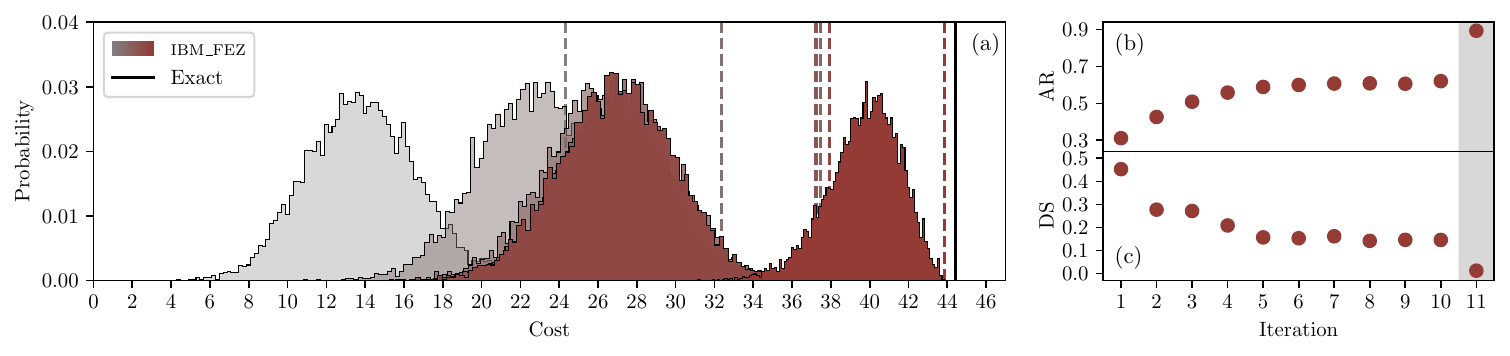}%
    \caption{\textbf{156-qubit MAX W-3-SAT HUBO results on \textsc{ibm\_fez}.} \textbf{a,} Cost function distributions for iterations 1, 3, 5, 7, 9 and 11; from grey to maroon respectively. The continuous line corresponds to the exact ground-state energy $E_0$ [\supnote{2}]. The cost function is in units of $\max(\omega_{ijk})$ in~\eqlabel{eq:mw3s}. The dashed lines correspond to the maxima of the distributions. \textbf{b-c,} Approximation ratios and distances to solution obtained for each iteration. The shaded area indicates that the bias update is changed from unsigned to signed in the last iteration. See~\tablabel{tab:mw3s_resources}.}\label{fig:ibm_wm3s}
\end{figure*}%

Figure~\ref{fig:comparison} illustrates that BF-DCQO is capable of outperforming other established methods on current noisy quantum platforms, returning an energy distribution whose minimum outcome is very close to the exact ground state energy. In particular, we have obtained a $26.7\%$ enhancement of the AR and $92.3\%$ of the DS metrics with respect to D-Wave and a $71.7\%$ enhancement of DS for SA approach but a small $-8.0\%$ lowering for the AR metric. Despite the fact that three layers were chosen for CVaR-QAOA to recover the expected behaviour of BF-DCQO with $n_\text{trot}=3$, the obtained results show the worst performance, even in comparison with D-Wave. This finding is in accordance with the results presented in Ref.~\cite{pelofske2024short-depth}. Furthermore, it should be noted that BF-DCQO and QAOA do not suffer from the inherent qubit overhead required for mapping HUBO into QUBO problems, as D-Wave does. In addition, as the order of many-body terms in the problem Hamiltonian $H_f$ increases, the qubit overhead increases and more resources are required. In particular, regular quadratization methods require a number of ancillary qubits proportional to $n_{k\text{-body}}(k-2)$, with $n_{k\text{-body}}$ the number of $k$-body terms of $H_f$~\cite{ishikawa2011transformation, anthony2017quadratic, dattani2019quadratizationdiscreteoptimizationquantum}. Therefore, a worse performance is expected.
Other known methods such as Gurobi~\cite{gurobi} and CPLEX~\cite{cplex} also suffer from this drawback. For the results presented in~\figlabel{fig:comparison}, 334 variables were needed for D-Wave. 
This feature is absent in BF-DCQO, making it a suitable technique for working with large-scale optimization problems with minimal resource requirements.

Up to this point, we have compared the performance of BF-DCQO to other methods by simulating the entire BF-DCQO routine with MPS and running the circuit obtained in the last iteration on hardware. In the following section, the impact of running the entire routine under noisy environments is examined. To this end, larger values of $\theta_\text{cutoff}$ will be set in order to guarantee implementability. We note that considering a larger number of measurements might allow us to partially recover the lost expressivity due to the increase of~$\theta_\text{cutoff}$.

\subsection*{Hardware implementation on IBM}

We first make use of the 156-qubit superconducting quantum platform \textsc{ibm\_fez} with the native gate set $\{X, \sqrt{X}, \text{RZ}(\theta), \text{CZ}\}$ (see~\supnote{3}). The linearity of its heavy-hex qubit coupling map encourages us to work with Hamiltonians whose many-body terms are nearest neighbours, thereby avoiding an overhead of entangling gates for coupling distant qubits. With this in mind, we solve the same 156-qubit NN HUBO instance that was previously tested in~\figlabel{fig:comparison}, but we now run the entire routine on \textsc{ibm\_fez}. The computation is thus conducted in a purely quantum manner. We use $n_\text{shots}=10000$, $\alpha=0.01$, $\theta_\text{cutoff}=0.13$ and a total of eleven iterations, considering an unsigned-bias update for the first ten and a signed bias for the last one. The results are presented in~\figlabel{fig:ibm_nn}, which depicts the iterative procedure of BF-DCQO and how the obtained outcomes move towards the desired solution. The amount of resources needed per iteration is given in~\tablabel{tab:nn_resources}. 

As a final remark, the final iteration of the purely quantum BF-DCQO routine exhibits a slightly worse performance when compared to the results of~\figlabel{fig:comparison}, which employed a noiseless simulation with a considerably lower gate threshold. The purely quantum BF-DCQO still outperforms the optimal solution obtained on the D-Wave platform, where we obtained $34.1\%$ of AR enhancement and $66.1\%$ of DS enhancement with respect the experiment run on D-Wave in~\figlabel{fig:comparison}.
\begin{table}[!tb]
    \caption{\textbf{156-qubit NN HUBO results and resources needed on \textsc{ibm\_fez}.} We consider $n_\text{shots}=10000$, $\alpha=0.01$ and $\theta_\text{cutoff}=0.13$. At the last iteration, we performed signed bias.}\label{tab:nn_resources}
    \begin{ruledtabular}\begin{tabular}{rrrrrrrr}
       \multicolumn{1}{c}{Iteration}  & \multicolumn{1}{c}{$X$} & \multicolumn{1}{c}{$\sqrt{X}$} & \multicolumn{1}{c}{$\text{RZ}(\theta)$} & \multicolumn{1}{c}{CZ} & \multicolumn{1}{c}{Depth} & \multicolumn{1}{c}{AR} & \multicolumn{1}{c}{DS} \\ \midrule
        1 & 115 & 1020 & 1030 & 477 & 129 & 0.274 & 0.397 \\
        2 & 86 & 969 & 916 & 430 & 126 & 0.372 & 0.344 \\
        3 & 92 & 958 & 869 & 424 & 129 & 0.415 & 0.261 \\
        4 & 85 & 937 & 861 & 410 & 124 & 0.470 & 0.225 \\
        5 & 93 & 938 & 862 & 410 & 127 & 0.494 & 0.176 \\
        6 & 83 & 941 & 878 & 410 & 124 & 0.458 & 0.227 \\
        7 & 85 & 940 & 878 & 410 & 125 & 0.499 & 0.188 \\
        8 & 79 & 899 & 834 & 386 & 90 & 0.514 & 0.203 \\
        9 & 79 & 903 & 835 & 386 & 93 & 0.484 & 0.245 \\
        10 & 79 & 941 & 859 & 410 & 130 & 0.494 & 0.218 \\
        11 & 3 & 330 & 180 & 13 & 39 & 0.815 & 0.038
    \end{tabular}\end{ruledtabular}
\end{table}%
\begin{table}[!tb]
    \caption{\textbf{156-qubit NN MAX W-3-SAT HUBO results and resources needed on \textsc{ibm\_fez}.} We consider $n_\text{shots}=10000$, $\alpha=0.01$, and $\theta_\text{cutoff}=0.06$. In the last iteration, we apply a signed bias, where no entangling gates are used.}\label{tab:mw3s_resources}
    \begin{ruledtabular}\begin{tabular}{rrrrrrrr}
       \multicolumn{1}{c}{Iteration}  & \multicolumn{1}{c}{$X$} & \multicolumn{1}{c}{$\sqrt{X}$} & \multicolumn{1}{c}{$\text{RZ}(\theta)$} & \multicolumn{1}{c}{CZ} & \multicolumn{1}{c}{Depth} & \multicolumn{1}{c}{AR} & \multicolumn{1}{c}{DS} \\ \midrule
        1 & 88 & 1359 & 1309 & 495 & 189 & 0.311 & 0.453 \\
        2 & 39 & 949 & 808 & 285 & 103 & 0.426 & 0.277 \\ 
        3 & 2 & 450 & 296 & 64 & 23 & 0.509 & 0.271 \\
        4 & 3 & 427 & 271 & 56 & 23 & 0.559 & 0.209 \\
        5 & 0 & 421 & 267 & 50 & 23 & 0.589 & 0.157 \\
        6 & 2 & 411 & 265 & 50 & 19 & 0.600 & 0.153 \\
        7 & 2 & 416 & 267 & 50 & 23 & 0.608 & 0.162 \\
        8 & 2 & 417 & 263 & 50 & 23 & 0.609 & 0.142 \\
        9 & 2 & 412 & 252 & 50 & 18 & 0.606 & 0.146 \\
        10 & 2 & 417 & 261 & 50 & 22 & 0.621 & 0.146 \\
        11 & - & 311 & 156 & - & 4 & 0.894 & 0.013 \\
    \end{tabular}\end{ruledtabular}
\end{table}
\begin{figure*}[!tb]
    \centering
    \includegraphics[width=\linewidth]{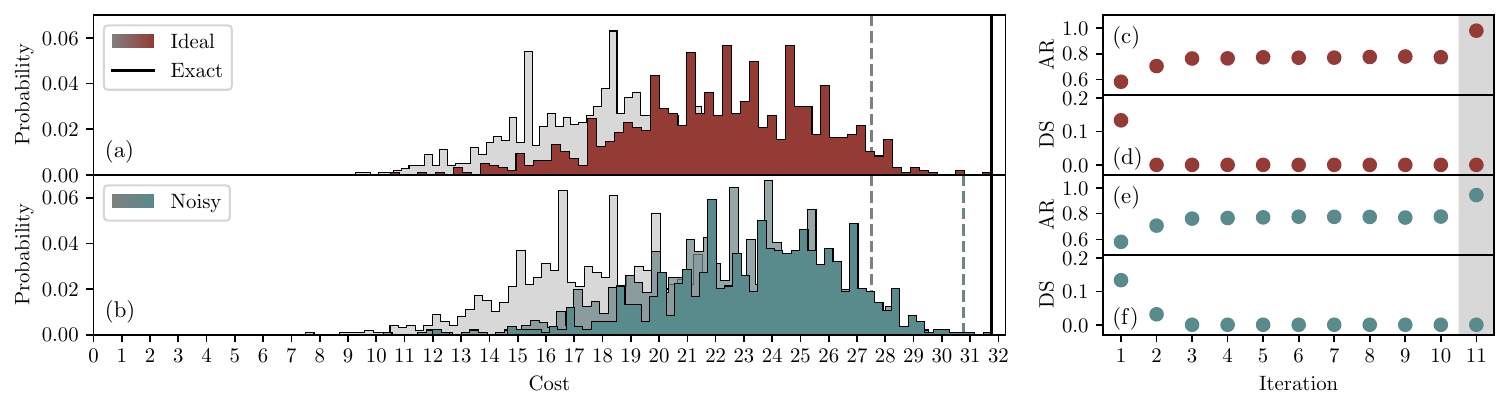}%
    \caption{\textbf{25-qubit dense MAX 3-SAT HUBO results on \textsc{IonQ Aria 1} emulator.} \textbf{a-b,} Cost function distributions for the first two and three iterations without noise (grey to maroon) and including an \textsc{IonQ Aria 1} noise model profile plus debiased error mitigation (grey to teal), respectively. The exact ground state (continuous line) is reached for both scenarios. The dashed lines correspond to the maximum cost obtained from the distributions. The cost function is in units of the coupling $\omega_{ijk}$ in~\eqlabel{eq:mw3s}, which here is chosen as uniform. \textbf{c-f,} approximation ratios and distance to solution for both ideal and noisy emulations, respectively. Shaded areas indicate that the bias updating is changed from unsigned to signed bias at the last iteration. See~\tablabel{tab:m3s_ionq_resources}.}\label{fig:ionq_m3s}
\end{figure*}%
\begin{table*}[!tb]
    \caption{\textbf{25-qubit dense MAX 3-SAT HUBO results and resources needed on \textsc{IonQ Aria 1} emulator.} Noiseless (left) and \textsc{IonQ Aria 1}-noise profile with debiased error mitigation (right) simulations are considered with $n_\text{shots}=1000$, $\alpha=0.01$ and $\theta_\text{cutoff}=0.06$. At the last iteration, we performed signed bias, where no entangling gates were used.}\label{tab:m3s_ionq_resources}
    \begin{ruledtabular}\begin{tabular}{rrrrrrrrrrrrr}
       \multicolumn{1}{c}{Iteration}  & \multicolumn{2}{c}{$\text{MS}(\phi_0,\phi_1,\theta)$} & \multicolumn{2}{c}{$\text{GPi}(\phi)$} & \multicolumn{2}{c}{$\text{GPi2}(\phi)$} & \multicolumn{2}{c}{Depth} & \multicolumn{2}{c}{AR} & \multicolumn{2}{c}{DS} \\ \midrule
        1 & 96 & 96 & 282 & 282 & 948 & 948 & 219 & 219 & 0.582 & 0.580 & 0.134 & 0.134 \\
        2 & 96 & 96 & 282 & 282 & 948 & 948 & 219 & 219 & 0.704 & 0.707 & 0.000 & 0.031 \\
        3 & 96 & 96 & 282 & 282 & 948 & 948 & 219 & 219 & 0.763 & 0.761 & 0.000 & 0.000 \\
        4 & 96 & 96 & 282 & 282 & 948 & 948 & 219 & 219 & 0.764 & 0.766 & 0.000 & 0.000 \\
        5 & 96 & 96 & 282 & 282 & 948 & 948 & 219 & 219 & 0.772 & 0.770 & 0.000 & 0.000 \\
        6 & 96 & 96 & 282 & 282 & 948 & 948 & 219 & 219 & 0.769 & 0.776 & 0.000 & 0.000 \\
        7 & 96 & 96 & 282 & 282 & 948 & 948 & 219 & 219 & 0.769 & 0.775 & 0.000 & 0.000 \\
        8 & 96 & 96 & 282 & 282 & 948 & 948 & 219 & 219 & 0.775 & 0.774 & 0.000 & 0.000 \\
        9 & 96 & 96 & 282 & 282 & 948 & 948 & 219 & 219 & 0.778 & 0.769 & 0.000 & 0.000 \\
        10 & 96 & 96 & 282 & 282 & 948 & 948 & 219 & 219 & 0.773 & 0.777 & 0.000 & 0.000 \\
        11 & - & - & 26 & 26 & 52 & 52 & 7 & 7& 0.979 & 0.943 & 0.000 & 0.000
    \end{tabular}\end{ruledtabular}
\end{table*}

With the aim of solving more industrially relevant and denser large-scale problems, we also solve a 156-qubit NN MAX W-3-SAT instance on \textsc{ibm\_fez}, described by~\eqlabel{eq:mw3s} with only nearest-neighbour terms. We use the same parameters as for the previous instance but change $\theta_\text{cutoff}=0.06$ for convenience. The results are shown in~\figlabel{fig:ibm_wm3s}, and the amount of resources needed per iteration is listed in~\tablabel{tab:mw3s_resources}. For this particular problem, the best outcome obtained is quite close to the optimal solution despite a purely quantum routine, which highlights the efficacy of BF-DCQO under noisy environments. Additionally, we solved the same HUBO instance on D-Wave using $n_\text{shots}=100000$. The purely quantum BF-DCQO again outperforms the D-Wave results: we obtain a $43.0\%$ enhancement of the AR and an $88.5\%$ enhancement of the DS with respect the experiment run on the D-Wave platform. 

\subsection*{Dense problems with all-to-all connectivity}\label{sec:dense}

The sparsity and high locality of the terms involved in the instances considered on IBM render the problems amenable to classical simulation. To evaluate the efficacy of BF-DCQO in more intricate scenarios with long-range couplings, we extend our analysis to denser versions of MAX 3-SAT. In particular, we set as clause-to-variable ratio the critical density for 3-SAT, $\alpha_C=4.3$. Dense MAX 3-SAT instances with uniform couplings equal to unity are then generated randomly with the number of clauses given by $\lceil N\alpha_C \rceil$. It is important to note that the expected optimal solutions of MAX $k$-SAT instances, where clauses are uniformly weighted, can be either a product state or a low-entangled superposition of few bitstrings. As a last remark, optimal solution degeneracy was not present in our previous studies and it may compromise the performance.

We employ the 25-qubit \textsc{IonQ Aria 1} emulator with native gate set given by $\{\text{MS}(\phi_0,\phi_1,\theta), \text{GPi}(\phi), \text{GPi2}(\phi)\}$ (see~\supnote{3}). It mimics an all-to-all connected trapped-ion platform and encompasses both noiseless and noisy simulations, applying debiased error mitigation~\cite{maksymov2023enhancingquantumcomputerperformance} with a noise profile that matches the \textsc{Aria 1} expected performance. The system size of the problem allows to obtain the exact optimal solution to be obtained by our protocol. The results for $n_\text{shots}=1000$, $\alpha=0.01$ and $\theta_\text{cutoff}=0.06$ can be seen in~\figlabel{fig:ionq_m3s}, where the amount of resources required at each iteration presented in~\tablabel{tab:m3s_ionq_resources}. The data shows that the BF-DCQO protocol is able to reach the exact optimal state in both noiseless and noisy environments at the second iteration, which serves to further validate the efficacy of the method.

\subsection*{Entanglement in the iterative optimization}\label{sec:entanglement}

The role of entanglement dynamics in quantum algorithms is an interesting and nontrivial question. In adiabatic quantum optimization, both the initial state and the desired final state are product states or low-entangled multipartite states, while the adiabatic path may explore highly entangled states~\cite{bauer2015entanglement,hauke2015probing}. 
To gain insight into our observation that the required number of entangling gates is reduced with increasing iterations, we consider the flow of entanglement along the dynamical process. 

We compute the entanglement entropy of~\eqlabel{eq:vN_entropy} for the 433-qubit NN and the 25-qubit dense MAX 3-SAT HUBO instances solved previously in Figs.~\ref{fig:ibm_nn_osprey} and~\ref{fig:ionq_m3s}. As mentioned before, the first case corresponds to a sparse problem with a nondegenerate optimal solution and the second one to a dense problem with a degenerate optimal solution. For the second case, we only consider the noisy simulation including debiased error mitigation [\figlabel{fig:ionq_m3s}b]. We show, in~\figlabel{fig:entanglement}, how the entanglement entropy decays with the subsequent BF-DCQO iterations for both problems. Due to the dense nonlocal couplings, the average entanglement entropy generated in the second instance is larger than in the first. The numbers of entangling gates used here are listed in~Tables~\ref{tab:nn_osprey_resources} and~\ref{tab:m3s_ionq_resources}. We can see that for the second instance, in the first three iterations, the average entanglement entropy decreases even when the number of entangling gates remains the same. Similar to the AR and DS in Figs.~\ref{fig:ibm_nn_osprey} and~\ref{fig:ionq_m3s}, the entropy then nearly saturates up to the last iteration. The signed bias applied in the last iteration significantly reduces the number of entangling gates and the entanglement entropy drops to a value close to zero. This indicates that the iteration protocol produces a classical ground state as expected.
\begin{figure}[!tb]
    \centering
    \includegraphics[width=\linewidth]{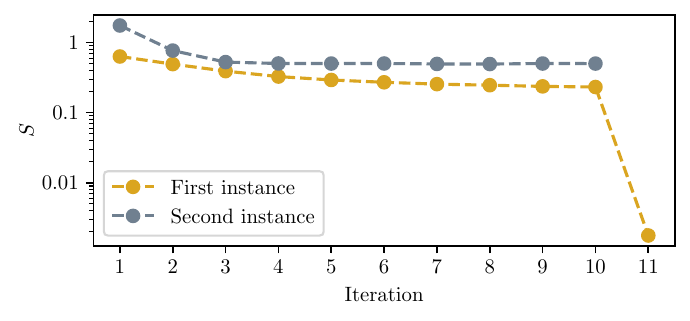}%
    \caption{\textbf{Average entanglement entropy $S$ as a function of the iteration.} The first instance corresponds to 433-qubit NN HUBO problem and the second instance to a 25-qubit dense MAX 3-SAT HUBO problem. Signed bias is applied at the last iteration, significantly reducing the number of entangling gates used and consequently the entanglement entropy. For the second instance, the entanglement entropy drops to zero at the last iteration as there are no entangling gates. We therefore leave out the last data point.}\label{fig:entanglement}%
\end{figure}%

We note that while the entanglement entropy analyzed here is naturally obtained from the MPS representation, it is not a straightforward quantity to measure experimentally. In order to characterize entanglement in experiments run on quantum hardware, for instance in the case of highly entangled states not amenable to tensor network simulations, alternative measures would be required, such as correlator functions by means of quantum measurements~\cite{rossini2020quantum, Srivastava2024}. 

\section*{Conclusions}\label{sec:conclusion}

In this study, we investigate and evaluate the performance of variants of the recently proposed BF-DCQO algorithm~\cite{cadavid2024biasfielddigitizedcounterdiabaticquantum} by solving large-scale HUBO instances on IBM quantum processors and IonQ noisy emulators. The superiority of our BF-DCQO variant over well-established classical techniques and optimization on the D-Wave \textsc{Advantage\_system4.1} is verified through computing the approximation ratio and the distance to solution in an iterative process. We obtain better optimal solutions even while the circuits after each iteration demand less resources. This is achieved by means of an iterative learning procedure, where the optimal solutions obtained from the last iteration are used for the subsequent one, providing several techniques to guarantee convergence under different scenarios. Furthermore, in order to address not only the current but also the near-term hardware, we examine the expected performance of BF-DCQO on the upcoming IBM Osprey platform with 433 qubits. Simulating 433-qubit HUBO instances without noise yields promising results. The BF-DCQO algorithm does not require any classical optimization subroutines, thus avoiding potential trainability drawbacks, nor does it require extra qubits for mapping the initial HUBO into a QUBO problem.

We provide numerical evidence of how our protocol disentangles the states obtained after each iteration, one of the key signatures of our method. This feature enables an iterative reduction in the number of resources required, thus making subsequent circuits more classically tractable and opening the door to employing hybrid quantum-classical strategies. Therefore, the BF-DCQO algorithm emerges as a suitable algorithm for solving higher-order binary optimization problems on both current and near-term quantum platforms, paving the path towards solving large-scale optimization problems with status-quo and next-generation quantum processors.

\section*{Acknowledgments} 

We would like to thank Shubham Kumar for preparing the schematic figure. We also thank Gaurav Dev and Bhargava A. Balaganchi for fruitful discussions. We acknowledge the use of IBM Quantum services for this work. The views expressed are those of the authors and do not reflect the official policy or position of IBM or the IBM Quantum team.

\noindent\textbf{Data availability.} Data is available from the corresponding author upon reasonable request.

\noindent\textbf{Code availability.} The codes used to generate data for this paper are available from the corresponding author upon reasonable request.

\noindent\textbf{Competing interests.} The authors declare no competing interests.

\noindent\textbf{Author contributions.} E.S. and N.N.H. conceived and supervised the work. S.V.R., A.-M.V., A.G.C. and N.N.H. developed the theoretical framework. S.V.R., A.-M.V., A.G.C. and A.S. developed the codes for obtaining the benchmark and hardware results. S.V.R. and N.N.H. analyzed the results. All authors participated in the discussion of results and writing the manuscript.

\bibliography{bibfile}

\begin{thebibliography}{102}%
\makeatletter
\providecommand \@ifxundefined [1]{%
 \@ifx{#1\undefined}
}%
\providecommand \@ifnum [1]{%
 \ifnum #1\expandafter \@firstoftwo
 \else \expandafter \@secondoftwo
 \fi
}%
\providecommand \@ifx [1]{%
 \ifx #1\expandafter \@firstoftwo
 \else \expandafter \@secondoftwo
 \fi
}%
\providecommand \natexlab [1]{#1}%
\providecommand \enquote  [1]{``#1''}%
\providecommand \bibnamefont  [1]{#1}%
\providecommand \bibfnamefont [1]{#1}%
\providecommand \citenamefont [1]{#1}%
\providecommand \href@noop [0]{\@secondoftwo}%
\providecommand \href [0]{\begingroup \@sanitize@url \@href}%
\providecommand \@href[1]{\@@startlink{#1}\@@href}%
\providecommand \@@href[1]{\endgroup#1\@@endlink}%
\providecommand \@sanitize@url [0]{\catcode `\\12\catcode `\$12\catcode `\&12\catcode `\#12\catcode `\^12\catcode `\_12\catcode `\%12\relax}%
\providecommand \@@startlink[1]{}%
\providecommand \@@endlink[0]{}%
\providecommand \url  [0]{\begingroup\@sanitize@url \@url }%
\providecommand \@url [1]{\endgroup\@href {#1}{\urlprefix }}%
\providecommand \urlprefix  [0]{URL }%
\providecommand \Eprint [0]{\href }%
\providecommand \doibase [0]{https://doi.org/}%
\providecommand \selectlanguage [0]{\@gobble}%
\providecommand \bibinfo  [0]{\@secondoftwo}%
\providecommand \bibfield  [0]{\@secondoftwo}%
\providecommand \translation [1]{[#1]}%
\providecommand \BibitemOpen [0]{}%
\providecommand \bibitemStop [0]{}%
\providecommand \bibitemNoStop [0]{.\EOS\space}%
\providecommand \EOS [0]{\spacefactor3000\relax}%
\providecommand \BibitemShut  [1]{\csname bibitem#1\endcsname}%
\let\auto@bib@innerbib\@empty
%</preamble>
\bibitem [{\citenamefont {Fu}\ and\ \citenamefont {Anderson}(1986)}]{fu1986application}%
  \BibitemOpen
  \bibfield  {author} {\bibinfo {author} {\bibfnamefont {Y.}~\bibnamefont {Fu}}\ and\ \bibinfo {author} {\bibfnamefont {P.~W.}\ \bibnamefont {Anderson}},\ }\bibfield  {title} {\bibinfo {title} {{Application of statistical mechanics to NP-complete problems in combinatorial optimisation}},\ }\href {https://doi.org/10.1088/0305-4470/19/9/033} {\bibfield  {journal} {\bibinfo  {journal} {Journal of Physics A: Mathematical and General}\ }\textbf {\bibinfo {volume} {19}},\ \bibinfo {pages} {1605} (\bibinfo {year} {1986})}\BibitemShut {NoStop}%
\bibitem [{\citenamefont {Lucas}(2014)}]{lucas2014ising}%
  \BibitemOpen
  \bibfield  {author} {\bibinfo {author} {\bibfnamefont {A.}~\bibnamefont {Lucas}},\ }\bibfield  {title} {\bibinfo {title} {{Ising formulations of many NP problems}},\ }\href {https://doi.org/10.3389/fphy.2014.00005} {\bibfield  {journal} {\bibinfo  {journal} {Front. Phys.}\ }\textbf {\bibinfo {volume} {2}},\ \bibinfo {pages} {5} (\bibinfo {year} {2014})}\BibitemShut {NoStop}%
\bibitem [{\citenamefont {Albash}\ and\ \citenamefont {Lidar}(2018)}]{albash2018adiabatic}%
  \BibitemOpen
  \bibfield  {author} {\bibinfo {author} {\bibfnamefont {T.}~\bibnamefont {Albash}}\ and\ \bibinfo {author} {\bibfnamefont {D.~A.}\ \bibnamefont {Lidar}},\ }\bibfield  {title} {\bibinfo {title} {Adiabatic quantum computation},\ }\href {https://doi.org/10.1103/RevModPhys.90.015002} {\bibfield  {journal} {\bibinfo  {journal} {Rev. Mod. Phys.}\ }\textbf {\bibinfo {volume} {90}},\ \bibinfo {pages} {015002} (\bibinfo {year} {2018})}\BibitemShut {NoStop}%
\bibitem [{\citenamefont {Farhi}\ \emph {et~al.}(2014)\citenamefont {Farhi}, \citenamefont {Goldstone},\ and\ \citenamefont {Gutmann}}]{farhi2014quantumapproximateoptimizationalgorithm}%
  \BibitemOpen
  \bibfield  {author} {\bibinfo {author} {\bibfnamefont {E.}~\bibnamefont {Farhi}}, \bibinfo {author} {\bibfnamefont {J.}~\bibnamefont {Goldstone}},\ and\ \bibinfo {author} {\bibfnamefont {S.}~\bibnamefont {Gutmann}},\ }\href {https://arxiv.org/abs/1411.4028} {\bibinfo {title} {{A Quantum Approximate Optimization Algorithm}}} (\bibinfo {year} {2014}),\ \Eprint {https://arxiv.org/abs/1411.4028} {arXiv:1411.4028 [quant-ph]} \BibitemShut {NoStop}%
\bibitem [{\citenamefont {Barends}\ \emph {et~al.}(2016)\citenamefont {Barends}, \citenamefont {Shabani}, \citenamefont {Lamata}, \citenamefont {Kelly}, \citenamefont {Mezzacapo}, \citenamefont {Heras}, \citenamefont {Babbush}, \citenamefont {Fowler}, \citenamefont {Campbell}, \citenamefont {Chen} \emph {et~al.}}]{barends2016digitized}%
  \BibitemOpen
  \bibfield  {author} {\bibinfo {author} {\bibfnamefont {R.}~\bibnamefont {Barends}}, \bibinfo {author} {\bibfnamefont {A.}~\bibnamefont {Shabani}}, \bibinfo {author} {\bibfnamefont {L.}~\bibnamefont {Lamata}}, \bibinfo {author} {\bibfnamefont {J.}~\bibnamefont {Kelly}}, \bibinfo {author} {\bibfnamefont {A.}~\bibnamefont {Mezzacapo}}, \bibinfo {author} {\bibfnamefont {U.~L.}\ \bibnamefont {Heras}}, \bibinfo {author} {\bibfnamefont {R.}~\bibnamefont {Babbush}}, \bibinfo {author} {\bibfnamefont {A.~G.}\ \bibnamefont {Fowler}}, \bibinfo {author} {\bibfnamefont {B.}~\bibnamefont {Campbell}}, \bibinfo {author} {\bibfnamefont {Y.}~\bibnamefont {Chen}}, \emph {et~al.},\ }\bibfield  {title} {\bibinfo {title} {Digitized adiabatic quantum computing with a superconducting circuit},\ }\href {https://doi.org/10.1038/nature17658} {\bibfield  {journal} {\bibinfo  {journal} {Nature}\ }\textbf {\bibinfo {volume} {534}},\ \bibinfo {pages} {222–226} (\bibinfo {year} {2016})}\BibitemShut {NoStop}%
\bibitem [{\citenamefont {Durr}\ and\ \citenamefont {Hoyer}(1999)}]{durr1999quantumalgorithmfindingminimum}%
  \BibitemOpen
  \bibfield  {author} {\bibinfo {author} {\bibfnamefont {C.}~\bibnamefont {Durr}}\ and\ \bibinfo {author} {\bibfnamefont {P.}~\bibnamefont {Hoyer}},\ }\href {https://arxiv.org/abs/quant-ph/9607014} {\bibinfo {title} {{A Quantum Algorithm for Finding the Minimum}}} (\bibinfo {year} {1999}),\ \Eprint {https://arxiv.org/abs/quant-ph/9607014} {arXiv:quant-ph/9607014 [quant-ph]} \BibitemShut {NoStop}%
\bibitem [{\citenamefont {Montanaro}(2018)}]{montanaro2018quantum-walk}%
  \BibitemOpen
  \bibfield  {author} {\bibinfo {author} {\bibfnamefont {A.}~\bibnamefont {Montanaro}},\ }\bibfield  {title} {\bibinfo {title} {{Quantum-Walk Speedup of Backtracking Algorithms}},\ }\href {https://doi.org/10.4086/toc.2018.v014a015} {\bibfield  {journal} {\bibinfo  {journal} {Theory of Computing}\ }\textbf {\bibinfo {volume} {14}},\ \bibinfo {pages} {1} (\bibinfo {year} {2018})}\BibitemShut {NoStop}%
\bibitem [{\citenamefont {Montanaro}(2020)}]{montanaro2020quantumspeedup}%
  \BibitemOpen
  \bibfield  {author} {\bibinfo {author} {\bibfnamefont {A.}~\bibnamefont {Montanaro}},\ }\bibfield  {title} {\bibinfo {title} {Quantum speedup of branch-and-bound algorithms},\ }\href {https://doi.org/10.1103/PhysRevResearch.2.013056} {\bibfield  {journal} {\bibinfo  {journal} {Phys. Rev. Res.}\ }\textbf {\bibinfo {volume} {2}},\ \bibinfo {pages} {013056} (\bibinfo {year} {2020})}\BibitemShut {NoStop}%
\bibitem [{\citenamefont {Chakrabarti}\ \emph {et~al.}(2022)\citenamefont {Chakrabarti}, \citenamefont {Minssen}, \citenamefont {Yalovetzky},\ and\ \citenamefont {Pistoia}}]{chakrabarti2022universalquantumspeedupbranchandbound}%
  \BibitemOpen
  \bibfield  {author} {\bibinfo {author} {\bibfnamefont {S.}~\bibnamefont {Chakrabarti}}, \bibinfo {author} {\bibfnamefont {P.}~\bibnamefont {Minssen}}, \bibinfo {author} {\bibfnamefont {R.}~\bibnamefont {Yalovetzky}},\ and\ \bibinfo {author} {\bibfnamefont {M.}~\bibnamefont {Pistoia}},\ }\href {https://arxiv.org/abs/2210.03210} {\bibinfo {title} {{Universal Quantum Speedup for Branch-and-Bound, Branch-and-Cut, and Tree-Search Algorithms}}} (\bibinfo {year} {2022}),\ \Eprint {https://arxiv.org/abs/2210.03210} {arXiv:2210.03210 [quant-ph]} \BibitemShut {NoStop}%
\bibitem [{\citenamefont {Somma}\ \emph {et~al.}(2008)\citenamefont {Somma}, \citenamefont {Boixo}, \citenamefont {Barnum},\ and\ \citenamefont {Knill}}]{somma2008quantum}%
  \BibitemOpen
  \bibfield  {author} {\bibinfo {author} {\bibfnamefont {R.~D.}\ \bibnamefont {Somma}}, \bibinfo {author} {\bibfnamefont {S.}~\bibnamefont {Boixo}}, \bibinfo {author} {\bibfnamefont {H.}~\bibnamefont {Barnum}},\ and\ \bibinfo {author} {\bibfnamefont {E.}~\bibnamefont {Knill}},\ }\bibfield  {title} {\bibinfo {title} {{Quantum Simulations of Classical Annealing Processes}},\ }\href {https://doi.org/10.1103/PhysRevLett.101.130504} {\bibfield  {journal} {\bibinfo  {journal} {Phys. Rev. Lett.}\ }\textbf {\bibinfo {volume} {101}},\ \bibinfo {pages} {130504} (\bibinfo {year} {2008})}\BibitemShut {NoStop}%
\bibitem [{\citenamefont {Wocjan}\ and\ \citenamefont {Abeyesinghe}(2008)}]{wocjan2008speedup}%
  \BibitemOpen
  \bibfield  {author} {\bibinfo {author} {\bibfnamefont {P.}~\bibnamefont {Wocjan}}\ and\ \bibinfo {author} {\bibfnamefont {A.}~\bibnamefont {Abeyesinghe}},\ }\bibfield  {title} {\bibinfo {title} {Speedup via quantum sampling},\ }\href {https://doi.org/10.1103/PhysRevA.78.042336} {\bibfield  {journal} {\bibinfo  {journal} {Phys. Rev. A}\ }\textbf {\bibinfo {volume} {78}},\ \bibinfo {pages} {042336} (\bibinfo {year} {2008})}\BibitemShut {NoStop}%
\bibitem [{\citenamefont {Hastings}(2018)}]{hastings2018shortpathquantum}%
  \BibitemOpen
  \bibfield  {author} {\bibinfo {author} {\bibfnamefont {M.~B.}\ \bibnamefont {Hastings}},\ }\bibfield  {title} {\bibinfo {title} {A {S}hort {P}ath {Q}uantum {A}lgorithm for {E}xact {O}ptimization},\ }\href {https://doi.org/10.22331/q-2018-07-26-78} {\bibfield  {journal} {\bibinfo  {journal} {{Quantum}}\ }\textbf {\bibinfo {volume} {2}},\ \bibinfo {pages} {78} (\bibinfo {year} {2018})}\BibitemShut {NoStop}%
\bibitem [{\citenamefont {Dalzell}\ \emph {et~al.}(2023)\citenamefont {Dalzell}, \citenamefont {Pancotti}, \citenamefont {Campbell},\ and\ \citenamefont {Brand\~{a}o}}]{dalzell2023mind}%
  \BibitemOpen
  \bibfield  {author} {\bibinfo {author} {\bibfnamefont {A.~M.}\ \bibnamefont {Dalzell}}, \bibinfo {author} {\bibfnamefont {N.}~\bibnamefont {Pancotti}}, \bibinfo {author} {\bibfnamefont {E.~T.}\ \bibnamefont {Campbell}},\ and\ \bibinfo {author} {\bibfnamefont {F.~G.}\ \bibnamefont {Brand\~{a}o}},\ }\bibfield  {title} {\bibinfo {title} {{Mind the Gap: Achieving a Super-Grover Quantum Speedup by Jumping to the End}},\ }in\ \href {https://doi.org/10.1145/3564246.3585203} {\emph {\bibinfo {booktitle} {Proceedings of the 55th Annual ACM Symposium on Theory of Computing}}},\ \bibinfo {series and number} {STOC 2023}\ (\bibinfo  {publisher} {Association for Computing Machinery},\ \bibinfo {address} {New York, NY, USA},\ \bibinfo {year} {2023})\ p.\ \bibinfo {pages} {1131–1144}\BibitemShut {NoStop}%
\bibitem [{\citenamefont {Boulebnane}\ and\ \citenamefont {Montanaro}(2024)}]{boulebnane2022solvingbooleansatisfiabilityproblems}%
  \BibitemOpen
  \bibfield  {author} {\bibinfo {author} {\bibfnamefont {S.}~\bibnamefont {Boulebnane}}\ and\ \bibinfo {author} {\bibfnamefont {A.}~\bibnamefont {Montanaro}},\ }\bibfield  {title} {\bibinfo {title} {{Solving Boolean Satisfiability Problems With The Quantum Approximate Optimization Algorithm}},\ }\href {https://doi.org/10.1103/PRXQuantum.5.030348} {\bibfield  {journal} {\bibinfo  {journal} {PRX Quantum}\ }\textbf {\bibinfo {volume} {5}},\ \bibinfo {pages} {030348} (\bibinfo {year} {2024})}\BibitemShut {NoStop}%
\bibitem [{\citenamefont {Boehmer}(1967)}]{boehmer1967binary}%
  \BibitemOpen
  \bibfield  {author} {\bibinfo {author} {\bibfnamefont {A.}~\bibnamefont {Boehmer}},\ }\bibfield  {title} {\bibinfo {title} {Binary pulse compression codes},\ }\href {https://doi.org/10.1109/TIT.1967.1053969} {\bibfield  {journal} {\bibinfo  {journal} {IEEE Transactions on Information Theory}\ }\textbf {\bibinfo {volume} {13}},\ \bibinfo {pages} {156} (\bibinfo {year} {1967})}\BibitemShut {NoStop}%
\bibitem [{\citenamefont {Schroeder}(1970)}]{schroeder1970synthesis}%
  \BibitemOpen
  \bibfield  {author} {\bibinfo {author} {\bibfnamefont {M.}~\bibnamefont {Schroeder}},\ }\bibfield  {title} {\bibinfo {title} {{Synthesis of low-peak-factor signals and binary sequences with low autocorrelation (Corresp.)}},\ }\href {https://doi.org/10.1109/TIT.1970.1054411} {\bibfield  {journal} {\bibinfo  {journal} {IEEE Transactions on Information Theory}\ }\textbf {\bibinfo {volume} {16}},\ \bibinfo {pages} {85} (\bibinfo {year} {1970})}\BibitemShut {NoStop}%
\bibitem [{\citenamefont {Shaydulin}\ \emph {et~al.}(2024)\citenamefont {Shaydulin}, \citenamefont {Li}, \citenamefont {Chakrabarti}, \citenamefont {DeCross}, \citenamefont {Herman}, \citenamefont {Kumar}, \citenamefont {Larson}, \citenamefont {Lykov}, \citenamefont {Minssen}, \citenamefont {Sun} \emph {et~al.}}]{shaydulin2024evidence}%
  \BibitemOpen
  \bibfield  {author} {\bibinfo {author} {\bibfnamefont {R.}~\bibnamefont {Shaydulin}}, \bibinfo {author} {\bibfnamefont {C.}~\bibnamefont {Li}}, \bibinfo {author} {\bibfnamefont {S.}~\bibnamefont {Chakrabarti}}, \bibinfo {author} {\bibfnamefont {M.}~\bibnamefont {DeCross}}, \bibinfo {author} {\bibfnamefont {D.}~\bibnamefont {Herman}}, \bibinfo {author} {\bibfnamefont {N.}~\bibnamefont {Kumar}}, \bibinfo {author} {\bibfnamefont {J.}~\bibnamefont {Larson}}, \bibinfo {author} {\bibfnamefont {D.}~\bibnamefont {Lykov}}, \bibinfo {author} {\bibfnamefont {P.}~\bibnamefont {Minssen}}, \bibinfo {author} {\bibfnamefont {Y.}~\bibnamefont {Sun}}, \emph {et~al.},\ }\bibfield  {title} {\bibinfo {title} {Evidence of scaling advantage for the quantum approximate optimization algorithm on a classically intractable problem},\ }\href {https://doi.org/10.1126/sciadv.adm6761} {\bibfield  {journal} {\bibinfo  {journal} {Science Advances}\ }\textbf {\bibinfo {volume} {10}},\ \bibinfo {pages} {eadm6761} (\bibinfo {year} {2024})}\BibitemShut {NoStop}%
\bibitem [{\citenamefont {Orús}(2014)}]{orus2014practical}%
  \BibitemOpen
  \bibfield  {author} {\bibinfo {author} {\bibfnamefont {R.}~\bibnamefont {Orús}},\ }\bibfield  {title} {\bibinfo {title} {A practical introduction to tensor networks: Matrix product states and projected entangled pair states},\ }\href {https://doi.org/https://doi.org/10.1016/j.aop.2014.06.013} {\bibfield  {journal} {\bibinfo  {journal} {Annals of Physics}\ }\textbf {\bibinfo {volume} {349}},\ \bibinfo {pages} {117} (\bibinfo {year} {2014})}\BibitemShut {NoStop}%
\bibitem [{\citenamefont {Silvi}\ \emph {et~al.}(2019)\citenamefont {Silvi}, \citenamefont {Tschirsich}, \citenamefont {Gerster}, \citenamefont {Jünemann}, \citenamefont {Jaschke}, \citenamefont {Rizzi},\ and\ \citenamefont {Montangero}}]{silvi2019tensor}%
  \BibitemOpen
  \bibfield  {author} {\bibinfo {author} {\bibfnamefont {P.}~\bibnamefont {Silvi}}, \bibinfo {author} {\bibfnamefont {F.}~\bibnamefont {Tschirsich}}, \bibinfo {author} {\bibfnamefont {M.}~\bibnamefont {Gerster}}, \bibinfo {author} {\bibfnamefont {J.}~\bibnamefont {Jünemann}}, \bibinfo {author} {\bibfnamefont {D.}~\bibnamefont {Jaschke}}, \bibinfo {author} {\bibfnamefont {M.}~\bibnamefont {Rizzi}},\ and\ \bibinfo {author} {\bibfnamefont {S.}~\bibnamefont {Montangero}},\ }\bibfield  {title} {\bibinfo {title} {{The Tensor Networks Anthology: Simulation techniques for many-body quantum lattice systems}},\ }\href {https://doi.org/10.21468/SciPostPhysLectNotes.8} {\bibfield  {journal} {\bibinfo  {journal} {SciPost Phys. Lect. Notes}\ ,\ \bibinfo {pages} {8}} (\bibinfo {year} {2019})}\BibitemShut {NoStop}%
\bibitem [{\citenamefont {Ran}\ \emph {et~al.}(2020)\citenamefont {Ran}, \citenamefont {Tirrito}, \citenamefont {Peng}, \citenamefont {Chen}, \citenamefont {Tagliacozzo}, \citenamefont {Su},\ and\ \citenamefont {Lewenstein}}]{ran2020tensor}%
  \BibitemOpen
  \bibfield  {author} {\bibinfo {author} {\bibfnamefont {S.-J.}\ \bibnamefont {Ran}}, \bibinfo {author} {\bibfnamefont {E.}~\bibnamefont {Tirrito}}, \bibinfo {author} {\bibfnamefont {C.}~\bibnamefont {Peng}}, \bibinfo {author} {\bibfnamefont {X.}~\bibnamefont {Chen}}, \bibinfo {author} {\bibfnamefont {L.}~\bibnamefont {Tagliacozzo}}, \bibinfo {author} {\bibfnamefont {G.}~\bibnamefont {Su}},\ and\ \bibinfo {author} {\bibfnamefont {M.}~\bibnamefont {Lewenstein}},\ }\href {https://doi.org/10.1007/978-3-030-34489-4} {\emph {\bibinfo {title} {Tensor network contractions: methods and applications to quantum many-body systems}}},\ Vol.\ \bibinfo {volume} {964}\ (\bibinfo  {publisher} {Springer Nature},\ \bibinfo {year} {2020})\BibitemShut {NoStop}%
\bibitem [{\citenamefont {Cirac}\ \emph {et~al.}(2021)\citenamefont {Cirac}, \citenamefont {P\'erez-Garc\'{\i}a}, \citenamefont {Schuch},\ and\ \citenamefont {Verstraete}}]{cirac2021matrix}%
  \BibitemOpen
  \bibfield  {author} {\bibinfo {author} {\bibfnamefont {J.~I.}\ \bibnamefont {Cirac}}, \bibinfo {author} {\bibfnamefont {D.}~\bibnamefont {P\'erez-Garc\'{\i}a}}, \bibinfo {author} {\bibfnamefont {N.}~\bibnamefont {Schuch}},\ and\ \bibinfo {author} {\bibfnamefont {F.}~\bibnamefont {Verstraete}},\ }\bibfield  {title} {\bibinfo {title} {Matrix product states and projected entangled pair states: Concepts, symmetries, theorems},\ }\href {https://doi.org/10.1103/RevModPhys.93.045003} {\bibfield  {journal} {\bibinfo  {journal} {Rev. Mod. Phys.}\ }\textbf {\bibinfo {volume} {93}},\ \bibinfo {pages} {045003} (\bibinfo {year} {2021})}\BibitemShut {NoStop}%
\bibitem [{\citenamefont {Bañuls}(2023)}]{banuls2023tensor}%
  \BibitemOpen
  \bibfield  {author} {\bibinfo {author} {\bibfnamefont {M.~C.}\ \bibnamefont {Bañuls}},\ }\bibfield  {title} {\bibinfo {title} {Tensor network algorithms: A route map},\ }\href {https://doi.org/https://doi.org/10.1146/annurev-conmatphys-040721-022705} {\bibfield  {journal} {\bibinfo  {journal} {Annual Review of Condensed Matter Physics}\ }\textbf {\bibinfo {volume} {14}},\ \bibinfo {pages} {173} (\bibinfo {year} {2023})}\BibitemShut {NoStop}%
\bibitem [{\citenamefont {Or{\'u}s}(2019)}]{orus2019tensor}%
  \BibitemOpen
  \bibfield  {author} {\bibinfo {author} {\bibfnamefont {R.}~\bibnamefont {Or{\'u}s}},\ }\bibfield  {title} {\bibinfo {title} {Tensor networks for complex quantum systems},\ }\href {https://doi.org/10.1038/s42254-019-0086-7} {\bibfield  {journal} {\bibinfo  {journal} {Nature Reviews Physics}\ }\textbf {\bibinfo {volume} {1}},\ \bibinfo {pages} {538} (\bibinfo {year} {2019})}\BibitemShut {NoStop}%
\bibitem [{\citenamefont {Lin}\ \emph {et~al.}(2021)\citenamefont {Lin}, \citenamefont {Dilip}, \citenamefont {Green}, \citenamefont {Smith},\ and\ \citenamefont {Pollmann}}]{lin2021real}%
  \BibitemOpen
  \bibfield  {author} {\bibinfo {author} {\bibfnamefont {S.-H.}\ \bibnamefont {Lin}}, \bibinfo {author} {\bibfnamefont {R.}~\bibnamefont {Dilip}}, \bibinfo {author} {\bibfnamefont {A.~G.}\ \bibnamefont {Green}}, \bibinfo {author} {\bibfnamefont {A.}~\bibnamefont {Smith}},\ and\ \bibinfo {author} {\bibfnamefont {F.}~\bibnamefont {Pollmann}},\ }\bibfield  {title} {\bibinfo {title} {Real- and imaginary-time evolution with compressed quantum circuits},\ }\href {https://doi.org/10.1103/PRXQuantum.2.010342} {\bibfield  {journal} {\bibinfo  {journal} {PRX Quantum}\ }\textbf {\bibinfo {volume} {2}},\ \bibinfo {pages} {010342} (\bibinfo {year} {2021})}\BibitemShut {NoStop}%
\bibitem [{\citenamefont {Rudolph}\ \emph {et~al.}(2023)\citenamefont {Rudolph}, \citenamefont {Miller}, \citenamefont {Motlagh}, \citenamefont {Chen}, \citenamefont {Acharya},\ and\ \citenamefont {Perdomo-Ortiz}}]{rudolph2023synergistic}%
  \BibitemOpen
  \bibfield  {author} {\bibinfo {author} {\bibfnamefont {M.~S.}\ \bibnamefont {Rudolph}}, \bibinfo {author} {\bibfnamefont {J.}~\bibnamefont {Miller}}, \bibinfo {author} {\bibfnamefont {D.}~\bibnamefont {Motlagh}}, \bibinfo {author} {\bibfnamefont {J.}~\bibnamefont {Chen}}, \bibinfo {author} {\bibfnamefont {A.}~\bibnamefont {Acharya}},\ and\ \bibinfo {author} {\bibfnamefont {A.}~\bibnamefont {Perdomo-Ortiz}},\ }\bibfield  {title} {\bibinfo {title} {Synergistic pretraining of parametrized quantum circuits via tensor networks},\ }\href {https://doi.org/10.1038/s41467-023-43908-6} {\bibfield  {journal} {\bibinfo  {journal} {Nature Communications}\ }\textbf {\bibinfo {volume} {14}},\ \bibinfo {pages} {8367} (\bibinfo {year} {2023})}\BibitemShut {NoStop}%
\bibitem [{\citenamefont {Anselme~Martin}\ \emph {et~al.}(2024)\citenamefont {Anselme~Martin}, \citenamefont {Ayral}, \citenamefont {Jamet}, \citenamefont {Ran\v{c}i\'{c}},\ and\ \citenamefont {Simon}}]{martin2024combining}%
  \BibitemOpen
  \bibfield  {author} {\bibinfo {author} {\bibfnamefont {B.}~\bibnamefont {Anselme~Martin}}, \bibinfo {author} {\bibfnamefont {T.}~\bibnamefont {Ayral}}, \bibinfo {author} {\bibfnamefont {F.}~\bibnamefont {Jamet}}, \bibinfo {author} {\bibfnamefont {M.~J.}\ \bibnamefont {Ran\v{c}i\'{c}}},\ and\ \bibinfo {author} {\bibfnamefont {P.}~\bibnamefont {Simon}},\ }\bibfield  {title} {\bibinfo {title} {Combining matrix product states and noisy quantum computers for quantum simulation},\ }\href {https://doi.org/10.1103/PhysRevA.109.062437} {\bibfield  {journal} {\bibinfo  {journal} {Phys. Rev. A}\ }\textbf {\bibinfo {volume} {109}},\ \bibinfo {pages} {062437} (\bibinfo {year} {2024})}\BibitemShut {NoStop}%
\bibitem [{\citenamefont {Berry}\ \emph {et~al.}(2025)\citenamefont {Berry}, \citenamefont {Tong}, \citenamefont {Khattar}, \citenamefont {White}, \citenamefont {Kim}, \citenamefont {Low}, \citenamefont {Boixo}, \citenamefont {Ding}, \citenamefont {Lin}, \citenamefont {Lee} \emph {et~al.}}]{berry2024rapid}%
  \BibitemOpen
  \bibfield  {author} {\bibinfo {author} {\bibfnamefont {D.~W.}\ \bibnamefont {Berry}}, \bibinfo {author} {\bibfnamefont {Y.}~\bibnamefont {Tong}}, \bibinfo {author} {\bibfnamefont {T.}~\bibnamefont {Khattar}}, \bibinfo {author} {\bibfnamefont {A.}~\bibnamefont {White}}, \bibinfo {author} {\bibfnamefont {T.~I.}\ \bibnamefont {Kim}}, \bibinfo {author} {\bibfnamefont {G.~H.}\ \bibnamefont {Low}}, \bibinfo {author} {\bibfnamefont {S.}~\bibnamefont {Boixo}}, \bibinfo {author} {\bibfnamefont {Z.}~\bibnamefont {Ding}}, \bibinfo {author} {\bibfnamefont {L.}~\bibnamefont {Lin}}, \bibinfo {author} {\bibfnamefont {S.}~\bibnamefont {Lee}}, \emph {et~al.},\ }\bibfield  {title} {\bibinfo {title} {{Rapid Initial-State Preparation for the Quantum Simulation of Strongly Correlated Molecules}},\ }\href {https://doi.org/10.1103/PRXQuantum.6.020327} {\bibfield  {journal} {\bibinfo  {journal} {PRX Quantum}\ }\textbf {\bibinfo {volume} {6}},\ \bibinfo {pages} {020327} (\bibinfo {year} {2025})}\BibitemShut {NoStop}%
\bibitem [{\citenamefont {Kim}\ \emph {et~al.}(2023)\citenamefont {Kim}, \citenamefont {Eddins}, \citenamefont {Anand}, \citenamefont {Wei}, \citenamefont {van~den Berg}, \citenamefont {Rosenblatt}, \citenamefont {Nayfeh}, \citenamefont {Wu}, \citenamefont {Zaletel}, \citenamefont {Temme},\ and\ \citenamefont {Kandala}}]{kim2023evidence}%
  \BibitemOpen
  \bibfield  {author} {\bibinfo {author} {\bibfnamefont {Y.}~\bibnamefont {Kim}}, \bibinfo {author} {\bibfnamefont {A.}~\bibnamefont {Eddins}}, \bibinfo {author} {\bibfnamefont {S.}~\bibnamefont {Anand}}, \bibinfo {author} {\bibfnamefont {K.~X.}\ \bibnamefont {Wei}}, \bibinfo {author} {\bibfnamefont {E.}~\bibnamefont {van~den Berg}}, \bibinfo {author} {\bibfnamefont {S.}~\bibnamefont {Rosenblatt}}, \bibinfo {author} {\bibfnamefont {H.}~\bibnamefont {Nayfeh}}, \bibinfo {author} {\bibfnamefont {Y.}~\bibnamefont {Wu}}, \bibinfo {author} {\bibfnamefont {M.}~\bibnamefont {Zaletel}}, \bibinfo {author} {\bibfnamefont {K.}~\bibnamefont {Temme}},\ and\ \bibinfo {author} {\bibfnamefont {A.}~\bibnamefont {Kandala}},\ }\bibfield  {title} {\bibinfo {title} {Evidence for the utility of quantum computing before fault tolerance},\ }\href {https://doi.org/10.1038/s41586-023-06096-3} {\bibfield  {journal} {\bibinfo  {journal} {Nature}\ }\textbf {\bibinfo {volume} {618}},\ \bibinfo {pages} {500–505} (\bibinfo {year} {2023})}\BibitemShut {NoStop}%
\bibitem [{\citenamefont {Tindall}\ \emph {et~al.}(2024)\citenamefont {Tindall}, \citenamefont {Fishman}, \citenamefont {Stoudenmire},\ and\ \citenamefont {Sels}}]{tindall2024efficient}%
  \BibitemOpen
  \bibfield  {author} {\bibinfo {author} {\bibfnamefont {J.}~\bibnamefont {Tindall}}, \bibinfo {author} {\bibfnamefont {M.}~\bibnamefont {Fishman}}, \bibinfo {author} {\bibfnamefont {E.~M.}\ \bibnamefont {Stoudenmire}},\ and\ \bibinfo {author} {\bibfnamefont {D.}~\bibnamefont {Sels}},\ }\bibfield  {title} {\bibinfo {title} {{Efficient Tensor Network Simulation of IBM's Eagle Kicked Ising Experiment}},\ }\href {https://doi.org/10.1103/PRXQuantum.5.010308} {\bibfield  {journal} {\bibinfo  {journal} {PRX Quantum}\ }\textbf {\bibinfo {volume} {5}},\ \bibinfo {pages} {010308} (\bibinfo {year} {2024})}\BibitemShut {NoStop}%
\bibitem [{\citenamefont {Begušić}\ \emph {et~al.}(2024)\citenamefont {Begušić}, \citenamefont {Gray},\ and\ \citenamefont {Chan}}]{begusic2024fast}%
  \BibitemOpen
  \bibfield  {author} {\bibinfo {author} {\bibfnamefont {T.}~\bibnamefont {Begušić}}, \bibinfo {author} {\bibfnamefont {J.}~\bibnamefont {Gray}},\ and\ \bibinfo {author} {\bibfnamefont {G.~K.-L.}\ \bibnamefont {Chan}},\ }\bibfield  {title} {\bibinfo {title} {Fast and converged classical simulations of evidence for the utility of quantum computing before fault tolerance},\ }\href {https://doi.org/10.1126/sciadv.adk4321} {\bibfield  {journal} {\bibinfo  {journal} {Science Advances}\ }\textbf {\bibinfo {volume} {10}},\ \bibinfo {pages} {eadk4321} (\bibinfo {year} {2024})}\BibitemShut {NoStop}%
\bibitem [{\citenamefont {Patra}\ \emph {et~al.}(2024)\citenamefont {Patra}, \citenamefont {Jahromi}, \citenamefont {Singh},\ and\ \citenamefont {Or\'us}}]{patra2024efficient}%
  \BibitemOpen
  \bibfield  {author} {\bibinfo {author} {\bibfnamefont {S.}~\bibnamefont {Patra}}, \bibinfo {author} {\bibfnamefont {S.~S.}\ \bibnamefont {Jahromi}}, \bibinfo {author} {\bibfnamefont {S.}~\bibnamefont {Singh}},\ and\ \bibinfo {author} {\bibfnamefont {R.}~\bibnamefont {Or\'us}},\ }\bibfield  {title} {\bibinfo {title} {{Efficient tensor network simulation of IBM's largest quantum processors}},\ }\href {https://doi.org/10.1103/PhysRevResearch.6.013326} {\bibfield  {journal} {\bibinfo  {journal} {Phys. Rev. Res.}\ }\textbf {\bibinfo {volume} {6}},\ \bibinfo {pages} {013326} (\bibinfo {year} {2024})}\BibitemShut {NoStop}%
\bibitem [{\citenamefont {Liao}\ \emph {et~al.}(2023)\citenamefont {Liao}, \citenamefont {Wang}, \citenamefont {Zhou}, \citenamefont {Zhang},\ and\ \citenamefont {Xiang}}]{liao2023simulation}%
  \BibitemOpen
  \bibfield  {author} {\bibinfo {author} {\bibfnamefont {H.-J.}\ \bibnamefont {Liao}}, \bibinfo {author} {\bibfnamefont {K.}~\bibnamefont {Wang}}, \bibinfo {author} {\bibfnamefont {Z.-S.}\ \bibnamefont {Zhou}}, \bibinfo {author} {\bibfnamefont {P.}~\bibnamefont {Zhang}},\ and\ \bibinfo {author} {\bibfnamefont {T.}~\bibnamefont {Xiang}},\ }\href {https://arxiv.org/abs/2308.03082} {\bibinfo {title} {Simulation of {IBM}'s kicked {I}sing experiment with {P}rojected {E}ntangled {P}air {O}perator}} (\bibinfo {year} {2023}),\ \Eprint {https://arxiv.org/abs/2308.03082} {arXiv:2308.03082 [quant-ph]} \BibitemShut {NoStop}%
\bibitem [{\citenamefont {King}\ \emph {et~al.}(2023)\citenamefont {King}, \citenamefont {Raymond}, \citenamefont {Lanting}, \citenamefont {Harris}, \citenamefont {Zucca}, \citenamefont {Altomare}, \citenamefont {Berkley}, \citenamefont {Boothby}, \citenamefont {Ejtemaee}, \citenamefont {Enderud} \emph {et~al.}}]{king2023quantum}%
  \BibitemOpen
  \bibfield  {author} {\bibinfo {author} {\bibfnamefont {A.~D.}\ \bibnamefont {King}}, \bibinfo {author} {\bibfnamefont {J.}~\bibnamefont {Raymond}}, \bibinfo {author} {\bibfnamefont {T.}~\bibnamefont {Lanting}}, \bibinfo {author} {\bibfnamefont {R.}~\bibnamefont {Harris}}, \bibinfo {author} {\bibfnamefont {A.}~\bibnamefont {Zucca}}, \bibinfo {author} {\bibfnamefont {F.}~\bibnamefont {Altomare}}, \bibinfo {author} {\bibfnamefont {A.~J.}\ \bibnamefont {Berkley}}, \bibinfo {author} {\bibfnamefont {K.}~\bibnamefont {Boothby}}, \bibinfo {author} {\bibfnamefont {S.}~\bibnamefont {Ejtemaee}}, \bibinfo {author} {\bibfnamefont {C.}~\bibnamefont {Enderud}}, \emph {et~al.},\ }\bibfield  {title} {\bibinfo {title} {Quantum critical dynamics in a 5,000-qubit programmable spin glass},\ }\href {https://doi.org/10.1038/s41586-023-05867-2} {\bibfield  {journal} {\bibinfo  {journal} {Nature}\ }\textbf {\bibinfo {volume} {617}},\ \bibinfo {pages} {61–66} (\bibinfo {year} {2023})}\BibitemShut {NoStop}%
\bibitem [{\citenamefont {King}\ \emph {et~al.}(2025)\citenamefont {King}, \citenamefont {Nocera}, \citenamefont {Rams}, \citenamefont {Dziarmaga}, \citenamefont {Wiersema}, \citenamefont {Bernoudy}, \citenamefont {Raymond}, \citenamefont {Kaushal}, \citenamefont {Heinsdorf}, \citenamefont {Harris} \emph {et~al.}}]{king2024computationalsupremacyquantumsimulation}%
  \BibitemOpen
  \bibfield  {author} {\bibinfo {author} {\bibfnamefont {A.~D.}\ \bibnamefont {King}}, \bibinfo {author} {\bibfnamefont {A.}~\bibnamefont {Nocera}}, \bibinfo {author} {\bibfnamefont {M.~M.}\ \bibnamefont {Rams}}, \bibinfo {author} {\bibfnamefont {J.}~\bibnamefont {Dziarmaga}}, \bibinfo {author} {\bibfnamefont {R.}~\bibnamefont {Wiersema}}, \bibinfo {author} {\bibfnamefont {W.}~\bibnamefont {Bernoudy}}, \bibinfo {author} {\bibfnamefont {J.}~\bibnamefont {Raymond}}, \bibinfo {author} {\bibfnamefont {N.}~\bibnamefont {Kaushal}}, \bibinfo {author} {\bibfnamefont {N.}~\bibnamefont {Heinsdorf}}, \bibinfo {author} {\bibfnamefont {R.}~\bibnamefont {Harris}}, \emph {et~al.},\ }\bibfield  {title} {\bibinfo {title} {Beyond-classical computation in quantum simulation},\ }\href {https://doi.org/10.1126/science.ado6285} {\bibfield  {journal} {\bibinfo  {journal} {Science}\ }\textbf {\bibinfo {volume} {388}},\ \bibinfo {pages} {199} (\bibinfo {year} {2025})}\BibitemShut {NoStop}%
\bibitem [{\citenamefont {Han}\ \emph {et~al.}(2018)\citenamefont {Han}, \citenamefont {Wang}, \citenamefont {Fan}, \citenamefont {Wang},\ and\ \citenamefont {Zhang}}]{han2018unsupervised}%
  \BibitemOpen
  \bibfield  {author} {\bibinfo {author} {\bibfnamefont {Z.-Y.}\ \bibnamefont {Han}}, \bibinfo {author} {\bibfnamefont {J.}~\bibnamefont {Wang}}, \bibinfo {author} {\bibfnamefont {H.}~\bibnamefont {Fan}}, \bibinfo {author} {\bibfnamefont {L.}~\bibnamefont {Wang}},\ and\ \bibinfo {author} {\bibfnamefont {P.}~\bibnamefont {Zhang}},\ }\bibfield  {title} {\bibinfo {title} {{Unsupervised Generative Modeling Using Matrix Product States}},\ }\href {https://doi.org/10.1103/PhysRevX.8.031012} {\bibfield  {journal} {\bibinfo  {journal} {Phys. Rev. X}\ }\textbf {\bibinfo {volume} {8}},\ \bibinfo {pages} {031012} (\bibinfo {year} {2018})}\BibitemShut {NoStop}%
\bibitem [{\citenamefont {Lopez-Piqueres}\ \emph {et~al.}(2023)\citenamefont {Lopez-Piqueres}, \citenamefont {Chen},\ and\ \citenamefont {Perdomo-Ortiz}}]{lopez2023symmetric}%
  \BibitemOpen
  \bibfield  {author} {\bibinfo {author} {\bibfnamefont {J.}~\bibnamefont {Lopez-Piqueres}}, \bibinfo {author} {\bibfnamefont {J.}~\bibnamefont {Chen}},\ and\ \bibinfo {author} {\bibfnamefont {A.}~\bibnamefont {Perdomo-Ortiz}},\ }\bibfield  {title} {\bibinfo {title} {Symmetric tensor networks for generative modeling and constrained combinatorial optimization},\ }\href {https://doi.org/10.1088/2632-2153/ace0f5} {\bibfield  {journal} {\bibinfo  {journal} {Machine Learning: Science and Technology}\ }\textbf {\bibinfo {volume} {4}},\ \bibinfo {pages} {035009} (\bibinfo {year} {2023})}\BibitemShut {NoStop}%
\bibitem [{\citenamefont {Alcazar}\ \emph {et~al.}(2024)\citenamefont {Alcazar}, \citenamefont {Ghazi~Vakili}, \citenamefont {Kalayci},\ and\ \citenamefont {Perdomo-Ortiz}}]{alcazar2024enhancing}%
  \BibitemOpen
  \bibfield  {author} {\bibinfo {author} {\bibfnamefont {J.}~\bibnamefont {Alcazar}}, \bibinfo {author} {\bibfnamefont {M.}~\bibnamefont {Ghazi~Vakili}}, \bibinfo {author} {\bibfnamefont {C.~B.}\ \bibnamefont {Kalayci}},\ and\ \bibinfo {author} {\bibfnamefont {A.}~\bibnamefont {Perdomo-Ortiz}},\ }\bibfield  {title} {\bibinfo {title} {Enhancing combinatorial optimization with classical and quantum generative models},\ }\href {https://doi.org/10.1038/s41467-024-46959-5} {\bibfield  {journal} {\bibinfo  {journal} {Nature Communications}\ }\textbf {\bibinfo {volume} {15}},\ \bibinfo {pages} {2761} (\bibinfo {year} {2024})}\BibitemShut {NoStop}%
\bibitem [{\citenamefont {Mugel}\ \emph {et~al.}(2022)\citenamefont {Mugel}, \citenamefont {Kuchkovsky}, \citenamefont {S\'anchez}, \citenamefont {Fern\'andez-Lorenzo}, \citenamefont {Luis-Hita}, \citenamefont {Lizaso},\ and\ \citenamefont {Or\'us}}]{mugel2022dynamic}%
  \BibitemOpen
  \bibfield  {author} {\bibinfo {author} {\bibfnamefont {S.}~\bibnamefont {Mugel}}, \bibinfo {author} {\bibfnamefont {C.}~\bibnamefont {Kuchkovsky}}, \bibinfo {author} {\bibfnamefont {E.}~\bibnamefont {S\'anchez}}, \bibinfo {author} {\bibfnamefont {S.}~\bibnamefont {Fern\'andez-Lorenzo}}, \bibinfo {author} {\bibfnamefont {J.}~\bibnamefont {Luis-Hita}}, \bibinfo {author} {\bibfnamefont {E.}~\bibnamefont {Lizaso}},\ and\ \bibinfo {author} {\bibfnamefont {R.}~\bibnamefont {Or\'us}},\ }\bibfield  {title} {\bibinfo {title} {Dynamic portfolio optimization with real datasets using quantum processors and quantum-inspired tensor networks},\ }\href {https://doi.org/10.1103/PhysRevResearch.4.013006} {\bibfield  {journal} {\bibinfo  {journal} {Phys. Rev. Res.}\ }\textbf {\bibinfo {volume} {4}},\ \bibinfo {pages} {013006} (\bibinfo {year} {2022})}\BibitemShut {NoStop}%
\bibitem [{\citenamefont {Hao}\ \emph {et~al.}(2022)\citenamefont {Hao}, \citenamefont {Huang}, \citenamefont {Jia},\ and\ \citenamefont {Peng}}]{hao2022quantum}%
  \BibitemOpen
  \bibfield  {author} {\bibinfo {author} {\bibfnamefont {T.}~\bibnamefont {Hao}}, \bibinfo {author} {\bibfnamefont {X.}~\bibnamefont {Huang}}, \bibinfo {author} {\bibfnamefont {C.}~\bibnamefont {Jia}},\ and\ \bibinfo {author} {\bibfnamefont {C.}~\bibnamefont {Peng}},\ }\bibfield  {title} {\bibinfo {title} {{A Quantum-Inspired Tensor Network Algorithm for Constrained Combinatorial Optimization Problems}},\ }\bibfield  {journal} {\bibinfo  {journal} {Frontiers in Physics}\ }\textbf {\bibinfo {volume} {10}},\ \href {https://doi.org/10.3389/fphy.2022.906590} {10.3389/fphy.2022.906590} (\bibinfo {year} {2022})\BibitemShut {NoStop}%
\bibitem [{\citenamefont {Patra}\ \emph {et~al.}(2025)\citenamefont {Patra}, \citenamefont {Singh},\ and\ \citenamefont {Or\'us}}]{patra2024projected}%
  \BibitemOpen
  \bibfield  {author} {\bibinfo {author} {\bibfnamefont {S.}~\bibnamefont {Patra}}, \bibinfo {author} {\bibfnamefont {S.}~\bibnamefont {Singh}},\ and\ \bibinfo {author} {\bibfnamefont {R.}~\bibnamefont {Or\'us}},\ }\bibfield  {title} {\bibinfo {title} {Projected entangled pair states with flexible geometry},\ }\href {https://doi.org/10.1103/PhysRevResearch.7.L012002} {\bibfield  {journal} {\bibinfo  {journal} {Phys. Rev. Res.}\ }\textbf {\bibinfo {volume} {7}},\ \bibinfo {pages} {L012002} (\bibinfo {year} {2025})}\BibitemShut {NoStop}%
\bibitem [{\citenamefont {Pelofske}\ \emph {et~al.}(2023)\citenamefont {Pelofske}, \citenamefont {B{\"a}rtschi},\ and\ \citenamefont {Eidenbenz}}]{pelofske2023quantum}%
  \BibitemOpen
  \bibfield  {author} {\bibinfo {author} {\bibfnamefont {E.}~\bibnamefont {Pelofske}}, \bibinfo {author} {\bibfnamefont {A.}~\bibnamefont {B{\"a}rtschi}},\ and\ \bibinfo {author} {\bibfnamefont {S.}~\bibnamefont {Eidenbenz}},\ }\bibfield  {title} {\bibinfo {title} {{Quantum Annealing vs. QAOA: 127 Qubit Higher-Order Ising Problems on NISQ Computers}},\ }in\ \href {https://doi.org/10.1007/978-3-031-32041-5_13} {\emph {\bibinfo {booktitle} {{High Performance Computing: 38th International Conference, ISC High Performance 2023, Hamburg, Germany, May 21--25, 2023, Proceedings}}}},\ \bibinfo {series} {Lecture Notes in Computer Science}, Vol.\ \bibinfo {volume} {13944}\ (\bibinfo  {publisher} {Springer Nature Switzerland},\ \bibinfo {address} {Cham},\ \bibinfo {year} {2023})\ pp.\ \bibinfo {pages} {240--258}\BibitemShut {NoStop}%
\bibitem [{\citenamefont {Pelofske}\ \emph {et~al.}(2024)\citenamefont {Pelofske}, \citenamefont {Bärtschi},\ and\ \citenamefont {Eidenbenz}}]{pelofske2024short-depth}%
  \BibitemOpen
  \bibfield  {author} {\bibinfo {author} {\bibfnamefont {E.}~\bibnamefont {Pelofske}}, \bibinfo {author} {\bibfnamefont {A.}~\bibnamefont {Bärtschi}},\ and\ \bibinfo {author} {\bibfnamefont {S.}~\bibnamefont {Eidenbenz}},\ }\bibfield  {title} {\bibinfo {title} {{Short-depth QAOA circuits and quantum annealing on higher-order ising models}},\ }\href {https://doi.org/10.1038/s41534-024-00825-w} {\bibfield  {journal} {\bibinfo  {journal} {npj Quantum Information}\ }\textbf {\bibinfo {volume} {10}},\ \bibinfo {pages} {1–19} (\bibinfo {year} {2024})}\BibitemShut {NoStop}%
\bibitem [{\citenamefont {Barron}\ \emph {et~al.}(2024)\citenamefont {Barron}, \citenamefont {Egger}, \citenamefont {Pelofske}, \citenamefont {Bärtschi}, \citenamefont {Eidenbenz}, \citenamefont {Lehmkuehler},\ and\ \citenamefont {Woerner}}]{barron2023provableboundsnoisefreeexpectation}%
  \BibitemOpen
  \bibfield  {author} {\bibinfo {author} {\bibfnamefont {S.~V.}\ \bibnamefont {Barron}}, \bibinfo {author} {\bibfnamefont {D.~J.}\ \bibnamefont {Egger}}, \bibinfo {author} {\bibfnamefont {E.}~\bibnamefont {Pelofske}}, \bibinfo {author} {\bibfnamefont {A.}~\bibnamefont {Bärtschi}}, \bibinfo {author} {\bibfnamefont {S.}~\bibnamefont {Eidenbenz}}, \bibinfo {author} {\bibfnamefont {M.}~\bibnamefont {Lehmkuehler}},\ and\ \bibinfo {author} {\bibfnamefont {S.}~\bibnamefont {Woerner}},\ }\bibfield  {title} {\bibinfo {title} {Provable bounds for noise-free expectation values computed from noisy samples},\ }\href {https://doi.org/10.1038/s43588-024-00709-1} {\bibfield  {journal} {\bibinfo  {journal} {Nature Computational Science}\ }\textbf {\bibinfo {volume} {4}},\ \bibinfo {pages} {865–875} (\bibinfo {year} {2024})}\BibitemShut {NoStop}%
\bibitem [{\citenamefont {Sachdeva}\ \emph {et~al.}(2024)\citenamefont {Sachdeva}, \citenamefont {Hartnett}, \citenamefont {Maity}, \citenamefont {Marsh}, \citenamefont {Wang}, \citenamefont {Winick}, \citenamefont {Dougherty}, \citenamefont {Canuto}, \citenamefont {Chong}, \citenamefont {Hush}, \citenamefont {Mundada}, \citenamefont {Bentley}, \citenamefont {Biercuk},\ and\ \citenamefont {Baum}}]{sachdeva2024quantum}%
  \BibitemOpen
  \bibfield  {author} {\bibinfo {author} {\bibfnamefont {N.}~\bibnamefont {Sachdeva}}, \bibinfo {author} {\bibfnamefont {G.~S.}\ \bibnamefont {Hartnett}}, \bibinfo {author} {\bibfnamefont {S.}~\bibnamefont {Maity}}, \bibinfo {author} {\bibfnamefont {S.}~\bibnamefont {Marsh}}, \bibinfo {author} {\bibfnamefont {Y.}~\bibnamefont {Wang}}, \bibinfo {author} {\bibfnamefont {A.}~\bibnamefont {Winick}}, \bibinfo {author} {\bibfnamefont {R.}~\bibnamefont {Dougherty}}, \bibinfo {author} {\bibfnamefont {D.}~\bibnamefont {Canuto}}, \bibinfo {author} {\bibfnamefont {Y.~Q.}\ \bibnamefont {Chong}}, \bibinfo {author} {\bibfnamefont {M.}~\bibnamefont {Hush}}, \bibinfo {author} {\bibfnamefont {P.~S.}\ \bibnamefont {Mundada}}, \bibinfo {author} {\bibfnamefont {C.~D.~B.}\ \bibnamefont {Bentley}}, \bibinfo {author} {\bibfnamefont {M.~J.}\ \bibnamefont {Biercuk}},\ and\ \bibinfo {author} {\bibfnamefont {Y.}~\bibnamefont {Baum}},\ }\href {https://arxiv.org/abs/2406.01743} {\bibinfo {title} {{Quantum optimization using a 127-qubit gate-model IBM quantum computer can outperform quantum annealers for nontrivial binary optimization problems}}} (\bibinfo {year} {2024}),\ \Eprint {https://arxiv.org/abs/2406.01743} {arXiv:2406.01743 [quant-ph]} \BibitemShut {NoStop}%
\bibitem [{\citenamefont {Demirplak}\ and\ \citenamefont {Rice}(2003)}]{demirplak2003adiabatic}%
  \BibitemOpen
  \bibfield  {author} {\bibinfo {author} {\bibfnamefont {M.}~\bibnamefont {Demirplak}}\ and\ \bibinfo {author} {\bibfnamefont {S.~A.}\ \bibnamefont {Rice}},\ }\bibfield  {title} {\bibinfo {title} {{Adiabatic Population Transfer with Control Fields}},\ }\href {https://doi.org/10.1021/jp030708a} {\bibfield  {journal} {\bibinfo  {journal} {The Journal of Physical Chemistry A}\ }\textbf {\bibinfo {volume} {107}},\ \bibinfo {pages} {9937} (\bibinfo {year} {2003})}\BibitemShut {NoStop}%
\bibitem [{\citenamefont {Berry}(2009)}]{berry2009transitionless}%
  \BibitemOpen
  \bibfield  {author} {\bibinfo {author} {\bibfnamefont {M.~V.}\ \bibnamefont {Berry}},\ }\bibfield  {title} {\bibinfo {title} {Transitionless quantum driving},\ }\href {https://doi.org/10.1088/1751-8113/42/36/365303} {\bibfield  {journal} {\bibinfo  {journal} {Journal of Physics A: Mathematical and Theoretical}\ }\textbf {\bibinfo {volume} {42}},\ \bibinfo {pages} {365303} (\bibinfo {year} {2009})}\BibitemShut {NoStop}%
\bibitem [{\citenamefont {Chen}\ \emph {et~al.}(2010)\citenamefont {Chen}, \citenamefont {Ruschhaupt}, \citenamefont {Schmidt}, \citenamefont {del Campo}, \citenamefont {Gu\'ery-Odelin},\ and\ \citenamefont {Muga}}]{chen2010fast}%
  \BibitemOpen
  \bibfield  {author} {\bibinfo {author} {\bibfnamefont {X.}~\bibnamefont {Chen}}, \bibinfo {author} {\bibfnamefont {A.}~\bibnamefont {Ruschhaupt}}, \bibinfo {author} {\bibfnamefont {S.}~\bibnamefont {Schmidt}}, \bibinfo {author} {\bibfnamefont {A.}~\bibnamefont {del Campo}}, \bibinfo {author} {\bibfnamefont {D.}~\bibnamefont {Gu\'ery-Odelin}},\ and\ \bibinfo {author} {\bibfnamefont {J.~G.}\ \bibnamefont {Muga}},\ }\bibfield  {title} {\bibinfo {title} {{Fast Optimal Frictionless Atom Cooling in Harmonic Traps: Shortcut to Adiabaticity}},\ }\href {https://doi.org/10.1103/PhysRevLett.104.063002} {\bibfield  {journal} {\bibinfo  {journal} {Phys. Rev. Lett.}\ }\textbf {\bibinfo {volume} {104}},\ \bibinfo {pages} {063002} (\bibinfo {year} {2010})}\BibitemShut {NoStop}%
\bibitem [{\citenamefont {del Campo}(2013)}]{campo2013shortcuts}%
  \BibitemOpen
  \bibfield  {author} {\bibinfo {author} {\bibfnamefont {A.}~\bibnamefont {del Campo}},\ }\bibfield  {title} {\bibinfo {title} {Shortcuts to adiabaticity by counterdiabatic driving},\ }\href {https://doi.org/10.1103/PhysRevLett.111.100502} {\bibfield  {journal} {\bibinfo  {journal} {Phys. Rev. Lett.}\ }\textbf {\bibinfo {volume} {111}},\ \bibinfo {pages} {100502} (\bibinfo {year} {2013})}\BibitemShut {NoStop}%
\bibitem [{\citenamefont {Chandarana}\ \emph {et~al.}(2022)\citenamefont {Chandarana}, \citenamefont {Hegade}, \citenamefont {Paul}, \citenamefont {Albarr\'an-Arriagada}, \citenamefont {Solano}, \citenamefont {del Campo},\ and\ \citenamefont {Chen}}]{chandarana2022digitized}%
  \BibitemOpen
  \bibfield  {author} {\bibinfo {author} {\bibfnamefont {P.}~\bibnamefont {Chandarana}}, \bibinfo {author} {\bibfnamefont {N.~N.}\ \bibnamefont {Hegade}}, \bibinfo {author} {\bibfnamefont {K.}~\bibnamefont {Paul}}, \bibinfo {author} {\bibfnamefont {F.}~\bibnamefont {Albarr\'an-Arriagada}}, \bibinfo {author} {\bibfnamefont {E.}~\bibnamefont {Solano}}, \bibinfo {author} {\bibfnamefont {A.}~\bibnamefont {del Campo}},\ and\ \bibinfo {author} {\bibfnamefont {X.}~\bibnamefont {Chen}},\ }\bibfield  {title} {\bibinfo {title} {Digitized-counterdiabatic quantum approximate optimization algorithm},\ }\href {https://doi.org/10.1103/PhysRevResearch.4.013141} {\bibfield  {journal} {\bibinfo  {journal} {Phys. Rev. Res.}\ }\textbf {\bibinfo {volume} {4}},\ \bibinfo {pages} {013141} (\bibinfo {year} {2022})}\BibitemShut {NoStop}%
\bibitem [{\citenamefont {Hegade}\ \emph {et~al.}(2022)\citenamefont {Hegade}, \citenamefont {Chen},\ and\ \citenamefont {Solano}}]{hegade2022digitized}%
  \BibitemOpen
  \bibfield  {author} {\bibinfo {author} {\bibfnamefont {N.~N.}\ \bibnamefont {Hegade}}, \bibinfo {author} {\bibfnamefont {X.}~\bibnamefont {Chen}},\ and\ \bibinfo {author} {\bibfnamefont {E.}~\bibnamefont {Solano}},\ }\bibfield  {title} {\bibinfo {title} {Digitized counterdiabatic quantum optimization},\ }\href {https://doi.org/10.1103/PhysRevResearch.4.L042030} {\bibfield  {journal} {\bibinfo  {journal} {Phys. Rev. Res.}\ }\textbf {\bibinfo {volume} {4}},\ \bibinfo {pages} {L042030} (\bibinfo {year} {2022})}\BibitemShut {NoStop}%
\bibitem [{\citenamefont {Cadavid}\ \emph {et~al.}(2025)\citenamefont {Cadavid}, \citenamefont {Dalal}, \citenamefont {Simen}, \citenamefont {Solano},\ and\ \citenamefont {Hegade}}]{cadavid2024biasfielddigitizedcounterdiabaticquantum}%
  \BibitemOpen
  \bibfield  {author} {\bibinfo {author} {\bibfnamefont {A.~G.}\ \bibnamefont {Cadavid}}, \bibinfo {author} {\bibfnamefont {A.}~\bibnamefont {Dalal}}, \bibinfo {author} {\bibfnamefont {A.}~\bibnamefont {Simen}}, \bibinfo {author} {\bibfnamefont {E.}~\bibnamefont {Solano}},\ and\ \bibinfo {author} {\bibfnamefont {N.~N.}\ \bibnamefont {Hegade}},\ }\bibfield  {title} {\bibinfo {title} {Bias-field digitized counterdiabatic quantum optimization},\ }\href {https://doi.org/10.1103/PhysRevResearch.7.L022010} {\bibfield  {journal} {\bibinfo  {journal} {Phys. Rev. Res.}\ }\textbf {\bibinfo {volume} {7}},\ \bibinfo {pages} {L022010} (\bibinfo {year} {2025})}\BibitemShut {NoStop}%
\bibitem [{\citenamefont {Battiti}(2009)}]{Battiti2009}%
  \BibitemOpen
  \bibfield  {author} {\bibinfo {author} {\bibfnamefont {R.}~\bibnamefont {Battiti}},\ }\bibinfo {title} {Maximum satisfiability problem},\ in\ \href {https://doi.org/10.1007/978-0-387-74759-0_364} {\emph {\bibinfo {booktitle} {Encyclopedia of Optimization}}},\ \bibinfo {editor} {edited by\ \bibinfo {editor} {\bibfnamefont {C.~A.}\ \bibnamefont {Floudas}}\ and\ \bibinfo {editor} {\bibfnamefont {P.~M.}\ \bibnamefont {Pardalos}}}\ (\bibinfo  {publisher} {Springer US},\ \bibinfo {address} {Boston, MA},\ \bibinfo {year} {2009})\ pp.\ \bibinfo {pages} {2035--2041}\BibitemShut {NoStop}%
\bibitem [{\citenamefont {Hegade}\ \emph {et~al.}(2021{\natexlab{a}})\citenamefont {Hegade}, \citenamefont {Paul}, \citenamefont {Albarr\'an-Arriagada}, \citenamefont {Chen},\ and\ \citenamefont {Solano}}]{hegade2021factorization}%
  \BibitemOpen
  \bibfield  {author} {\bibinfo {author} {\bibfnamefont {N.~N.}\ \bibnamefont {Hegade}}, \bibinfo {author} {\bibfnamefont {K.}~\bibnamefont {Paul}}, \bibinfo {author} {\bibfnamefont {F.}~\bibnamefont {Albarr\'an-Arriagada}}, \bibinfo {author} {\bibfnamefont {X.}~\bibnamefont {Chen}},\ and\ \bibinfo {author} {\bibfnamefont {E.}~\bibnamefont {Solano}},\ }\bibfield  {title} {\bibinfo {title} {Digitized adiabatic quantum factorization},\ }\href {https://doi.org/10.1103/PhysRevA.104.L050403} {\bibfield  {journal} {\bibinfo  {journal} {Phys. Rev. A}\ }\textbf {\bibinfo {volume} {104}},\ \bibinfo {pages} {L050403} (\bibinfo {year} {2021}{\natexlab{a}})}\BibitemShut {NoStop}%
\bibitem [{\citenamefont {Robert}\ \emph {et~al.}(2021)\citenamefont {Robert}, \citenamefont {Barkoutsos}, \citenamefont {Woerner},\ and\ \citenamefont {Tavernelli}}]{robert2021resource}%
  \BibitemOpen
  \bibfield  {author} {\bibinfo {author} {\bibfnamefont {A.}~\bibnamefont {Robert}}, \bibinfo {author} {\bibfnamefont {P.~K.}\ \bibnamefont {Barkoutsos}}, \bibinfo {author} {\bibfnamefont {S.}~\bibnamefont {Woerner}},\ and\ \bibinfo {author} {\bibfnamefont {I.}~\bibnamefont {Tavernelli}},\ }\bibfield  {title} {\bibinfo {title} {Resource-efficient quantum algorithm for protein folding},\ }\href {https://doi.org/10.1038/s41534-021-00368-4} {\bibfield  {journal} {\bibinfo  {journal} {npj Quantum Information}\ }\textbf {\bibinfo {volume} {7}},\ \bibinfo {pages} {1–5} (\bibinfo {year} {2021})}\BibitemShut {NoStop}%
\bibitem [{\citenamefont {Chandarana}\ \emph {et~al.}(2023)\citenamefont {Chandarana}, \citenamefont {Hegade}, \citenamefont {Montalban}, \citenamefont {Solano},\ and\ \citenamefont {Chen}}]{chandarana2024digitized}%
  \BibitemOpen
  \bibfield  {author} {\bibinfo {author} {\bibfnamefont {P.}~\bibnamefont {Chandarana}}, \bibinfo {author} {\bibfnamefont {N.~N.}\ \bibnamefont {Hegade}}, \bibinfo {author} {\bibfnamefont {I.}~\bibnamefont {Montalban}}, \bibinfo {author} {\bibfnamefont {E.}~\bibnamefont {Solano}},\ and\ \bibinfo {author} {\bibfnamefont {X.}~\bibnamefont {Chen}},\ }\bibfield  {title} {\bibinfo {title} {{Digitized Counterdiabatic Quantum Algorithm for Protein Folding}},\ }\href {https://doi.org/10.1103/PhysRevApplied.20.014024} {\bibfield  {journal} {\bibinfo  {journal} {Phys. Rev. Appl.}\ }\textbf {\bibinfo {volume} {20}},\ \bibinfo {pages} {014024} (\bibinfo {year} {2023})}\BibitemShut {NoStop}%
\bibitem [{\citenamefont {Gardner}(1985)}]{gardner1985spin}%
  \BibitemOpen
  \bibfield  {author} {\bibinfo {author} {\bibfnamefont {E.}~\bibnamefont {Gardner}},\ }\bibfield  {title} {\bibinfo {title} {Spin glasses with p-spin interactions},\ }\href {https://doi.org/https://doi.org/10.1016/0550-3213(85)90374-8} {\bibfield  {journal} {\bibinfo  {journal} {Nuclear Physics B}\ }\textbf {\bibinfo {volume} {257}},\ \bibinfo {pages} {747} (\bibinfo {year} {1985})}\BibitemShut {NoStop}%
\bibitem [{\citenamefont {Kolodrubetz}\ \emph {et~al.}(2017)\citenamefont {Kolodrubetz}, \citenamefont {Sels}, \citenamefont {Mehta},\ and\ \citenamefont {Polkovnikov}}]{kolodrubetz2017geometry}%
  \BibitemOpen
  \bibfield  {author} {\bibinfo {author} {\bibfnamefont {M.}~\bibnamefont {Kolodrubetz}}, \bibinfo {author} {\bibfnamefont {D.}~\bibnamefont {Sels}}, \bibinfo {author} {\bibfnamefont {P.}~\bibnamefont {Mehta}},\ and\ \bibinfo {author} {\bibfnamefont {A.}~\bibnamefont {Polkovnikov}},\ }\bibfield  {title} {\bibinfo {title} {Geometry and non-adiabatic response in quantum and classical systems},\ }\href {https://doi.org/https://doi.org/10.1016/j.physrep.2017.07.001} {\bibfield  {journal} {\bibinfo  {journal} {Physics Reports}\ }\textbf {\bibinfo {volume} {697}},\ \bibinfo {pages} {1} (\bibinfo {year} {2017})}\BibitemShut {NoStop}%
\bibitem [{\citenamefont {Sels}\ and\ \citenamefont {Polkovnikov}(2017)}]{sels2017minimizing}%
  \BibitemOpen
  \bibfield  {author} {\bibinfo {author} {\bibfnamefont {D.}~\bibnamefont {Sels}}\ and\ \bibinfo {author} {\bibfnamefont {A.}~\bibnamefont {Polkovnikov}},\ }\bibfield  {title} {\bibinfo {title} {Minimizing irreversible losses in quantum systems by local counterdiabatic driving},\ }\href {https://doi.org/10.1073/pnas.1619826114} {\bibfield  {journal} {\bibinfo  {journal} {Proceedings of the National Academy of Sciences}\ }\textbf {\bibinfo {volume} {114}},\ \bibinfo {pages} {E3909} (\bibinfo {year} {2017})}\BibitemShut {NoStop}%
\bibitem [{\citenamefont {Claeys}\ \emph {et~al.}(2019)\citenamefont {Claeys}, \citenamefont {Pandey}, \citenamefont {Sels},\ and\ \citenamefont {Polkovnikov}}]{claeys2019floquet}%
  \BibitemOpen
  \bibfield  {author} {\bibinfo {author} {\bibfnamefont {P.~W.}\ \bibnamefont {Claeys}}, \bibinfo {author} {\bibfnamefont {M.}~\bibnamefont {Pandey}}, \bibinfo {author} {\bibfnamefont {D.}~\bibnamefont {Sels}},\ and\ \bibinfo {author} {\bibfnamefont {A.}~\bibnamefont {Polkovnikov}},\ }\bibfield  {title} {\bibinfo {title} {{Floquet-Engineering Counterdiabatic Protocols in Quantum Many-Body Systems}},\ }\href {https://doi.org/10.1103/PhysRevLett.123.090602} {\bibfield  {journal} {\bibinfo  {journal} {Phys. Rev. Lett.}\ }\textbf {\bibinfo {volume} {123}},\ \bibinfo {pages} {090602} (\bibinfo {year} {2019})}\BibitemShut {NoStop}%
\bibitem [{\citenamefont {Hatomura}\ and\ \citenamefont {Takahashi}(2021)}]{hatomura2021controlling}%
  \BibitemOpen
  \bibfield  {author} {\bibinfo {author} {\bibfnamefont {T.}~\bibnamefont {Hatomura}}\ and\ \bibinfo {author} {\bibfnamefont {K.}~\bibnamefont {Takahashi}},\ }\bibfield  {title} {\bibinfo {title} {Controlling and exploring quantum systems by algebraic expression of adiabatic gauge potential},\ }\href {https://doi.org/10.1103/PhysRevA.103.012220} {\bibfield  {journal} {\bibinfo  {journal} {Phys. Rev. A}\ }\textbf {\bibinfo {volume} {103}},\ \bibinfo {pages} {012220} (\bibinfo {year} {2021})}\BibitemShut {NoStop}%
\bibitem [{\citenamefont {Takahashi}\ and\ \citenamefont {del Campo}(2024)}]{takahashi2024shortcuts}%
  \BibitemOpen
  \bibfield  {author} {\bibinfo {author} {\bibfnamefont {K.}~\bibnamefont {Takahashi}}\ and\ \bibinfo {author} {\bibfnamefont {A.}~\bibnamefont {del Campo}},\ }\bibfield  {title} {\bibinfo {title} {{Shortcuts to Adiabaticity in Krylov Space}},\ }\href {https://doi.org/10.1103/PhysRevX.14.011032} {\bibfield  {journal} {\bibinfo  {journal} {Phys. Rev. X}\ }\textbf {\bibinfo {volume} {14}},\ \bibinfo {pages} {011032} (\bibinfo {year} {2024})}\BibitemShut {NoStop}%
\bibitem [{\citenamefont {Romero}\ \emph {et~al.}(2024)\citenamefont {Romero}, \citenamefont {Chen}, \citenamefont {Platero},\ and\ \citenamefont {Ban}}]{romero2024optimizing}%
  \BibitemOpen
  \bibfield  {author} {\bibinfo {author} {\bibfnamefont {S.~V.}\ \bibnamefont {Romero}}, \bibinfo {author} {\bibfnamefont {X.}~\bibnamefont {Chen}}, \bibinfo {author} {\bibfnamefont {G.}~\bibnamefont {Platero}},\ and\ \bibinfo {author} {\bibfnamefont {Y.}~\bibnamefont {Ban}},\ }\bibfield  {title} {\bibinfo {title} {{Optimizing edge-state transfer in a Su-Schrieffer-Heeger chain via hybrid analog-digital strategies}},\ }\href {https://doi.org/10.1103/PhysRevApplied.21.034033} {\bibfield  {journal} {\bibinfo  {journal} {Phys. Rev. Appl.}\ }\textbf {\bibinfo {volume} {21}},\ \bibinfo {pages} {034033} (\bibinfo {year} {2024})}\BibitemShut {NoStop}%
\bibitem [{\citenamefont {Dalal}\ \emph {et~al.}(2024)\citenamefont {Dalal}, \citenamefont {Montalban}, \citenamefont {Hegade}, \citenamefont {Cadavid}, \citenamefont {Solano}, \citenamefont {Awasthi}, \citenamefont {Vodola}, \citenamefont {Jones}, \citenamefont {Weiss},\ and\ \citenamefont {F\"uchsel}}]{dalal2024digitizedcounterdiabaticquantumalgorithms}%
  \BibitemOpen
  \bibfield  {author} {\bibinfo {author} {\bibfnamefont {A.}~\bibnamefont {Dalal}}, \bibinfo {author} {\bibfnamefont {I.}~\bibnamefont {Montalban}}, \bibinfo {author} {\bibfnamefont {N.~N.}\ \bibnamefont {Hegade}}, \bibinfo {author} {\bibfnamefont {A.~G.}\ \bibnamefont {Cadavid}}, \bibinfo {author} {\bibfnamefont {E.}~\bibnamefont {Solano}}, \bibinfo {author} {\bibfnamefont {A.}~\bibnamefont {Awasthi}}, \bibinfo {author} {\bibfnamefont {D.}~\bibnamefont {Vodola}}, \bibinfo {author} {\bibfnamefont {C.}~\bibnamefont {Jones}}, \bibinfo {author} {\bibfnamefont {H.}~\bibnamefont {Weiss}},\ and\ \bibinfo {author} {\bibfnamefont {G.}~\bibnamefont {F\"uchsel}},\ }\bibfield  {title} {\bibinfo {title} {Digitized counterdiabatic quantum algorithms for logistics scheduling},\ }\href {https://doi.org/10.1103/PhysRevApplied.22.064068} {\bibfield  {journal} {\bibinfo  {journal} {Phys. Rev. Appl.}\ }\textbf {\bibinfo {volume} {22}},\ \bibinfo {pages} {064068} (\bibinfo {year} {2024})}\BibitemShut {NoStop}%
\bibitem [{\citenamefont {Hormozi}\ \emph {et~al.}(2017)\citenamefont {Hormozi}, \citenamefont {Brown}, \citenamefont {Carleo},\ and\ \citenamefont {Troyer}}]{hormozi2017nonstoquastic}%
  \BibitemOpen
  \bibfield  {author} {\bibinfo {author} {\bibfnamefont {L.}~\bibnamefont {Hormozi}}, \bibinfo {author} {\bibfnamefont {E.~W.}\ \bibnamefont {Brown}}, \bibinfo {author} {\bibfnamefont {G.}~\bibnamefont {Carleo}},\ and\ \bibinfo {author} {\bibfnamefont {M.}~\bibnamefont {Troyer}},\ }\bibfield  {title} {\bibinfo {title} {{Nonstoquastic Hamiltonians and quantum annealing of an Ising spin glass}},\ }\href {https://doi.org/10.1103/PhysRevB.95.184416} {\bibfield  {journal} {\bibinfo  {journal} {Phys. Rev. B}\ }\textbf {\bibinfo {volume} {95}},\ \bibinfo {pages} {184416} (\bibinfo {year} {2017})}\BibitemShut {NoStop}%
\bibitem [{\citenamefont {Hegade}\ \emph {et~al.}(2021{\natexlab{b}})\citenamefont {Hegade}, \citenamefont {Paul}, \citenamefont {Ding}, \citenamefont {Sanz}, \citenamefont {Albarr\'an-Arriagada}, \citenamefont {Solano},\ and\ \citenamefont {Chen}}]{hegade2021shortcuts}%
  \BibitemOpen
  \bibfield  {author} {\bibinfo {author} {\bibfnamefont {N.~N.}\ \bibnamefont {Hegade}}, \bibinfo {author} {\bibfnamefont {K.}~\bibnamefont {Paul}}, \bibinfo {author} {\bibfnamefont {Y.}~\bibnamefont {Ding}}, \bibinfo {author} {\bibfnamefont {M.}~\bibnamefont {Sanz}}, \bibinfo {author} {\bibfnamefont {F.}~\bibnamefont {Albarr\'an-Arriagada}}, \bibinfo {author} {\bibfnamefont {E.}~\bibnamefont {Solano}},\ and\ \bibinfo {author} {\bibfnamefont {X.}~\bibnamefont {Chen}},\ }\bibfield  {title} {\bibinfo {title} {{Shortcuts to Adiabaticity in Digitized Adiabatic Quantum Computing}},\ }\href {https://doi.org/10.1103/PhysRevApplied.15.024038} {\bibfield  {journal} {\bibinfo  {journal} {Phys. Rev. Appl.}\ }\textbf {\bibinfo {volume} {15}},\ \bibinfo {pages} {024038} (\bibinfo {year} {2021}{\natexlab{b}})}\BibitemShut {NoStop}%
\bibitem [{\citenamefont {Suzuki}(1990)}]{SUZUKI1990319}%
  \BibitemOpen
  \bibfield  {author} {\bibinfo {author} {\bibfnamefont {M.}~\bibnamefont {Suzuki}},\ }\bibfield  {title} {\bibinfo {title} {{Fractal decomposition of exponential operators with applications to many-body theories and Monte Carlo simulations}},\ }\href {https://doi.org/https://doi.org/10.1016/0375-9601(90)90962-N} {\bibfield  {journal} {\bibinfo  {journal} {Physics Letters A}\ }\textbf {\bibinfo {volume} {146}},\ \bibinfo {pages} {319} (\bibinfo {year} {1990})}\BibitemShut {NoStop}%
\bibitem [{\citenamefont {Sriluckshmy}\ \emph {et~al.}(2023)\citenamefont {Sriluckshmy}, \citenamefont {Pina-Canelles}, \citenamefont {Ponce}, \citenamefont {Algaba}, \citenamefont {Šimkovic IV},\ and\ \citenamefont {Leib}}]{Sriluckshmy_2023}%
  \BibitemOpen
  \bibfield  {author} {\bibinfo {author} {\bibfnamefont {P.~V.}\ \bibnamefont {Sriluckshmy}}, \bibinfo {author} {\bibfnamefont {V.}~\bibnamefont {Pina-Canelles}}, \bibinfo {author} {\bibfnamefont {M.}~\bibnamefont {Ponce}}, \bibinfo {author} {\bibfnamefont {M.~G.}\ \bibnamefont {Algaba}}, \bibinfo {author} {\bibfnamefont {F.}~\bibnamefont {Šimkovic IV}},\ and\ \bibinfo {author} {\bibfnamefont {M.}~\bibnamefont {Leib}},\ }\bibfield  {title} {\bibinfo {title} {{Optimal, hardware native decomposition of parameterized multi-qubit Pauli gates}},\ }\href {https://doi.org/10.1088/2058-9565/acfa20} {\bibfield  {journal} {\bibinfo  {journal} {Quantum Science and Technology}\ }\textbf {\bibinfo {volume} {8}},\ \bibinfo {pages} {045029} (\bibinfo {year} {2023})}\BibitemShut {NoStop}%
\bibitem [{\citenamefont {Algaba}\ \emph {et~al.}(2024)\citenamefont {Algaba}, \citenamefont {Sriluckshmy}, \citenamefont {Leib},\ and\ \citenamefont {{\v{S}}imkovic~IV}}]{Algaba2024lowdepthsimulations}%
  \BibitemOpen
  \bibfield  {author} {\bibinfo {author} {\bibfnamefont {M.~G.}\ \bibnamefont {Algaba}}, \bibinfo {author} {\bibfnamefont {P.~V.}\ \bibnamefont {Sriluckshmy}}, \bibinfo {author} {\bibfnamefont {M.}~\bibnamefont {Leib}},\ and\ \bibinfo {author} {\bibfnamefont {F.}~\bibnamefont {{\v{S}}imkovic~IV}},\ }\bibfield  {title} {\bibinfo {title} {Low-depth simulations of fermionic systems on square-grid quantum hardware},\ }\href {https://doi.org/10.22331/q-2024-04-30-1327} {\bibfield  {journal} {\bibinfo  {journal} {{Quantum}}\ }\textbf {\bibinfo {volume} {8}},\ \bibinfo {pages} {1327} (\bibinfo {year} {2024})}\BibitemShut {NoStop}%
\bibitem [{\citenamefont {Gra\ss{}}(2019)}]{grass2019quantum}%
  \BibitemOpen
  \bibfield  {author} {\bibinfo {author} {\bibfnamefont {T.}~\bibnamefont {Gra\ss{}}},\ }\bibfield  {title} {\bibinfo {title} {{Quantum Annealing with Longitudinal Bias Fields}},\ }\href {https://doi.org/10.1103/PhysRevLett.123.120501} {\bibfield  {journal} {\bibinfo  {journal} {Phys. Rev. Lett.}\ }\textbf {\bibinfo {volume} {123}},\ \bibinfo {pages} {120501} (\bibinfo {year} {2019})}\BibitemShut {NoStop}%
\bibitem [{\citenamefont {Grass}(2022)}]{grass2022quantum}%
  \BibitemOpen
  \bibfield  {author} {\bibinfo {author} {\bibfnamefont {T.}~\bibnamefont {Grass}},\ }\bibfield  {title} {\bibinfo {title} {{Quantum Annealing Sampling with a Bias Field}},\ }\href {https://doi.org/10.1103/PhysRevApplied.18.044036} {\bibfield  {journal} {\bibinfo  {journal} {Phys. Rev. Appl.}\ }\textbf {\bibinfo {volume} {18}},\ \bibinfo {pages} {044036} (\bibinfo {year} {2022})}\BibitemShut {NoStop}%
\bibitem [{\citenamefont {Egger}\ \emph {et~al.}(2021)\citenamefont {Egger}, \citenamefont {Mare{\v{c}}ek},\ and\ \citenamefont {Woerner}}]{Egger2021warmstartingquantum}%
  \BibitemOpen
  \bibfield  {author} {\bibinfo {author} {\bibfnamefont {D.~J.}\ \bibnamefont {Egger}}, \bibinfo {author} {\bibfnamefont {J.}~\bibnamefont {Mare{\v{c}}ek}},\ and\ \bibinfo {author} {\bibfnamefont {S.}~\bibnamefont {Woerner}},\ }\bibfield  {title} {\bibinfo {title} {Warm-starting quantum optimization},\ }\href {https://doi.org/10.22331/q-2021-06-17-479} {\bibfield  {journal} {\bibinfo  {journal} {{Quantum}}\ }\textbf {\bibinfo {volume} {5}},\ \bibinfo {pages} {479} (\bibinfo {year} {2021})}\BibitemShut {NoStop}%
\bibitem [{\citenamefont {Truger}\ \emph {et~al.}(2024)\citenamefont {Truger}, \citenamefont {Barzen}, \citenamefont {Bechtold}, \citenamefont {Beisel}, \citenamefont {Leymann}, \citenamefont {Mandl},\ and\ \citenamefont {Yussupov}}]{truger2024warm}%
  \BibitemOpen
  \bibfield  {author} {\bibinfo {author} {\bibfnamefont {F.}~\bibnamefont {Truger}}, \bibinfo {author} {\bibfnamefont {J.}~\bibnamefont {Barzen}}, \bibinfo {author} {\bibfnamefont {M.}~\bibnamefont {Bechtold}}, \bibinfo {author} {\bibfnamefont {M.}~\bibnamefont {Beisel}}, \bibinfo {author} {\bibfnamefont {F.}~\bibnamefont {Leymann}}, \bibinfo {author} {\bibfnamefont {A.}~\bibnamefont {Mandl}},\ and\ \bibinfo {author} {\bibfnamefont {V.}~\bibnamefont {Yussupov}},\ }\bibfield  {title} {\bibinfo {title} {{Warm-Starting and Quantum Computing: A Systematic Mapping Study}},\ }\href {https://doi.org/10.1145/3652510} {\bibfield  {journal} {\bibinfo  {journal} {ACM Comput. Surv.}\ }\textbf {\bibinfo {volume} {56}},\ \bibinfo {pages} {1} (\bibinfo {year} {2024})}\BibitemShut {NoStop}%
\bibitem [{\citenamefont {Barkoutsos}\ \emph {et~al.}(2020)\citenamefont {Barkoutsos}, \citenamefont {Nannicini}, \citenamefont {Robert}, \citenamefont {Tavernelli},\ and\ \citenamefont {Woerner}}]{Barkoutsos2020improving}%
  \BibitemOpen
  \bibfield  {author} {\bibinfo {author} {\bibfnamefont {P.~K.}\ \bibnamefont {Barkoutsos}}, \bibinfo {author} {\bibfnamefont {G.}~\bibnamefont {Nannicini}}, \bibinfo {author} {\bibfnamefont {A.}~\bibnamefont {Robert}}, \bibinfo {author} {\bibfnamefont {I.}~\bibnamefont {Tavernelli}},\ and\ \bibinfo {author} {\bibfnamefont {S.}~\bibnamefont {Woerner}},\ }\bibfield  {title} {\bibinfo {title} {Improving {V}ariational {Q}uantum {O}ptimization using {CV}a{R}},\ }\href {https://doi.org/10.22331/q-2020-04-20-256} {\bibfield  {journal} {\bibinfo  {journal} {{Quantum}}\ }\textbf {\bibinfo {volume} {4}},\ \bibinfo {pages} {256} (\bibinfo {year} {2020})}\BibitemShut {NoStop}%
\bibitem [{\citenamefont {Gardner}\ and\ \citenamefont {Derrida}(1985)}]{gardner1985zero}%
  \BibitemOpen
  \bibfield  {author} {\bibinfo {author} {\bibfnamefont {E.}~\bibnamefont {Gardner}}\ and\ \bibinfo {author} {\bibfnamefont {B.}~\bibnamefont {Derrida}},\ }\bibfield  {title} {\bibinfo {title} {Zero temperature magnetization of a one-dimensional spin glass},\ }\href {https://doi.org/10.1007/BF01018668} {\bibfield  {journal} {\bibinfo  {journal} {Journal of Statistical Physics}\ }\textbf {\bibinfo {volume} {39}},\ \bibinfo {pages} {367–377} (\bibinfo {year} {1985})}\BibitemShut {NoStop}%
\bibitem [{\citenamefont {Chamberland}\ \emph {et~al.}(2020)\citenamefont {Chamberland}, \citenamefont {Zhu}, \citenamefont {Yoder}, \citenamefont {Hertzberg},\ and\ \citenamefont {Cross}}]{chamberland2020topological}%
  \BibitemOpen
  \bibfield  {author} {\bibinfo {author} {\bibfnamefont {C.}~\bibnamefont {Chamberland}}, \bibinfo {author} {\bibfnamefont {G.}~\bibnamefont {Zhu}}, \bibinfo {author} {\bibfnamefont {T.~J.}\ \bibnamefont {Yoder}}, \bibinfo {author} {\bibfnamefont {J.~B.}\ \bibnamefont {Hertzberg}},\ and\ \bibinfo {author} {\bibfnamefont {A.~W.}\ \bibnamefont {Cross}},\ }\bibfield  {title} {\bibinfo {title} {{Topological and Subsystem Codes on Low-Degree Graphs with Flag Qubits}},\ }\href {https://doi.org/10.1103/PhysRevX.10.011022} {\bibfield  {journal} {\bibinfo  {journal} {Phys. Rev. X}\ }\textbf {\bibinfo {volume} {10}},\ \bibinfo {pages} {011022} (\bibinfo {year} {2020})}\BibitemShut {NoStop}%
\bibitem [{\citenamefont {Impagliazzo}\ and\ \citenamefont {Paturi}(2001)}]{impagliazzo2001complexity}%
  \BibitemOpen
  \bibfield  {author} {\bibinfo {author} {\bibfnamefont {R.}~\bibnamefont {Impagliazzo}}\ and\ \bibinfo {author} {\bibfnamefont {R.}~\bibnamefont {Paturi}},\ }\bibfield  {title} {\bibinfo {title} {{On the Complexity of k-SAT}},\ }\href {https://doi.org/10.1006/jcss.2000.1727} {\bibfield  {journal} {\bibinfo  {journal} {J. Comput. Syst. Sci.}\ }\textbf {\bibinfo {volume} {62}},\ \bibinfo {pages} {367–375} (\bibinfo {year} {2001})}\BibitemShut {NoStop}%
\bibitem [{\citenamefont {Philathong}\ \emph {et~al.}(2021)\citenamefont {Philathong}, \citenamefont {Akshay}, \citenamefont {Samburskaya},\ and\ \citenamefont {Biamonte}}]{Philathong_2021}%
  \BibitemOpen
  \bibfield  {author} {\bibinfo {author} {\bibfnamefont {H.}~\bibnamefont {Philathong}}, \bibinfo {author} {\bibfnamefont {V.}~\bibnamefont {Akshay}}, \bibinfo {author} {\bibfnamefont {K.}~\bibnamefont {Samburskaya}},\ and\ \bibinfo {author} {\bibfnamefont {J.}~\bibnamefont {Biamonte}},\ }\bibfield  {title} {\bibinfo {title} {{Computational phase transitions: benchmarking Ising machines and quantum optimisers}},\ }\href {https://doi.org/10.1088/2632-072X/abdadc} {\bibfield  {journal} {\bibinfo  {journal} {Journal of Physics: Complexity}\ }\textbf {\bibinfo {volume} {2}},\ \bibinfo {pages} {011002} (\bibinfo {year} {2021})}\BibitemShut {NoStop}%
\bibitem [{\citenamefont {Zhang}\ \emph {et~al.}(2022)\citenamefont {Zhang}, \citenamefont {Sone},\ and\ \citenamefont {Zhuang}}]{zhang2022quantum}%
  \BibitemOpen
  \bibfield  {author} {\bibinfo {author} {\bibfnamefont {B.}~\bibnamefont {Zhang}}, \bibinfo {author} {\bibfnamefont {A.}~\bibnamefont {Sone}},\ and\ \bibinfo {author} {\bibfnamefont {Q.}~\bibnamefont {Zhuang}},\ }\bibfield  {title} {\bibinfo {title} {Quantum computational phase transition in combinatorial problems},\ }\href {https://doi.org/10.1038/s41534-022-00596-2} {\bibfield  {journal} {\bibinfo  {journal} {npj Quantum Information}\ }\textbf {\bibinfo {volume} {8}},\ \bibinfo {pages} {87} (\bibinfo {year} {2022})}\BibitemShut {NoStop}%
\bibitem [{\citenamefont {Philathong}\ \emph {et~al.}(2023)\citenamefont {Philathong}, \citenamefont {Akshay}, \citenamefont {Zacharov},\ and\ \citenamefont {Biamonte}}]{philathong2023computational}%
  \BibitemOpen
  \bibfield  {author} {\bibinfo {author} {\bibfnamefont {H.}~\bibnamefont {Philathong}}, \bibinfo {author} {\bibfnamefont {V.}~\bibnamefont {Akshay}}, \bibinfo {author} {\bibfnamefont {I.}~\bibnamefont {Zacharov}},\ and\ \bibinfo {author} {\bibfnamefont {J.}~\bibnamefont {Biamonte}},\ }\bibfield  {title} {\bibinfo {title} {{Computational phase transition signature in Gibbs sampling}},\ }\href {https://doi.org/10.1088/2632-072X/ad1410} {\bibfield  {journal} {\bibinfo  {journal} {Journal of Physics: Complexity}\ }\textbf {\bibinfo {volume} {4}},\ \bibinfo {pages} {045010} (\bibinfo {year} {2023})}\BibitemShut {NoStop}%
\bibitem [{\citenamefont {Hauke}\ \emph {et~al.}(2015)\citenamefont {Hauke}, \citenamefont {Bonnes}, \citenamefont {Heyl},\ and\ \citenamefont {Lechner}}]{hauke2015probing}%
  \BibitemOpen
  \bibfield  {author} {\bibinfo {author} {\bibfnamefont {P.}~\bibnamefont {Hauke}}, \bibinfo {author} {\bibfnamefont {L.}~\bibnamefont {Bonnes}}, \bibinfo {author} {\bibfnamefont {M.}~\bibnamefont {Heyl}},\ and\ \bibinfo {author} {\bibfnamefont {W.}~\bibnamefont {Lechner}},\ }\bibfield  {title} {\bibinfo {title} {Probing entanglement in adiabatic quantum optimization with trapped ions},\ }\bibfield  {journal} {\bibinfo  {journal} {Frontiers in Physics}\ }\textbf {\bibinfo {volume} {3}},\ \href {https://doi.org/10.3389/fphy.2015.00021} {10.3389/fphy.2015.00021} (\bibinfo {year} {2015})\BibitemShut {NoStop}%
\bibitem [{\citenamefont {Bauer}\ \emph {et~al.}(2015)\citenamefont {Bauer}, \citenamefont {Wang}, \citenamefont {Pižorn},\ and\ \citenamefont {Troyer}}]{bauer2015entanglement}%
  \BibitemOpen
  \bibfield  {author} {\bibinfo {author} {\bibfnamefont {B.}~\bibnamefont {Bauer}}, \bibinfo {author} {\bibfnamefont {L.}~\bibnamefont {Wang}}, \bibinfo {author} {\bibfnamefont {I.}~\bibnamefont {Pižorn}},\ and\ \bibinfo {author} {\bibfnamefont {M.}~\bibnamefont {Troyer}},\ }\href {https://arxiv.org/abs/1501.06914} {\bibinfo {title} {Entanglement as a resource in adiabatic quantum optimization}} (\bibinfo {year} {2015}),\ \Eprint {https://arxiv.org/abs/1501.06914} {arXiv:1501.06914 [cond-mat.dis-nn]} \BibitemShut {NoStop}%
\bibitem [{\citenamefont {Dupont}\ \emph {et~al.}(2022{\natexlab{a}})\citenamefont {Dupont}, \citenamefont {Didier}, \citenamefont {Hodson}, \citenamefont {Moore},\ and\ \citenamefont {Reagor}}]{dupont2022entanglement}%
  \BibitemOpen
  \bibfield  {author} {\bibinfo {author} {\bibfnamefont {M.}~\bibnamefont {Dupont}}, \bibinfo {author} {\bibfnamefont {N.}~\bibnamefont {Didier}}, \bibinfo {author} {\bibfnamefont {M.~J.}\ \bibnamefont {Hodson}}, \bibinfo {author} {\bibfnamefont {J.~E.}\ \bibnamefont {Moore}},\ and\ \bibinfo {author} {\bibfnamefont {M.~J.}\ \bibnamefont {Reagor}},\ }\bibfield  {title} {\bibinfo {title} {Entanglement perspective on the quantum approximate optimization algorithm},\ }\href {https://doi.org/10.1103/PhysRevA.106.022423} {\bibfield  {journal} {\bibinfo  {journal} {Phys. Rev. A}\ }\textbf {\bibinfo {volume} {106}},\ \bibinfo {pages} {022423} (\bibinfo {year} {2022}{\natexlab{a}})}\BibitemShut {NoStop}%
\bibitem [{\citenamefont {Dupont}\ \emph {et~al.}(2022{\natexlab{b}})\citenamefont {Dupont}, \citenamefont {Didier}, \citenamefont {Hodson}, \citenamefont {Moore},\ and\ \citenamefont {Reagor}}]{dupont2022calibrating}%
  \BibitemOpen
  \bibfield  {author} {\bibinfo {author} {\bibfnamefont {M.}~\bibnamefont {Dupont}}, \bibinfo {author} {\bibfnamefont {N.}~\bibnamefont {Didier}}, \bibinfo {author} {\bibfnamefont {M.~J.}\ \bibnamefont {Hodson}}, \bibinfo {author} {\bibfnamefont {J.~E.}\ \bibnamefont {Moore}},\ and\ \bibinfo {author} {\bibfnamefont {M.~J.}\ \bibnamefont {Reagor}},\ }\bibfield  {title} {\bibinfo {title} {{Calibrating the Classical Hardness of the Quantum Approximate Optimization Algorithm}},\ }\href {https://doi.org/10.1103/PRXQuantum.3.040339} {\bibfield  {journal} {\bibinfo  {journal} {PRX Quantum}\ }\textbf {\bibinfo {volume} {3}},\ \bibinfo {pages} {040339} (\bibinfo {year} {2022}{\natexlab{b}})}\BibitemShut {NoStop}%
\bibitem [{\citenamefont {Sreedhar}\ \emph {et~al.}(2022)\citenamefont {Sreedhar}, \citenamefont {Vikstål}, \citenamefont {Svensson}, \citenamefont {Ask}, \citenamefont {Johansson},\ and\ \citenamefont {García-Álvarez}}]{sreedhar2022quantumapproximate}%
  \BibitemOpen
  \bibfield  {author} {\bibinfo {author} {\bibfnamefont {R.}~\bibnamefont {Sreedhar}}, \bibinfo {author} {\bibfnamefont {P.}~\bibnamefont {Vikstål}}, \bibinfo {author} {\bibfnamefont {M.}~\bibnamefont {Svensson}}, \bibinfo {author} {\bibfnamefont {A.}~\bibnamefont {Ask}}, \bibinfo {author} {\bibfnamefont {G.}~\bibnamefont {Johansson}},\ and\ \bibinfo {author} {\bibfnamefont {L.}~\bibnamefont {García-Álvarez}},\ }\href {https://arxiv.org/abs/2207.03404} {\bibinfo {title} {{The Quantum Approximate Optimization Algorithm performance with low entanglement and high circuit depth}}} (\bibinfo {year} {2022}),\ \Eprint {https://arxiv.org/abs/2207.03404} {arXiv:2207.03404 [quant-ph]} \BibitemShut {NoStop}%
\bibitem [{\citenamefont {Nakhl}\ \emph {et~al.}(2024)\citenamefont {Nakhl}, \citenamefont {Quella},\ and\ \citenamefont {Usman}}]{nakhl2024calibrating}%
  \BibitemOpen
  \bibfield  {author} {\bibinfo {author} {\bibfnamefont {A.~C.}\ \bibnamefont {Nakhl}}, \bibinfo {author} {\bibfnamefont {T.}~\bibnamefont {Quella}},\ and\ \bibinfo {author} {\bibfnamefont {M.}~\bibnamefont {Usman}},\ }\bibfield  {title} {\bibinfo {title} {Calibrating the role of entanglement in variational quantum circuits},\ }\href {https://doi.org/10.1103/PhysRevA.109.032413} {\bibfield  {journal} {\bibinfo  {journal} {Phys. Rev. A}\ }\textbf {\bibinfo {volume} {109}},\ \bibinfo {pages} {032413} (\bibinfo {year} {2024})}\BibitemShut {NoStop}%
\bibitem [{\citenamefont {Santra}\ \emph {et~al.}(2025)\citenamefont {Santra}, \citenamefont {Roy}, \citenamefont {Egger},\ and\ \citenamefont {Hauke}}]{santra2024genuine}%
  \BibitemOpen
  \bibfield  {author} {\bibinfo {author} {\bibfnamefont {G.~C.}\ \bibnamefont {Santra}}, \bibinfo {author} {\bibfnamefont {S.~S.}\ \bibnamefont {Roy}}, \bibinfo {author} {\bibfnamefont {D.~J.}\ \bibnamefont {Egger}},\ and\ \bibinfo {author} {\bibfnamefont {P.}~\bibnamefont {Hauke}},\ }\bibfield  {title} {\bibinfo {title} {Genuine multipartite entanglement in quantum optimization},\ }\href {https://doi.org/10.1103/PhysRevA.111.022434} {\bibfield  {journal} {\bibinfo  {journal} {Phys. Rev. A}\ }\textbf {\bibinfo {volume} {111}},\ \bibinfo {pages} {022434} (\bibinfo {year} {2025})}\BibitemShut {NoStop}%
\bibitem [{\citenamefont {Schollw\"ock}(2011)}]{Schollwock2011}%
  \BibitemOpen
  \bibfield  {author} {\bibinfo {author} {\bibfnamefont {U.}~\bibnamefont {Schollw\"ock}},\ }\bibfield  {title} {\bibinfo {title} {The density-matrix renormalization group in the age of matrix product states},\ }\href {https://doi.org/https://doi.org/10.1016/j.aop.2010.09.012} {\bibfield  {journal} {\bibinfo  {journal} {Annals of Physics}\ }\textbf {\bibinfo {volume} {326}},\ \bibinfo {pages} {96} (\bibinfo {year} {2011})}\BibitemShut {NoStop}%
\bibitem [{\citenamefont {Nielsen}\ and\ \citenamefont {Chuang}(2012)}]{nielsen2001quantum}%
  \BibitemOpen
  \bibfield  {author} {\bibinfo {author} {\bibfnamefont {M.~A.}\ \bibnamefont {Nielsen}}\ and\ \bibinfo {author} {\bibfnamefont {I.~L.}\ \bibnamefont {Chuang}},\ }\href {https://doi.org/10.1017/CBO9780511976667} {\emph {\bibinfo {title} {{Quantum Computation and Quantum Information: 10th Anniversary Edition}}}},\ \bibinfo {edition} {1st}\ ed.\ (\bibinfo  {publisher} {Cambridge University Press},\ \bibinfo {year} {2012})\BibitemShut {NoStop}%
\bibitem [{\citenamefont {Eisert}\ \emph {et~al.}(2010)\citenamefont {Eisert}, \citenamefont {Cramer},\ and\ \citenamefont {Plenio}}]{eisert2010colloquium}%
  \BibitemOpen
  \bibfield  {author} {\bibinfo {author} {\bibfnamefont {J.}~\bibnamefont {Eisert}}, \bibinfo {author} {\bibfnamefont {M.}~\bibnamefont {Cramer}},\ and\ \bibinfo {author} {\bibfnamefont {M.~B.}\ \bibnamefont {Plenio}},\ }\bibfield  {title} {\bibinfo {title} {{Colloquium: Area laws for the entanglement entropy}},\ }\href {https://doi.org/10.1103/RevModPhys.82.277} {\bibfield  {journal} {\bibinfo  {journal} {Rev. Mod. Phys.}\ }\textbf {\bibinfo {volume} {82}},\ \bibinfo {pages} {277} (\bibinfo {year} {2010})}\BibitemShut {NoStop}%
\bibitem [{\citenamefont {Srivastava}\ \emph {et~al.}(2024)\citenamefont {Srivastava}, \citenamefont {M{\"u}ller-Rigat}, \citenamefont {Lewenstein},\ and\ \citenamefont {Rajchel-Mieldzio{\'{c}}}}]{Srivastava2024}%
  \BibitemOpen
  \bibfield  {author} {\bibinfo {author} {\bibfnamefont {A.~K.}\ \bibnamefont {Srivastava}}, \bibinfo {author} {\bibfnamefont {G.}~\bibnamefont {M{\"u}ller-Rigat}}, \bibinfo {author} {\bibfnamefont {M.}~\bibnamefont {Lewenstein}},\ and\ \bibinfo {author} {\bibfnamefont {G.}~\bibnamefont {Rajchel-Mieldzio{\'{c}}}},\ }\bibinfo {title} {Introduction to quantum entanglement in many-body systems},\ in\ \href {https://doi.org/10.1007/978-3-031-55657-9_4} {\emph {\bibinfo {booktitle} {New Trends and Platforms for Quantum Technologies}}},\ \bibinfo {editor} {edited by\ \bibinfo {editor} {\bibfnamefont {R.}~\bibnamefont {Aguado}}, \bibinfo {editor} {\bibfnamefont {R.}~\bibnamefont {Citro}}, \bibinfo {editor} {\bibfnamefont {M.}~\bibnamefont {Lewenstein}},\ and\ \bibinfo {editor} {\bibfnamefont {M.}~\bibnamefont {Stern}}}\ (\bibinfo  {publisher} {Springer Nature Switzerland},\ \bibinfo {address} {Cham},\ \bibinfo {year} {2024})\ pp.\ \bibinfo {pages} {225--285}\BibitemShut {NoStop}%
\bibitem [{\citenamefont {Kirkpatrick}\ \emph {et~al.}(1983)\citenamefont {Kirkpatrick}, \citenamefont {Gelatt},\ and\ \citenamefont {Vecchi}}]{kirkpatrick1983optimization}%
  \BibitemOpen
  \bibfield  {author} {\bibinfo {author} {\bibfnamefont {S.}~\bibnamefont {Kirkpatrick}}, \bibinfo {author} {\bibfnamefont {C.~D.}\ \bibnamefont {Gelatt}},\ and\ \bibinfo {author} {\bibfnamefont {M.~P.}\ \bibnamefont {Vecchi}},\ }\bibfield  {title} {\bibinfo {title} {{Optimization by Simulated Annealing}},\ }\href {https://doi.org/10.1126/science.220.4598.671} {\bibfield  {journal} {\bibinfo  {journal} {Science}\ }\textbf {\bibinfo {volume} {220}},\ \bibinfo {pages} {671} (\bibinfo {year} {1983})}\BibitemShut {NoStop}%
\bibitem [{\citenamefont {Metropolis}\ \emph {et~al.}(1953)\citenamefont {Metropolis}, \citenamefont {Rosenbluth}, \citenamefont {Rosenbluth}, \citenamefont {Teller},\ and\ \citenamefont {Teller}}]{metropolis1953equation}%
  \BibitemOpen
  \bibfield  {author} {\bibinfo {author} {\bibfnamefont {N.}~\bibnamefont {Metropolis}}, \bibinfo {author} {\bibfnamefont {A.~W.}\ \bibnamefont {Rosenbluth}}, \bibinfo {author} {\bibfnamefont {M.~N.}\ \bibnamefont {Rosenbluth}}, \bibinfo {author} {\bibfnamefont {A.~H.}\ \bibnamefont {Teller}},\ and\ \bibinfo {author} {\bibfnamefont {E.}~\bibnamefont {Teller}},\ }\bibfield  {title} {\bibinfo {title} {{Equation of State Calculations by Fast Computing Machines}},\ }\href {https://doi.org/10.1063/1.1699114} {\bibfield  {journal} {\bibinfo  {journal} {The Journal of Chemical Physics}\ }\textbf {\bibinfo {volume} {21}},\ \bibinfo {pages} {1087} (\bibinfo {year} {1953})}\BibitemShut {NoStop}%
\bibitem [{\citenamefont {Hastings}(1970)}]{hastings1970monte}%
  \BibitemOpen
  \bibfield  {author} {\bibinfo {author} {\bibfnamefont {W.~K.}\ \bibnamefont {Hastings}},\ }\bibfield  {title} {\bibinfo {title} {Monte carlo sampling methods using markov chains and their applications},\ }\href {https://doi.org/10.1093/biomet/57.1.97} {\bibfield  {journal} {\bibinfo  {journal} {Biometrika}\ }\textbf {\bibinfo {volume} {57}},\ \bibinfo {pages} {97} (\bibinfo {year} {1970})}\BibitemShut {NoStop}%
\bibitem [{\citenamefont {Glover}\ \emph {et~al.}(1993)\citenamefont {Glover}, \citenamefont {Taillard},\ and\ \citenamefont {de~Werra}}]{tabu}%
  \BibitemOpen
  \bibfield  {author} {\bibinfo {author} {\bibfnamefont {F.}~\bibnamefont {Glover}}, \bibinfo {author} {\bibfnamefont {E.}~\bibnamefont {Taillard}},\ and\ \bibinfo {author} {\bibfnamefont {D.}~\bibnamefont {de~Werra}},\ }\bibfield  {title} {\bibinfo {title} {{A User's Guide to Tabu Search}},\ }\href {https://doi.org/10.1007/BF02078647} {\bibfield  {journal} {\bibinfo  {journal} {Annals of Operations Research}\ }\textbf {\bibinfo {volume} {41}},\ \bibinfo {pages} {3} (\bibinfo {year} {1993})}\BibitemShut {NoStop}%
\bibitem [{\citenamefont {Fishman}\ \emph {et~al.}(2022)\citenamefont {Fishman}, \citenamefont {White},\ and\ \citenamefont {Stoudenmire}}]{itensor}%
  \BibitemOpen
  \bibfield  {author} {\bibinfo {author} {\bibfnamefont {M.}~\bibnamefont {Fishman}}, \bibinfo {author} {\bibfnamefont {S.~R.}\ \bibnamefont {White}},\ and\ \bibinfo {author} {\bibfnamefont {E.~M.}\ \bibnamefont {Stoudenmire}},\ }\bibfield  {title} {\bibinfo {title} {{The ITensor Software Library for Tensor Network Calculations}},\ }\href {https://doi.org/10.21468/SciPostPhysCodeb.4} {\bibfield  {journal} {\bibinfo  {journal} {SciPost Phys. Codebases}\ ,\ \bibinfo {pages} {4}} (\bibinfo {year} {2022})}\BibitemShut {NoStop}%
\bibitem [{\citenamefont {Ishikawa}(2011)}]{ishikawa2011transformation}%
  \BibitemOpen
  \bibfield  {author} {\bibinfo {author} {\bibfnamefont {H.}~\bibnamefont {Ishikawa}},\ }\bibfield  {title} {\bibinfo {title} {{Transformation of General Binary MRF Minimization to the First-Order Case}},\ }\href {https://doi.org/10.1109/TPAMI.2010.91} {\bibfield  {journal} {\bibinfo  {journal} {IEEE Transactions on Pattern Analysis and Machine Intelligence}\ }\textbf {\bibinfo {volume} {33}},\ \bibinfo {pages} {1234} (\bibinfo {year} {2011})}\BibitemShut {NoStop}%
\bibitem [{\citenamefont {Anthony}\ \emph {et~al.}(2017)\citenamefont {Anthony}, \citenamefont {Boros}, \citenamefont {Crama},\ and\ \citenamefont {Gruber}}]{anthony2017quadratic}%
  \BibitemOpen
  \bibfield  {author} {\bibinfo {author} {\bibfnamefont {M.}~\bibnamefont {Anthony}}, \bibinfo {author} {\bibfnamefont {E.}~\bibnamefont {Boros}}, \bibinfo {author} {\bibfnamefont {Y.}~\bibnamefont {Crama}},\ and\ \bibinfo {author} {\bibfnamefont {A.}~\bibnamefont {Gruber}},\ }\bibfield  {title} {\bibinfo {title} {Quadratic reformulations of nonlinear binary optimization problems},\ }\href {https://doi.org/10.1007/s10107-016-1032-4} {\bibfield  {journal} {\bibinfo  {journal} {Mathematical Programming}\ }\textbf {\bibinfo {volume} {162}},\ \bibinfo {pages} {115–144} (\bibinfo {year} {2017})}\BibitemShut {NoStop}%
\bibitem [{\citenamefont {Dattani}(2019)}]{dattani2019quadratizationdiscreteoptimizationquantum}%
  \BibitemOpen
  \bibfield  {author} {\bibinfo {author} {\bibfnamefont {N.}~\bibnamefont {Dattani}},\ }\href {https://arxiv.org/abs/1901.04405} {\bibinfo {title} {Quadratization in discrete optimization and quantum mechanics}} (\bibinfo {year} {2019}),\ \Eprint {https://arxiv.org/abs/1901.04405} {arXiv:1901.04405 [quant-ph]} \BibitemShut {NoStop}%
\bibitem [{\citenamefont {{Gurobi Optimization, LLC}}(2024)}]{gurobi}%
  \BibitemOpen
  \bibfield  {author} {\bibinfo {author} {\bibnamefont {{Gurobi Optimization, LLC}}},\ }\href {https://www.gurobi.com} {\bibinfo {title} {{Gurobi Optimizer Reference Manual}}} (\bibinfo {year} {2024})\BibitemShut {NoStop}%
\bibitem [{\citenamefont {Cplex}(2009)}]{cplex}%
  \BibitemOpen
  \bibfield  {author} {\bibinfo {author} {\bibfnamefont {I.~I.}\ \bibnamefont {Cplex}},\ }\bibfield  {title} {\bibinfo {title} {{V12.10.0: User's Manual for CPLEX}},\ }\href@noop {} {\bibfield  {journal} {\bibinfo  {journal} {International Business Machines Corporation}\ }\textbf {\bibinfo {volume} {46}},\ \bibinfo {pages} {157} (\bibinfo {year} {2009})}\BibitemShut {NoStop}%
\bibitem [{\citenamefont {Maksymov}\ \emph {et~al.}(2023)\citenamefont {Maksymov}, \citenamefont {Nguyen}, \citenamefont {Nam},\ and\ \citenamefont {Markov}}]{maksymov2023enhancingquantumcomputerperformance}%
  \BibitemOpen
  \bibfield  {author} {\bibinfo {author} {\bibfnamefont {A.}~\bibnamefont {Maksymov}}, \bibinfo {author} {\bibfnamefont {J.}~\bibnamefont {Nguyen}}, \bibinfo {author} {\bibfnamefont {Y.}~\bibnamefont {Nam}},\ and\ \bibinfo {author} {\bibfnamefont {I.}~\bibnamefont {Markov}},\ }\href {https://arxiv.org/abs/2301.07233} {\bibinfo {title} {Enhancing quantum computer performance via symmetrization}} (\bibinfo {year} {2023}),\ \Eprint {https://arxiv.org/abs/2301.07233} {arXiv:2301.07233 [quant-ph]} \BibitemShut {NoStop}%
\bibitem [{\citenamefont {Rossini}\ \emph {et~al.}(2020)\citenamefont {Rossini}, \citenamefont {Andolina}, \citenamefont {Rosa}, \citenamefont {Carrega},\ and\ \citenamefont {Polini}}]{rossini2020quantum}%
  \BibitemOpen
  \bibfield  {author} {\bibinfo {author} {\bibfnamefont {D.}~\bibnamefont {Rossini}}, \bibinfo {author} {\bibfnamefont {G.~M.}\ \bibnamefont {Andolina}}, \bibinfo {author} {\bibfnamefont {D.}~\bibnamefont {Rosa}}, \bibinfo {author} {\bibfnamefont {M.}~\bibnamefont {Carrega}},\ and\ \bibinfo {author} {\bibfnamefont {M.}~\bibnamefont {Polini}},\ }\bibfield  {title} {\bibinfo {title} {{Quantum Advantage in the Charging Process of Sachdev-Ye-Kitaev Batteries}},\ }\href {https://doi.org/10.1103/PhysRevLett.125.236402} {\bibfield  {journal} {\bibinfo  {journal} {Phys. Rev. Lett.}\ }\textbf {\bibinfo {volume} {125}},\ \bibinfo {pages} {236402} (\bibinfo {year} {2020})}\BibitemShut {NoStop}%
\end{thebibliography}%


\begin{thebibliography}{11}%
\makeatletter
\providecommand \@ifxundefined [1]{%
 \@ifx{#1\undefined}
}%
\providecommand \@ifnum [1]{%
 \ifnum #1\expandafter \@firstoftwo
 \else \expandafter \@secondoftwo
 \fi
}%
\providecommand \@ifx [1]{%
 \ifx #1\expandafter \@firstoftwo
 \else \expandafter \@secondoftwo
 \fi
}%
\providecommand \natexlab [1]{#1}%
\providecommand \enquote  [1]{``#1''}%
\providecommand \bibnamefont  [1]{#1}%
\providecommand \bibfnamefont [1]{#1}%
\providecommand \citenamefont [1]{#1}%
\providecommand \href@noop [0]{\@secondoftwo}%
\providecommand \href [0]{\begingroup \@sanitize@url \@href}%
\providecommand \@href[1]{\@@startlink{#1}\@@href}%
\providecommand \@@href[1]{\endgroup#1\@@endlink}%
\providecommand \@sanitize@url [0]{\catcode `\\12\catcode `\$12\catcode `\&12\catcode `\#12\catcode `\^12\catcode `\_12\catcode `\%12\relax}%
\providecommand \@@startlink[1]{}%
\providecommand \@@endlink[0]{}%
\providecommand \url  [0]{\begingroup\@sanitize@url \@url }%
\providecommand \@url [1]{\endgroup\@href {#1}{\urlprefix }}%
\providecommand \urlprefix  [0]{URL }%
\providecommand \Eprint [0]{\href }%
\providecommand \doibase [0]{https://doi.org/}%
\providecommand \selectlanguage [0]{\@gobble}%
\providecommand \bibinfo  [0]{\@secondoftwo}%
\providecommand \bibfield  [0]{\@secondoftwo}%
\providecommand \translation [1]{[#1]}%
\providecommand \BibitemOpen [0]{}%
\providecommand \bibitemStop [0]{}%
\providecommand \bibitemNoStop [0]{.\EOS\space}%
\providecommand \EOS [0]{\spacefactor3000\relax}%
\providecommand \BibitemShut  [1]{\csname bibitem#1\endcsname}%
\let\auto@bib@innerbib\@empty
%</preamble>
\bibitem [{\citenamefont {Gardner}\ and\ \citenamefont {Derrida}(1985)}]{gardner1985zero}%
  \BibitemOpen
  \bibfield  {author} {\bibinfo {author} {\bibfnamefont {E.}~\bibnamefont {Gardner}}\ and\ \bibinfo {author} {\bibfnamefont {B.}~\bibnamefont {Derrida}},\ }\bibfield  {title} {\bibinfo {title} {Zero temperature magnetization of a one-dimensional spin glass},\ }\href {https://doi.org/10.1007/BF01018668} {\bibfield  {journal} {\bibinfo  {journal} {Journal of Statistical Physics}\ }\textbf {\bibinfo {volume} {39}},\ \bibinfo {pages} {367–377} (\bibinfo {year} {1985})}\BibitemShut {NoStop}%
\bibitem [{\citenamefont {Kardar}(2007)}]{kardar2007statistical}%
  \BibitemOpen
  \bibfield  {author} {\bibinfo {author} {\bibfnamefont {M.}~\bibnamefont {Kardar}},\ }\href {https://doi.org/10.1017/CBO9780511815881} {\emph {\bibinfo {title} {{Statistical Physics of Fields}}}},\ \bibinfo {edition} {1st}\ ed.\ (\bibinfo  {publisher} {Cambridge University Press},\ \bibinfo {year} {2007})\BibitemShut {NoStop}%
\bibitem [{\citenamefont {Turban}(2016)}]{turban2016multispin}%
  \BibitemOpen
  \bibfield  {author} {\bibinfo {author} {\bibfnamefont {L.}~\bibnamefont {Turban}},\ }\bibfield  {title} {\bibinfo {title} {One-dimensional ising model with multispin interactions},\ }\href {https://doi.org/10.1088/1751-8113/49/35/355002} {\bibfield  {journal} {\bibinfo  {journal} {Journal of Physics A: Mathematical and Theoretical}\ }\textbf {\bibinfo {volume} {49}},\ \bibinfo {pages} {355002} (\bibinfo {year} {2016})}\BibitemShut {NoStop}%
\bibitem [{Note1()}]{Note1}%
  \BibitemOpen
  \bibinfo {note} {See~\protect \href {https://ionq.com/docs/getting-started-with-native-gates}{https://ionq.com/docs/getting-started-with-native-gates}.}\BibitemShut {Stop}%
\bibitem [{\citenamefont {Schollw\"ock}(2011)}]{Schollwock2011}%
  \BibitemOpen
  \bibfield  {author} {\bibinfo {author} {\bibfnamefont {U.}~\bibnamefont {Schollw\"ock}},\ }\bibfield  {title} {\bibinfo {title} {The density-matrix renormalization group in the age of matrix product states},\ }\href {https://doi.org/https://doi.org/10.1016/j.aop.2010.09.012} {\bibfield  {journal} {\bibinfo  {journal} {Annals of Physics}\ }\textbf {\bibinfo {volume} {326}},\ \bibinfo {pages} {96} (\bibinfo {year} {2011})}\BibitemShut {NoStop}%
\bibitem [{\citenamefont {Eisert}\ \emph {et~al.}(2010)\citenamefont {Eisert}, \citenamefont {Cramer},\ and\ \citenamefont {Plenio}}]{eisert2010colloquium}%
  \BibitemOpen
  \bibfield  {author} {\bibinfo {author} {\bibfnamefont {J.}~\bibnamefont {Eisert}}, \bibinfo {author} {\bibfnamefont {M.}~\bibnamefont {Cramer}},\ and\ \bibinfo {author} {\bibfnamefont {M.~B.}\ \bibnamefont {Plenio}},\ }\bibfield  {title} {\bibinfo {title} {{Colloquium: Area laws for the entanglement entropy}},\ }\href {https://doi.org/10.1103/RevModPhys.82.277} {\bibfield  {journal} {\bibinfo  {journal} {Rev. Mod. Phys.}\ }\textbf {\bibinfo {volume} {82}},\ \bibinfo {pages} {277} (\bibinfo {year} {2010})}\BibitemShut {NoStop}%
\bibitem [{\citenamefont {Fishman}\ \emph {et~al.}(2022)\citenamefont {Fishman}, \citenamefont {White},\ and\ \citenamefont {Stoudenmire}}]{itensor}%
  \BibitemOpen
  \bibfield  {author} {\bibinfo {author} {\bibfnamefont {M.}~\bibnamefont {Fishman}}, \bibinfo {author} {\bibfnamefont {S.~R.}\ \bibnamefont {White}},\ and\ \bibinfo {author} {\bibfnamefont {E.~M.}\ \bibnamefont {Stoudenmire}},\ }\bibfield  {title} {\bibinfo {title} {{The ITensor Software Library for Tensor Network Calculations}},\ }\href {https://doi.org/10.21468/SciPostPhysCodeb.4} {\bibfield  {journal} {\bibinfo  {journal} {SciPost Phys. Codebases}\ ,\ \bibinfo {pages} {4}} (\bibinfo {year} {2022})}\BibitemShut {NoStop}%
\bibitem [{\citenamefont {White}(2005)}]{white2005density}%
  \BibitemOpen
  \bibfield  {author} {\bibinfo {author} {\bibfnamefont {S.~R.}\ \bibnamefont {White}},\ }\bibfield  {title} {\bibinfo {title} {Density matrix renormalization group algorithms with a single center site},\ }\href {https://doi.org/10.1103/PhysRevB.72.180403} {\bibfield  {journal} {\bibinfo  {journal} {Phys. Rev. B}\ }\textbf {\bibinfo {volume} {72}},\ \bibinfo {pages} {180403} (\bibinfo {year} {2005})}\BibitemShut {NoStop}%
\bibitem [{\citenamefont {Rodríguez-Laguna}(2007{\natexlab{a}})}]{rodriguez-laguna2007quantum}%
  \BibitemOpen
  \bibfield  {author} {\bibinfo {author} {\bibfnamefont {J.}~\bibnamefont {Rodríguez-Laguna}},\ }\bibfield  {title} {\bibinfo {title} {Quantum wavefunction annealing of spin glasses on ladders},\ }\href {https://doi.org/10.1088/1742-5468/2007/05/P05008} {\bibfield  {journal} {\bibinfo  {journal} {Journal of Statistical Mechanics: Theory and Experiment}\ }\textbf {\bibinfo {volume} {2007}},\ \bibinfo {pages} {P05008} (\bibinfo {year} {2007}{\natexlab{a}})}\BibitemShut {NoStop}%
\bibitem [{\citenamefont {Rodríguez-Laguna}(2007{\natexlab{b}})}]{rodriguez-laguna2007density}%
  \BibitemOpen
  \bibfield  {author} {\bibinfo {author} {\bibfnamefont {J.}~\bibnamefont {Rodríguez-Laguna}},\ }\bibfield  {title} {\bibinfo {title} {Density matrix renormalization on random graphs and the quantum spin-glass transition},\ }\href {https://doi.org/10.1088/1751-8113/40/40/003} {\bibfield  {journal} {\bibinfo  {journal} {Journal of Physics A: Mathematical and Theoretical}\ }\textbf {\bibinfo {volume} {40}},\ \bibinfo {pages} {12043} (\bibinfo {year} {2007}{\natexlab{b}})}\BibitemShut {NoStop}%
\bibitem [{\citenamefont {Patra}\ \emph {et~al.}(2025)\citenamefont {Patra}, \citenamefont {Singh},\ and\ \citenamefont {Or\'us}}]{patra2024projected}%
  \BibitemOpen
  \bibfield  {author} {\bibinfo {author} {\bibfnamefont {S.}~\bibnamefont {Patra}}, \bibinfo {author} {\bibfnamefont {S.}~\bibnamefont {Singh}},\ and\ \bibinfo {author} {\bibfnamefont {R.}~\bibnamefont {Or\'us}},\ }\bibfield  {title} {\bibinfo {title} {Projected entangled pair states with flexible geometry},\ }\href {https://doi.org/10.1103/PhysRevResearch.7.L012002} {\bibfield  {journal} {\bibinfo  {journal} {Phys. Rev. Res.}\ }\textbf {\bibinfo {volume} {7}},\ \bibinfo {pages} {L012002} (\bibinfo {year} {2025})}\BibitemShut {NoStop}%
\end{thebibliography}%

\end{document}

\title{\texorpdfstring{Supplementary Information for:\\``Bias-field digitized counterdiabatic quantum algorithm for higher-order binary optimization''}{Supplementary Information for: ``Bias-field digitized counterdiabatic quantum algorithm for higher-order binary optimization''}}
\author{Sebastián V. Romero$^{\orcidlink{0000-0002-4675-4452}}$}
\affiliation{Kipu Quantum GmbH, Greifswalderstrasse 212, 10405 Berlin, Germany}
\affiliation{Department of Physical Chemistry, University of the Basque Country UPV/EHU, Apartado 644, 48080 Bilbao, Spain}

\author{Anne-Maria Visuri$^{\orcidlink{0000-0002-4167-7769}}$}
\affiliation{Kipu Quantum GmbH, Greifswalderstrasse 212, 10405 Berlin, Germany}

\author{Alejandro Gomez Cadavid$^{\orcidlink{0000-0003-3271-4684}}$}
\affiliation{Kipu Quantum GmbH, Greifswalderstrasse 212, 10405 Berlin, Germany}
\affiliation{Department of Physical Chemistry, University of the Basque Country UPV/EHU, Apartado 644, 48080 Bilbao, Spain}

\author{Anton Simen$^{\orcidlink{0000-0001-8863-4806}}$}
\affiliation{Kipu Quantum GmbH, Greifswalderstrasse 212, 10405 Berlin, Germany}
\affiliation{Department of Physical Chemistry, University of the Basque Country UPV/EHU, Apartado 644, 48080 Bilbao, Spain}

\author{Enrique Solano$^{\orcidlink{0000-0002-8602-1181}}$}
\affiliation{Kipu Quantum GmbH, Greifswalderstrasse 212, 10405 Berlin, Germany}

\author{Narendra N. Hegade$^{\orcidlink{0000-0002-9673-2833}}$}
\email[]{narendrahegade5@gmail.com}
\affiliation{Kipu Quantum GmbH, Greifswalderstrasse 212, 10405 Berlin, Germany}

\renewcommand{\thetable}{S\arabic{table}}
\renewcommand{\theequation}{S\arabic{equation}}

\renewcommand{\figurename}{SUPPLEMENTARY FIGURE}
\renewcommand{\bibnumfmt}[1]{[S#1]}
\renewcommand{\citenumfont}[1]{S#1}

\maketitle

\section*{Supplementary Note 1: First-order nested-commutator calculation}\label{sec:1nc}

In order to include the counterdiabatic driving contribution into our protocol, we have to perform the map $H_\text{ad}(\lambda) \mapsto H_\text{ad}(\lambda) + i\dot{\lambda}\alpha_1(\lambda)O_1(\lambda)$ with $O_k=[H_\text{ad}(\lambda), O_{k-1}(\lambda)]$ starting from $O_1(\lambda)=[H_\text{ad}(\lambda), \partial_\lambda H_\text{ad}(\lambda)]=[H_i, H_f]$ and $\alpha_1(\lambda)=-\Gamma_1(\lambda)/\Gamma_2(\lambda)$ with $\Gamma_k(\lambda)=\trace\big[O_k^\dagger(\lambda)O_k(\lambda)\big]$. We start by computing 
\begin{equation}
    \begin{split}
        O_1(\lambda) &=-2i \sum_{i} h_{i}^{x} h_{i}^{z} \sigma^y_{i}-2 i \sum_{i<j} J_{i j}(h_{i}^{x} \sigma^y_{i} \sigma^z_{j}+h_{j}^{x} \sigma^z_{i} \sigma^y_{j}) -2i\sum_{i<j<k}K_{ijk}(h_i^x\sigma^y_{i}\sigma^z_{j}\sigma^z_{k} + h_j^x\sigma^z_{i}\sigma^y_{j}\sigma^z_{k} + h_k^x\sigma^z_{i}\sigma^z_{j}\sigma^y_{k}).
    \end{split}
\end{equation}

Next, we compute $\Gamma_1(\lambda)$ as 
\begin{equation}
    \Gamma_1(\lambda) = \trace\big[O_1^\dagger(\lambda)O_1(\lambda)\big] = 4\Big[\sum_i (h_i^xh_i^z)^2 + \sum_{i\neq j}(h_i^x)^2J_{ij}^2 +\sum_{i<j<k}K_{ijk}^2[(h_i^x)^2 + (h_j^x)^2 + (h_k^x)^2]\Big]
\end{equation}
Following the same procedure for computing $\Gamma_2(\lambda)$ and finally taking the squared norm of each separated Pauli string, we obtain
\begin{equation}
    \begin{split}
        \Gamma_2(\lambda)&= 16\sum_i (h_i^x)^2\Big[(1-\lambda)h^b_ih_i^z+\lambda\Big((h_{i}^{z})^{2}+\sum_{j \neq i} J_{i j}^{2} +\sum_{j,k\neq i}K_{ijk}^2\Big)\Big]^2 \\
        &+16\sum_{i<j}(h_i^x)^2\Big[(1-\lambda)J_{ij}h^b_i + 2\lambda\Big(J_{ij}h_i^z + \sum_{i<k<j}K_{ikj}J_{ik} + \sum_{k>j}K_{ijk}J_{ik} +\sum_{k<i}K_{kij}J_{ki} \Big)\Big]^2 \\
        &+16 \sum_{i<j}(h_j^x)^2\Big[(1-\lambda)J_{ij}h_j^b + 2\lambda\Big(J_{ij}h_j^z +\sum_{i<k<j} K_{ikj}J_{kj} + \sum_{k>j}K_{ijk}J_{jk} +\sum_{k<i}K_{kij}J_{kj} \Big)\Big]^2 \\
        &+16(1-\lambda)^2\Big[\sum_{i}(h_{i}^{x})^4 (h_{i}^{z})^2  + \sum_{i<j}[(h_{i}^{x})^{2}+(h_{j}^{x})^{2}]^2 J_{i j}^2 +\sum_{i<j<k} K_{ijk}^2[(h_i^x)^2 + (h_j^x)^2 + (h_k^x)^2]^2\Big] \\
        &+64(1-\lambda)^2\Big[\sum_{i<j}(h_i^x)^2(h_j^x)^2J_{ij}^2 +\sum_{i<j<k} K_{ijk}^2[(h_i^x)^2(h_j^x)^2 + (h_i^x)^2(h_k^x)^2 + (h_j^x)^2(h_k^x)^2]\Big] \\%
        &\begin{split}
            +16\sum_{i<j<k} (h_i^x)^2 &\Big[(1-\lambda)K_{ijk}h^b_i +2\lambda\Big( J_{ij}J_{ik} +K_{ijk}h_i^z + \sum_{j<p<k}K_{ijp}K_{ipk} + \sum_{i<p<j}K_{ipj}K_{ipk} +\sum_{p<i}K_{pij}K_{pjk} +\sum_{p>k}K_{ijp}K_{ikp} \Big) \Big]^2
        \end{split} \\%
        &\begin{split}
            +16\sum_{i<j<k} (h_j^x)^2 &\Big[(1-\lambda)K_{ijk}h_j^b +2\lambda\Big( J_{j k} J_{i j} +K_{ijk}h_j^z+ \sum_{j<p<k}K_{ijp}K_{jpk} +\sum_{i<p<j}K_{ipj}K_{pjk} +\sum_{p<i}K_{pij}K_{pjk} +\sum_{p>k}K_{ijp}K_{jkp} \Big) \Big]^2
        \end{split} \\%
        &\begin{split}
            +16\sum_{i<j<k} (h_k^x)^2 &\Big[(1-\lambda)K_{ijk}h_k^b +2\lambda\Big( J_{i k} J_{j k} +K_{ijk}h_k^z +\sum_{j<p<k}K_{ipk}K_{jpk} +\sum_{i<p<j}K_{ipk}K_{pjk} +\sum_{p<i}K_{pik}K_{pjk} +\sum_{p>k}K_{ikp}K_{jkp} \Big) \Big]^2
        \end{split} \\%
        &\begin{split}
            +64\lambda^2\sum_{i<j<k<p}&[(h_i^x)^2(K_{ikp}J_{ij} + K_{ijp}J_{ik} + K_{ijk}J_{ip})^2 + (h_j^x)^2(K_{jkp}J_{ij} + K_{ijp}J_{jk} + K_{ijk}J_{jp})^2 \\
            &+(h_k^x)^2(K_{jkp}J_{ik} + K_{ikp}J_{jk} + K_{ijk}J_{kp})^2 + (h_p^x)^2(K_{jkp}J_{ip} + K_{ikp}J_{jp} + K_{ijp}J_{kp})^2]
        \end{split} \\%
        &\begin{split}
            +64\lambda^2\sum_{i<j<k<p<q}&[(h_i^x)^2(K_{ijk}K_{ipq} + K_{ikq}K_{ijp} + K_{ikp}K_{ijq})^2 +(h_j^x)^2(K_{ijk}K_{jpq} + K_{ijp}K_{jkq} + K_{ijq}K_{jkp})^2 \\
            &+(h_k^x)^2(K_{ijk}K_{kpq} + K_{ikp}K_{jkq} + K_{ikq}K_{jkp})^2 +(h_p^x)^2(K_{ijp}K_{kpq} + K_{ikp}K_{jpq} + K_{ipq}K_{jkp})^2 \\
            &+(h_q^x)^2(K_{ijq}K_{kpq} + K_{ikq}K_{jpq} + K_{ipq}K_{jkq})^2].
        \end{split}
    \end{split}
\end{equation}

This provides an analytical derivation of the first-order nested-commutator expansion of Eq.~(5) in the main text for a generic three-body spin-glass problem.

\section*{Supplementary Note 2: Iterative solution of the exact ground state}\label{app:exact}

To obtain numerically the exact ground state energies of the Hamiltonians of Eqs.~(13) and~(16) in the main text, we use an iterative algorithm similar to Ref.~\cite{gardner1985zero}. This method is similar to the zero-temperature limit of the transfer matrix method~\cite{kardar2007statistical, turban2016multispin}, and can be generalized to any (sufficiently small) range $r$ and locality of coupling terms. We summarize it here briefly for $r = 3$ and up to three-body interactions, locality of the Hamiltonians studied in this work. 

The energy contribution at site $i$ depends only on the local configuration $\bm{s}_i \coloneqq (s_{i-2},s_{i-1}, s_i)$, where $s_i = \pm 1$, for $3 \le i\le N$. This allows us to iteratively construct the ground state from the three-site subsystem configurations. The ground state is found by minimizing the consecutive energy increments 
\begin{equation}
    \Delta E_i = h_is_i + \sum^{r-1}_{j=1} J_{i-r+j,i}s_{i-r+j}s_i + \sum^{r-1}_{\substack{j,k=1\\j<k}} K_{i-r+j,i-r+k,i}s_{i-r+j}s_{i-r+k}s_i,
\end{equation}
with spin indices within $1, \dots,i$. At each step $i$, we generate all possible configurations $\bm{s}_i=\{\pm 1\}^r$, updating the previously computed state by retaining only the configuration with minimum total energy. Therefore, this method scales as $\mathcal{O}[2^rN]$, offering an efficient protocol for moderate $r$ values but converging to a brute force method when $r\to N$.

\section*{Supplementary Note 3: Circuit decomposition}\label{app:circuit}

An important aspect of preparing and running quantum circuits on hardware is to transpile the required quantum operations according to the corresponding native gate sets provided by the platform. These typically consist of an universal gate set containing several one-qubit gates and a single two-qubit entangling gate. For \textsc{ibm\_fez}, its native gate set is composed by
\begin{equation}
    X=\begin{bmatrix}0&1\\1&0\end{bmatrix},\qquad\sqrt{X}=\frac{1}{2}\begin{bmatrix}1+i & 1-i\\ 1-i& 1+i\end{bmatrix},\qquad \text{RZ}(\theta)=e^{-i\frac{\theta}{2}\sigma^z},
\end{equation}
with the entangling gate $\text{CZ}=\diag(1,1,1,-1)$. For the \textsc{IonQ Aria 1} platform, the native gate set is given by~\footnote{See~\href{https://ionq.com/docs/getting-started-with-native-gates}{https://ionq.com/docs/getting-started-with-native-gates}.}
\begin{equation}
    \text{GPi}(\phi) = \begin{bmatrix} 0 & e^{-i\phi} \\ e^{i\phi} & 0 \end{bmatrix}, \qquad\text{GPi2}(\phi) = \frac{1}{\sqrt{2}} \begin{bmatrix} 1 & -ie^{-i\phi} \\ -ie^{i\phi} & 1 \end{bmatrix},
\end{equation}
with entangling gate the so-called partially entangling M{\o}lmer-S{\o}renson gate
\begin{equation}
    \text{MS}(\phi_0,\phi_1,\theta) = \begin{bmatrix}
        \cos\frac{\theta}{2} & 0 & 0 & -ie^{-i(\phi_0 + \phi_1)}\sin\frac{\theta}{2} \\
        0 & \cos\frac{\theta}{2} & -ie^{-i(\phi_0 - \phi_1)}\sin\frac{\theta}{2} & 0 \\
        0 & -ie^{-i(\phi_0 - \phi_1)}\sin\frac{\theta}{2} & \cos\frac{\theta}{2} & 0 \\
        -ie^{-i(\phi_0 + \phi_1)}\sin\frac{\theta}{2} & 0 & 0 & \cos\frac{\theta}{2}
    \end{bmatrix}.
\end{equation}

The constrained connectivity among qubits of \textsc{ibm\_fez}, and quantum platforms with heavy-hex coupling maps in general, suggests to work with highly sparse problems. The linearity of the Hamiltonians considered naturally fits into the heavy-hex IBM qubit coupling maps, whose corresponding circuit decomposition can be seen in~\figlabel{fig:circ_nn}. To overcome this issue, higher connectivities between qubits are required. Among others, trapped ion-based platforms, like \textsc{IonQ Aria 1}, offer a natural architecture with which to acquire all-to-all connectivities with very precise gate fidelities. Therefore, these platforms can also be used for addressing denser problems.
\begin{figure}[!tb]
\centering
\includegraphics[width=.6\linewidth]{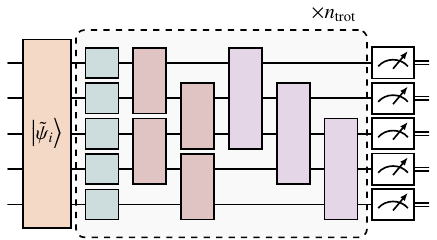}%
\caption{\textbf{Circuit decomposition of a five-qubit NN HUBO instance.} The first one-qubit operator layer prepares the initial ground state $\ket{\tilde{\psi}} = \bigotimes_{i=1}^N R_y(\theta_i)\ket{0}_i$. Afterwards, $n_\text{trot}$ Trotter steps are applied to implement the digitized time-evolution operator of Eq.~(7) in the main text. Finally, all qubits are measured, and their energy distribution is used for updating the bias field of the subsequent iteration.}\label{fig:circ_nn}
\end{figure}%

\section*{Supplementary Note 4: DMRG estimation of ground state energies}\label{app:dmrg}

To obtain optimal energies of the nearest-neighbour Hamiltonian of Eq.~(13) in the main text, we also test the DMRG algorithm~\cite{Schollwock2011}, which is an optimization method using a matrix product state (MPS) as a variational ansatz. Due to entanglement area laws~\cite{eisert2010colloquium}, DMRG works particularly well for one-dimensional systems described by MPSs and for local Hamiltonians, and could be used in cases where an exact solution is not available. We use the implementation in the ITensor Julia library~\cite{itensor}. As the initial state, we choose a random MPS with bond dimension $\chi > 1$ since setting the initial bond dimension to one results in higher energies of the final state. We apply a noise parameter in the DMRG sweeps to improve convergence in the presence of many local minima~\cite{white2005density}. For the value shown in Fig. 3 in the main text, we find convergence to a product state ($\chi = 1$) with 10 sweeps, a maximum initial bond dimension $10$, a truncation error cutoff of $10^{-10}$, and a noise parameter that is gradually reduced from $10^{-3}$ to zero through the sweeps. The energy $E_{\text{DMRG}} = -236.07$ shown in Fig.~3 in the main text is close to the exact ground-state energy $E_{\text{exact}} = -236.18$, indicating that DMRG converges to a low-lying excited state. Note that the random coupling constants are fixed here, i.e. there is no averaging over disorder configurations. 

The DMRG algorithm is known to be particularly efficient when the Hamiltonian has an exact MPO representation with a small bond dimension~\cite{Schollwock2011}. This occurs for Hamiltonians with short-range coupling terms such as considered here, or exponentially decaying ones. 
The convergence of DMRG, as a variational method, may however be prevented by the algorithm getting stuck in local minima.
In previous DMRG studies of classical spin glasses, a quantum wavefunction annealing method was found to improve convergence~\cite{rodriguez-laguna2007quantum, rodriguez-laguna2007density}.
As an alternative to DMRG, the ground state search can be performed by imaginary time evolution. This approach was recently used for solving QUBO problems via projected entangled pair states with flexible geometry~\cite{patra2024projected}.

\bibliography{bibfile}